\documentclass[a4paper,11pt]{article}

\usepackage[british]{babel}

\usepackage{graphicx}
\usepackage[dvips]{geometry}
\usepackage{float}
\usepackage{array}
\usepackage{amsmath}
\usepackage{amsfonts}
\usepackage{epstopdf}
\usepackage{psfrag}
\usepackage{graphicx}
\usepackage{amssymb}
\usepackage{amsmath}
\usepackage{rotating}
\usepackage{pstricks,pst-fill,pst-text,pst-node,pstricks-add}
\usepackage{lmodern}
\usepackage{tikz}
\usetikzlibrary{decorations.pathreplacing}
\usetikzlibrary{decorations.markings}

\usepackage[normalem]{ulem}

\usepackage[curve,matrix,arrow,color]{xy}
\usepackage{hyperref}

\usepackage[numbers,sort&compress]{natbib}
\nocite{TitlesOn}

\usepackage{color}

\makeatletter
\@addtoreset{equation}{section}
\makeatother

\newcommand{\Ket}[1]{| #1  \rangle}

\newcommand{\Braket}[1]{\langle #1  \rangle}

\newcommand\be            {\begin{equation}}
\newcommand\ee            {\end{equation}}

\newcommand{\half}{%
  \mathchoice{\ffrac{1}{2}}{\frac{1}{2}}{\frac{1}{2}}{\frac{1}{2}}}

\newcommand{\one}{\boldsymbol{1}}

\newcommand{\q}{\mathfrak{q}}
\newcommand{\ffrac}[2]{\mbox{\footnotesize$\displaystyle\frac{#1}{#2}$}}

\newcommand{\dimm}{\mathrm{dim}}
\newcommand{\BX}{\mathcal{X}}

\newcommand{\BW}{\mathcal{W}}

\newcommand{\brt}{\mathsf{br}}

\newcommand{\step}{\epsilon}

\usepackage{rotating}
\usepackage{multirow}

\newcommand{\bz}{\bar{z}}

\newcommand{\aSt}{\mathcal{W}}
\newcommand{\aX}{\mathcal{X}}

\newcommand{\ferm}{\theta}
\newcommand{\fermd}{\theta^{\dagger}}
\newcommand{\sferm}{\psi}
\newcommand{\sfermp}{\psi^{2}}
\newcommand{\sfermm}{\psi^{1}}
\newcommand{\bsfermp}{\bar{\psi}^{2}}
\newcommand{\bsfermm}{\bar{\psi}^{1}}
\newcommand{\phip}{\phi^{2}}
\newcommand{\phim}{\phi^{1}}

\newcommand{\nord}[1]{\boldsymbol{\colon}\!\!{#1}\boldsymbol{\colon}\!\!}

\newcommand{\Hbr}{H^{\brt}}
\newcommand{\Pbr}{P^{\brt}}
\newcommand{\ebr}{e^{\brt}}

\newcommand{\gl}{\mathfrak{gl}}

\newcommand{\fbr}{\mathbb{F}}

\newcommand{\oZ}{\mathbb{Z}}

\newcommand{\PTL}[1]{\mathsf{TL}^a_{#1}}
\newcommand{\TL}[1]{\mathsf{TL}_{#1}}

\newcommand{\rJTL}[1]{\mathsf{JTL}_{#1}}

\newcommand{\ATL}[1]{\mathsf{T}^a_{#1}}

\newcommand{\StTL}[1]{\mathcal{W}_{#1}}
\newcommand{\IrrTL}[1]{\mathcal{X}_{#1}}

\newcommand{\StJTL}[2]{\mathcal{W}_{#1,#2}}
\newcommand{\bStJTL}[2]{\overline{\mathcal{W}}_{#1,#2}}

\newcommand{\AStTL}[2]{\aSt_{#1,#2}}
\newcommand{\bAStTL}[2]{\overline{\mathcal{W}}_{#1,#2}}
\newcommand{\IrrJTL}[2]{\mathcal{X}_{#1,#2}}
\newcommand{\VirN}{\boldsymbol{\mathcal{V}}}
\newcommand{\Verma}[1]{\mathsf{V}_{#1}}
\newcommand{\IrrV}[1]{\mathsf{X}_{#1}}
\newcommand{\IrrVb}[1]{\overline{\mathsf{X}}_{#1}}

\newcommand{\bSt}{\BW}

\begin{document}

\title{On the correspondence between boundary and bulk lattice models and (logarithmic) conformal field theories }

\author{J. Bellet\^ete$^{1}$, A.M. Gainutdinov$^{2}$, J.L.  Jacobsen$^{1,3,4}$, H. Saleur$^{1,5}$, R. Vasseur$^{6,7}$\\
[3.0mm]
\footnotesize  ${}^1$Institut de Physique Th\'eorique, CEA Saclay,
  91191 Gif Sur Yvette, France \\
\footnotesize  ${}^2$Laboratoire de Math\'ematiques et Physique Th\'eorique CNRS,\\
\footnotesize Universit\'e de Tours, Parc de Grammont, 37200 Tours,  France\\
\footnotesize  ${}^3$Laboratoire de physique th\'eorique, D\'epartement de physique de l'ENS, \'Ecole normale sup\'erieure, \\
\footnotesize  UPMC Univ.~Paris 06, CNRS, PSL Research University, 75005 Paris, France \\
\footnotesize ${}^4$Sorbonne Universit\'es, UPMC Univ.~Paris 06, \'Ecole normale sup\'erieure, \\
\footnotesize  CNRS, Laboratoire de Physique Théorique (LPT ENS), 75005 Paris, France \\
\footnotesize  ${}^5$Department of Physics,
  University of Southern California, Los Angeles, CA 90089-0484\\
\footnotesize  ${}^6$Department of Physics, University of California, Berkeley, Berkeley CA 94720, USA \\
\footnotesize  ${}^7$Materials Science Division, Lawrence Berkeley National Laboratory, Berkeley CA 94720, USA}

\date{}

\maketitle

\begin{abstract}

The relationship between  bulk and boundary properties is one of the founding features of (rational) conformal field theory (CFT).
Our goal in this paper is to explore the possibility of having an  equivalent  relationship in the context of  lattice models.
We focus on models based on the  Temperley-Lieb algebra, and use  the concept of ``braid translation'', which is a natural way, in physical terms, to ``close'' an open spin chain by adding an interaction between the first and last spins using braiding to ``bring'' them next to each other. The interaction thus obtained is in general non-local, but has the key feature that it is expressed solely in terms of the algebra for the open spin chain -- the ``ordinary'' Temperley-Lieb algebra and its blob algebra generalization. This is in contrast with the usual periodic spin chains which involve only local interactions, and are described by the periodic Temperley-Lieb algebra. 
We show that for the Restricted Solid-On-Solid (RSOS) models, which are known to be described by minimal unitary CFTs (with central charge $c<1$) in the continuum limit, the braid translation in fact {\sl does provide} the ordinary periodic model starting from the open model with fixed (identical) boundary conditions on the two sides of the strip. This statement has a precise mathematical formulation, which is a pull-back map between irreducible modules of, respectively, the blob algebra and the affine Temperley-Lieb (ATL) algebra.
We then turn to the same kind of analysis for two models whose continuum limits are logarithmic CFTs (LCFTs) --- the alternating   $\mathfrak{gl}(1|1)$ and $\mathfrak{sl}(2|1)$  spin chains. We find that  the result for minimal models does not hold any longer: braid translation of the relevant (in that case, indecomposable but not irreducible) modules of the TL algebra {\sl does not} give rise to the modules known to be present in the periodic chains. In  the $\mathfrak{gl}(1|1)$ case, the content in terms of the irreducibles is the same, as well as the spectrum,  but the detailed structure (like logarithmic coupling) is profoundly different.  This carries over to the continuum limit. The situation is similar for the $\mathfrak{sl}(2|1)$ case. The problem of relating bulk and boundary lattice models for LCFTs thus remains open.

\end{abstract}

\newpage
\setcounter{tocdepth}{2}
\tableofcontents

% ---------------------------------------------------------------------------------------- %
%                                    INTRODUCTION
% ---------------------------------------------------------------------------------------- %

\section*{Introduction}

The properties of bulk and boundary  rational conformal field theories (RCFTs) are known to be profoundly related  \cite{ZuberPetkova, Fuchsetal1,Fuchsetal4,Fuchsetal5,Fuchsetali}. This relation can be understood at various levels. In \cite{ZuberPetkova}  for instance, the same graphs are found to classify both the torus and cylinder partition functions, a  direct consequence of the ubiquity of the $S$-matrix (encoding the modular transformation of the characters) which appears both in the fusion algebra of the bulk theory and in the classification of conformal boundary conditions via solutions of the Cardy equations.  A rigorous mathematical treatment of this fact is based on the use of the same modular tensor category for constructing both the boundary and bulk RCFTs, see more in~\cite{Fuchsetal5,Fuchsetal4}.

In simple terms, the relationship between bulk and boundary CFTs can be traced back to the fact that  boundary CFTs  explore the chiral sector of the  theories, while bulk CFTs are built up  properly combining  chiral and antichiral sectors: to make a caricature of it,  the bulk is obtained by  ``glueing''  two boundary theories. 

While here exposed very sketchily, the existence of this glueing relation is intriguing from a physical point of view. This is even more the case since work of the last many years has exhibited ever deeper similarities between properties of CFTs and their lattice regularizations.  These similarities   were used recently to accomplish significant progress~\cite{GJRSVreview} in the study of Logarithmic CFTs (LCFTs)~\cite{2013JPhA} where  direct, abstract approaches at the field-theoretic level encounter considerable difficulties, while a down-to-earth analysis of lattice models sheds light on questions such as indecomposable representations of the Virasoro algebra, values of the indecomposability parameters, or fusion of primary fields. In view of the recent progress in the understanding of boundary LCFTs \cite{RS3,PRZ}, a deeper understanding of a lattice analog of the boundary/bulk correspondence might indeed be able to clarify the still very difficult problem of understanding their bulk counterparts.

\begin{figure}
\centering
\begin{tikzpicture}[scale = 4/5, every node/.style = {scale = 4/5}]
%The chains
	\draw[black, line width = 3pt] (1,3) -- (6,3);
	\draw[black, line width = 3pt] (1,-3) -- (6,-3);
	\draw[black, dashed, line width = 3pt] (1,0) -- (2,0);
	\draw[black, dashed, line width = 2pt] (1,-3) .. controls (1,-4) and (6,-4) .. (6,-3);
%The lines
	\foreach \r in {1,...,5}{
		\draw[black, line width = 1pt] (\r,3) .. controls  (\r,2) and (\r+1,1) .. (\r +1 , 0) .. controls (\r + 1, -1) and (\r , -2) .. (\r, -3);
	\draw[decoration = {
				markings,
				mark= at position (9/10) with {\arrow[scale = 3]{>}}
				},decorate] (\r,3) .. controls  (\r,2) and (\r+1,1) .. (\r +1 , 0);
	\draw[decoration = {
				markings,
				mark= at position (9/10) with {\arrow[scale = 3]{>}}
				},decorate] (\r +1 , 0) .. controls (\r + 1, -1) and (\r , -2) .. (\r, -3);
	}
	\draw[white, line width = 4pt] (6,3) .. controls (6,2) and (1,1) .. (1,0) .. controls (1,-1) and (6,-2) .. (6,-3); 
	\draw[red, line width = 1pt] (6,3) .. controls (6,2) and (1,1) .. (1,0) .. controls (1,-1) and (6,-2) .. (6,-3); 
	\draw[decoration = {
				markings,
				mark= at position (9/10) with {\arrow[scale = 3,red]{>}}
				},decorate] (6,3) .. controls (6,2) and (1,1) .. (1,0);
	\draw[decoration = {
				markings,
				mark= at position (9/10) with {\arrow[scale = 3,red]{>}}
				},decorate] (1,0) .. controls (1,-1) and (6,-2) .. (6,-3);
%The nodes on the chains
	\foreach \r [count = \i] in {6,1,2,...,5}{
		\filldraw[white] (\r, 3) circle (5pt);
		\draw[line width = 1pt] (\r, 3) circle (5pt);
		\node at (\r, 3) {\small{\r}}; 
		\filldraw[white] (\r, -3) circle (5pt);
		\draw[line width = 1pt] (\r, -3) circle (5pt);
		\node at (\r, -3) {\small{\r}}; 
		\filldraw[white] (\i, 0) circle (5pt);
		\draw[line width = 1pt] (\i, 0) circle (5pt);
		\node at (\i, 0) {\small{\r}}; 
	};
% The arrow pointing to the ellipse
	\draw[black, line width = 3pt, |->] (7,0) -- (9,0);
        \node[anchor=south] at (8,0) {\large$\brt{}$};
%The closed spin chain
	\draw[black, line width = 2pt] (11,0) ellipse (40 pt and 20 pt);
	\foreach \r in {1,...,6}{
	\draw[decoration = {
		markings,
		mark= at position (\r/6) with \coordinate (z\r);,
		},decorate] (11,0) ellipse (40 pt and 20 pt);
	};
	\draw[white, line width = 12pt] (z1) -- (z6);
	\draw[black, dashed, line width = 2pt] (11,0) ellipse (40 pt and 20 pt);
	\foreach \r in {1,...,6}{
		\filldraw[white] (z\r) circle (5pt);
		\draw[line width = 1pt] (z\r) circle (5pt);
		\node at (z\r) {\small{\r}}; 
	};
%The original open chain, vertical
%	\draw[black, line width = 3pt] (-3,3) -- (-3,-3);
%	\foreach \r [count = \i]in {0,...,5}{
%		\filldraw[white] (-3,{-3 + (\r * 7)/6 }) circle (5pt);
%		\draw[line width = 1pt] (-3,{-3 + (\r * 7)/6 }) circle (5pt);
%		\node at (-3,{-3 + (\r * 7)/6 }) {\small{\i}}; 
%	};
% The arrow pointing from the open chain
%	\draw[black, line width = 3pt, |->] (-2,0) -- (0,0);
\end{tikzpicture}
%    \includegraphics[width=3cm]{cartoon1}
%    \hskip1cm\includegraphics[width=4cm]{cartoon2}
%    \hskip1cm\includegraphics[width=4cm]{cartoon3}
   \caption{
   The interaction between neighbours within the open chain are represented by the black lines. The blob adds an interaction between the first two positions; this new operator is then moved down to the last site by successive braidings, resulting in a closed chain.  This process of making the closed chain is called \textit{braid translation} and denoted by the arrow with $\brt$.
   }\label{cartoonbraidtrans.}
\end{figure}
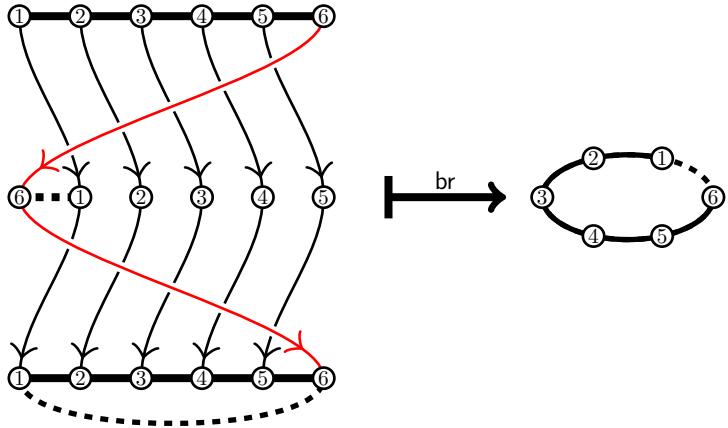

Our goal in this paper is therefore to explore in more detail the relationship between boundary and bulk CFTs on the lattice. While there might be ways to ``glue'' left and right sectors explicitly, we take a different route here, 
 based on the use of the ``braid translation''  which was introduced in~\cite{MartinSaleur}. To understand the general idea, recall that the focus of the ``algebraic associative approach''  to CFTs \cite{RS3} consists in studying the properties of the algebra of local energy densities for a finite lattice model, and to infer from these---based on, ultimately, a categorical
equivalence~\cite{GS-cat}---the properties  of the chiral algebra that appears in the continuum limit. In the simplest of all cases, to which we will restrict here, the lattice algebra for a system with boundaries is the Temperley-Lieb algebra (TL) (or its close cousin, the blob algebra) and  the continuum limit is the Virasoro algebra. It is easy to make up a lattice model for the bulk theory simply by eliminating boundaries, that is by focusing on closed chains instead of open ones. The corresponding lattice algebra is then, instead,  the periodic Temperley-Lieb algebra. While this cannot   be considered as a ``glueing'' of two ordinary Temperley-Lieb algebras, it is possible to obtain many of its representations by a sort of imaging procedure. This procedure exploits the fact that the TL generators can be used to  exchange (braid) in a consistent way the lattice degrees of freedom present on the sites of the chain. A periodic chain is obtained from an open one by introducing an interaction between the first and last sites: the ``braid translation'' provides such an interaction by taking a model where there is a boundary interaction between the first two sites (given by the blob generator) and successively braiding the lattice sites, only using the interactions between neighbors on the open chain. This is represented graphically in  Figure  \ref{cartoonbraidtrans.}.

The procedure may  at first seem artificial since the interaction thus obtained between the first and last lattice sites is built in a way that is different from the others, and apparently of a non-local nature. But it is in fact well known \cite{Braid1,Braid3} to ``reproduce'' the familiar periodic Ising and three-state Potts model chains starting from the corresponding open chains with suitable boundary conditions. We show in section \ref{sec:minmodels}  of this paper (after the necessary algebraic preliminaries in section \ref{sec:algebra}) that a similar results holds for {\sl all RSOS minimal models}. In algebraic terms, these models correspond to different irreducible representations of the periodic Temperley-Lieb algebra, and we demonstrate that these representations are isomorphic to those obtained by braid translation  of boundary RSOS models which are irreducible representations of the ordinary Temperley-Lieb and blob algebras.

Along the way, we establish some important results for which we did not find proofs in the literature, including a result on the decomposition of the RSOS periodic Temperley-Lieb algebra representations onto irreducibles, and also generating series for number of boundary RSOS paths.

We then discuss what happens in the continuum limit, and show that the equivalent process amounts to going from irreducible representations of a single Virasoro algebra to the  representations of the left and right Virasoro algebras that are the building blocks of the minimal models Hilbert spaces, hence realizing on the lattice a possible  equivalent of the left-right glueing of the continuum limit.

We next turn, in section \ref{sec:susy}  of the paper, to what happens in the logarithmic case, by studying in great details the $\mathfrak{gl}(1|1)$ spin chain made of alternating fundamental and anti-fundamental $\mathfrak{gl}(1|1)$-representations.  This spin chain leads~\cite{GRS1} to a symplectic fermions LCFT with $c=-2$ in the continuum limit. The picture we obtain is rather intriguing: the braid translation of the open spin chain (with free boundary conditions) ``almost'' gives the right result for the periodic model, in the sense that the spectrum of the Hamiltonian and  the degeneracies of the states are correct, {\em i.e.}, exactly those of the periodic $\mathfrak{gl}(1|1)$ spin chain. Nonetheless, the representations of the periodic Temperley-Lieb algebra thus obtained are {\sl not isomorphic} to those that occur in the genuine periodic spin chain. This leads, in the continuum limit, to a theory where the left and right parts of the Virasoro algebra are not properly glued: the structure of the modules is simply not the correct one \cite{GRS1,GRS2,GRS3}. This  has very direct physical consequences, leading for instance to incorrect values of the indecomposability parameters $\beta$ in one (left or right) of the chiral sectors of the theory. 

Like before, we establish along the way several interesting results. In particular, we provide analytical calculations of the indecomposability parameters $\beta$  in the symplectic fermions theory using the $\mathfrak{gl}(1|1)$ spin chain regularization. We also 
discuss  in subsection \ref{specialadvert}  the possibility of modifying the realization of the Virasoro algebra in terms of symplectic fermions to alter the value of these parameters, and provide a physical interpretation of the corresponding field theory. 

The observations on the braid translation in the $\mathfrak{gl}(1|1)$  case  are generalized to the case of the $\mathfrak{sl}(2|1)$ spin chain (which is related with the statistics of percolation hulls, and a LCFT with central charge $c=0$~\cite{SaleurPoly,MathieuRidout,SQHE,GRSV}) with totally similar results. 

These results, in hindsight, are somewhat natural. Braid translation works for minimal models because there are so few irreducible representations of the blob and periodic Temperley-Lieb algebras. In contrast, in the logarithmic case, non semi-simplicity guarantees a great variety of possible representations. The experience gathered in the last few years~\cite{GRS2,GRS3,GRSV}  suggests that the representations contributing to periodic models, in the logarithmic case, are considerably more complicated than those known so far to occur in the open case \cite{RS3}. It is thus not surprising that braid translation, in our present state of knowledge, does not allow us to connect straightforwardly models with open and periodic boundary conditions. On the other hand, it is important to emphasize that conformal boundary conditions for logarithmic CFTs are not fully classified: it might well be that there are more yet to be discovered, which would be related to bulk LCFTs by braid translation. This possibility and its potential consequences are discussed in more detail in the conclusion section. 

\medskip

The overview of the main text has been given above. In addition, we have relegated substantial material to the appendices. The detailed calculations for the braid translated $\mathfrak{gl}(1|1)$ spin chain can be found in Appendix~\ref{app:A}. We express the corresponding Hamiltonian and momentum operator in terms of fermionic mode operators and examine the difference between the naive and braid translated periodic generators. In Appendix~\ref{app:momentum-spin} we recall how to measure conformal data from the finite-size scaling of lattice quantities. The long Appendix~\ref{app:gl11} is dedicated to the periodic $\mathfrak{gl}(1|1)$ chain, for which we determine the indecomposability parameters (logarithmic couplings) in finite size, using both numerical and analytical approaches. This requires carefully fixing normalizations and extracting lattice versions of the Virasoro generators. In Appendix~\ref{app:comb} we present results on the enumeration of RSOS paths (Dyck paths between two walls), using generating function techniques. As a side result we obtain a few combinatorial identities that appear new. Finally, in Appendix~\ref{app:inv-brt} we discuss the construction of inverse braid translation, in which periodic $\mathfrak{gl}(1|1)$ and XXZ spin chains are transformed into open chains with a highly non-local expression for the blob operator.

\subsection*{Notations}

To help the reader navigate throughout the paper, we provide a partial list of notations (consistent with  other papers in the series \cite{GRS1,GRS2,GRS3,GRSV,GJSV}):
\begin{itemize}

\item$\TL{N}(\delta)$ --- the (ordinary) Temperley--Lieb algebra on $N$ sites with loop fugacity $\delta$,

\item $\mathcal{B}_N(\delta,y)$ --- the blob algebra on $N$ sites with the blob parameter $y$,

\item$\ATL{N}(\delta)$ --- the affine Temperley--Lieb algebra,

\item$\rJTL{N}(\delta)$ --- the Jones--Temperley--Lieb algebra, a finite-dimensional quotient of $\ATL{N}(\delta)$,

\item$\StTL{j}$ --- the standard modules over $\TL{N}(\delta)$, 

\item$\IrrTL{j}$ --- the simple modules over $\TL{N}(\delta)$, 

\item$\AStTL{j}{z^2={\rm e}^{2iK}}$ --- the standard modules over $\ATL{N}(\delta)$,

\item$\bAStTL{0}{\q^2}$ --- the standard module over $\rJTL{N}(\delta)$ for $j=0$,

\item $\IrrJTL{j}{z^2}$  --- the simple modules over $\ATL{N}(\delta)$,

\item $\BW^{b/u}_{j}$ ---  the standard modules over the blob algebra,
\item  $\BX^{b/u}_{j}$  --- the simple modules over the blob algebra,

\item $\Verma{r,s}$ --- Verma modules with the conformal weight $h_{r,s}$ where $r$ and $s$ are the Kac labels,
\item $\IrrV{r,s}$ --- irreducible Virasoro modules of the conformal dimension $h_{r,s}$,

\item$\VirN$  --- the product of left and right Virasoro algebras,

\item
$\IrrV{r,s}\otimes\IrrVb{r',s'}$ --- irreducible $\VirN$-modules,
\item$\chi_{r,s}:=\chi_{r,s}(q)$  --- characters of  $\IrrV{r,s}$ with the formal variable $q$,
 
 \item $F^{(0)}_{j,z^2}$  ---  characters or $(L_0,\bar{L}_0)$-graded dimensions of $\IrrJTL{j}{z^2}$ in the continuum limit.

\end{itemize}

\section{Algebraic preliminaries: some definitions}\label{sec:algebra}

The models we shall be interested in are related with various representations of the Temperley-Lieb algebra (TL) and its cousin the ``blob algebra%
\footnote{Also called one-boundary Temperley-Lieb algebra.}'' in the boundary case, and of the 
``periodic Temperley-Lieb algebra'' in the bulk case. There are strong reasons to believe that these algebras are the lattice analogs of the (chiral) Virasoro algebra and of the product of left and right Virasoro algebras respectively\footnote{More precisely for the bulk case, the periodic Temperley-Lieb algebra is the lattice analog of a certain extension of the left-right Virasoro algebra that is called the interchiral algebra in~\cite{GRS3}.}. Understanding the potential relationships between the ordinary TL algebra and the periodic TL algebra should shed thus some light on the situation in the corresponding conformal field theories.

\subsection{The boundary case}
\label{sec:blob-def}

The Temperley-Lieb algebra $\TL{N}(\delta)$ is fundamental in the definition of a large class of open integrable systems.
The algebra $\TL{N}(\delta)$  consists of all the words written with the $N-1$ generators $e_i$ ($1 \leq i \leq N-1$), subject to the relations
\begin{subequations} \label{TLdef}
\begin{eqnarray}
\left[ e_i , e_j \right] &=&0 \ (\left|i-j \right| \geq 2 )\\
e_i ^2 &=& \delta e_i\\
e_i e_{i \pm 1} e_i &=& e_i.
\end{eqnarray}
\end{subequations}
The algebra $\TL{N}(\delta)$ may be defined as the algebra of diagrams, with $N$ sites at the top
and $N$ sites at the bottom of a rectangle. The sites can be contracted in pairs, without crossing. The action of $e_i$ as an operator acting on $N$ strands has a well known graphical representation
where  the parameter %$n $
$\delta $ is interpreted as the weight carried by closed loops. Here are a few examples of this representation for $N=3$:
\begin{eqnarray}
e_{1}  &=&
\begin{tikzpicture}[scale=1/4, baseline = {(current bounding box.center)}]
	\draw[black, line width = 1pt] (0,0) .. controls (0,1) and (1,1) .. (1,0);
	\draw[black, line width = 1pt] (0,3) .. controls (0,2) and (1,2) .. (1,3);
	\draw[black, line width = 1pt] (2,0) -- (2,3);
	%I typically put borders aroung my diagrams but those already there didn't have them; de-commenting the commented lines in the diagrams will put back the borders.
%	\draw[black, line width = 2pt] (-.5,0) -- (2.5,0);
%	\draw[black, line width = 2pt] (-.5,3) -- (2.5,3);
%	\foreach \r in {0,...,2}{
%		\filldraw[red] (\r,0) circle (2 pt);
%		\filldraw[red] (\r,3) circle (2 pt);
%	};
\end{tikzpicture} \quad, \qquad \qquad \quad
e_{2} = 
\begin{tikzpicture}[scale=1/4, baseline = {(current bounding box.center)}]
	\draw[black, line width = 1pt] (1,0) .. controls (1,1) and (2,1) .. (2,0);
	\draw[black, line width = 1pt] (1,3) .. controls (1,2) and (2,2) .. (2,3);
	\draw[black, line width = 1pt] (0,0) -- (0,3);
%	\draw[black, line width = 2pt] (-.5,0) -- (2.5,0);
%	\draw[black, line width = 2pt] (-.5,3) -- (2.5,3);
%	\foreach \r in {0,...,2}{
%		\filldraw[red] (\r,0) circle (2 pt);
%		\filldraw[red] (\r,3) circle (2 pt);
%	};
\end{tikzpicture}\quad, \nonumber \\[2mm]
e_{1}e_{2}e_{1} &=&  
\begin{tikzpicture}[scale=1/4, baseline = {(current bounding box.center)}]
	\draw[black, line width = 1pt] (0,0) .. controls (0,1) and (1,1) .. (1,0);
	\draw[black, line width = 1pt] (0,3) .. controls (0,2) and (1,2) .. (1,3);
	\draw[black, line width = 1pt] (2,0) -- (2,3);
	\draw[black, line width = 1pt] (1,3) .. controls (1,4) and (2,4) .. (2,3);
	\draw[black, line width = 1pt] (1,6) .. controls (1,5) and (2,5) .. (2,6);
	\draw[black, line width = 1pt] (0,3) -- (0,6);
	\draw[black, line width = 1pt] (0,6) .. controls (0,7) and (1,7) .. (1,6);
	\draw[black, line width = 1pt] (0,9) .. controls (0,8) and (1,8) .. (1,9);
	\draw[black, line width = 1pt] (2,6) -- (2,9);
%	\foreach \s in {0,3,6,9}{
%		\draw[black, line width = 2pt] (-.5,\s) -- (2.5,\s);
%		\foreach \r in {0,...,2}{
%			\filldraw[red] (\r,\s) circle (2 pt);
%		};
%	};
\end{tikzpicture}
= e_{1}, \qquad
e_{2}e_{1}e_{2} =  
\begin{tikzpicture}[scale=1/4, baseline = {(current bounding box.center)}]
	\draw[black, line width = 1pt] (1,0) .. controls (1,1) and (2,1) .. (2,0);
	\draw[black, line width = 1pt] (1,3) .. controls (1,2) and (2,2) .. (2,3);
	\draw[black, line width = 1pt] (0,0) -- (0,3);
	\draw[black, line width = 1pt] (0,3) .. controls (0,4) and (1,4) .. (1,3);
	\draw[black, line width = 1pt] (0,6) .. controls (0,5) and (1,5) .. (1,6);
	\draw[black, line width = 1pt] (2,3) -- (2,6);
	\draw[black, line width = 1pt] (1,6) .. controls (1,7) and (2,7) .. (2,6);
	\draw[black, line width = 1pt] (1,9) .. controls (1,8) and (2,8) .. (2,9);
	\draw[black, line width = 1pt] (0,6) -- (0,9);
%	\foreach \s in {0,3,6,9}{
%		\draw[black, line width = 2pt] (-.5,\s) -- (2.5,\s);
%		\foreach \r in {0,...,2}{
%			\filldraw[red] (\r,\s) circle (2 pt);
%		};
%	};
\end{tikzpicture}
= e_{2}, \qquad
e_{1}e_{1} = 
\begin{tikzpicture}[scale=1/4, baseline = {(current bounding box.center)}]
	\draw[black, line width = 1pt] (0,0) .. controls (0,1) and (1,1) .. (1,0);
	\draw[black, line width = 1pt] (0,3) .. controls (0,2) and (1,2) .. (1,3);
	\draw[black, line width = 1pt] (2,0) -- (2,3);
	\draw[black, line width = 1pt] (0,3) .. controls (0,4) and (1,4) .. (1,3);
	\draw[black, line width = 1pt] (0,6) .. controls (0,5) and (1,5) .. (1,6);
	\draw[black, line width = 1pt] (2,3) -- (2,6);
%	\foreach \s in {0,3,6}{
%		\draw[black, line width = 2pt] (-.5,\s) -- (2.5,\s);
%		\foreach \r in {0,...,2}{
%			\filldraw[red] (\r,\s) circle (2 pt);
%		};
%	};
\end{tikzpicture}
= \delta e_{1}. \label{N3TLexamples}
\end{eqnarray}
Similarly, one can  define  in terms of diagrams a special class of TL representations -- the so-called standard modules
$\StTL{j}$, which are irreducible for $\q$ generic. They are  spanned by link diagrams with $2j$ through-lines
and the  action of the TL generators on these states is  interpreted as stacking
the corresponding diagrams, with the rule that contracting a pair of through-lines results in zero. 

To define the blob algebra $\mathcal{B}_N(\delta,y)$, consider now  all the words
written with the $N-1$ generators~$e_i$  and an extra generator $b$, subject to the additional relations
\begin{subequations} \label{Blobdef}
\begin{eqnarray}
b ^2 &=& b,\\
e_1 b e_1 &=& y e_1,\\
e_{i}b &=& b e_{i}, \quad  i>1.
\end{eqnarray}
\end{subequations}
The extra boundary
operator $b$ can be interpreted as  decorating lines with a ``blob'' 
$\bullet$.
 It gives to the corresponding ``blobbed loops''  a different weight $y$~\cite{MartinSaleur}.

In the following, we shall parametrize $\delta=\q+\q^{-1}$ where $\q$ is a complex number, and restrict to the case where $|\q|=1$. We also set $\q=\mathrm{e}^{i \gamma}$, with
  $\gamma = \frac{\pi}{x+1}$. It is then useful to represent $y$ as\footnote{In earlier papers \cite{JS}, the label $n$ was called  $r$ instead, because of its subsequent interpretation in terms of  the Kac label of Virasoro algebra representations in the continuum limit. We use another notation here, as this interpretation is different in the boundary and periodic cases.}     %
  \begin{equation}
\label{eqDefr}
\displaystyle y=\frac{\sin  (n+1) \gamma }{\sin  n \gamma } = \frac{ \left[ n+1 \right]_{\q}}{ \left[ n \right]_{\q}},
\end{equation}
where $n$ is a real number and  we   introduced  the $\q$-numbers
\begin{equation}
\displaystyle \left[ x\right]_{\q} = \dfrac{\q^{x}-\q^{-x}}{\q-\q^{-1}}.
\end{equation}

 \subsubsection{Standard modules of $\mathcal{B}_N(\delta,y)$}
The standard (or  generically irreducible) modules of the blob algebra are best described in terms of  link states, similarly to the TL standard modules $\StTL{j}$. They are parametrized by the number of through lines $2j$ (with $0\leq 2j\leq N$) propagating through the system, {\it i.e.}, extending from
the bottom to the top in the diagrammatic representation; there is $2j=1$ such line in the examples (\ref{N3TLexamples}). 
In addition, the standard modules carry a label $b$ or $u$ corresponding to the two orthogonal projectors --- the blob operator $b$ and the unblob operator $u=1-b$ --- and diagrammatically the lines can be decorated by blobs, with the leftmost through-line either ``blobbed'' or  ``unblobbed''. The diagrammatic meaning of these operators
was explained in~\cite{MartinSaleur,MartinWoodcock}.
 We then denote $\BW^b_0$ the standard module with no through lines, and $\BW^{b}_{j}$ ({\it resp.} $\BW^{u}_{j}$)
the standard module with $2j$ through lines in the blobbed ({\it resp.}, unblobbed) sector.
Following~\cite{MartinSaleur},
the action of the algebra on link states can be obtained geometrically by stacking the diagrams
on top of one another, just as for the Temperley-Lieb algebra, with the convention
that the TL generators $e_i$ cannot contract two through lines, otherwise the result of the action is zero.
The blobbed and unblobbed standard modules
have the same dimension
\begin{equation} \label{eqDimStdblob}
\hat d_j \equiv \mathrm{dim} \BW^{b/u}_{j} = \binom{N}{N/2 - j} \ .
\end{equation}

Degeneracies (or non-trivial submodules) in these representations occur when $n$ is integer and for any non-zero $\q$; if $\q$ is a root of unity the structure of submodules is quite involved. This is discussed in detail in \cite{MartinWoodcock,GJSV}, and the  features relevant for this paper will be recalled as needed.
When the standard modules $\bSt^{b/u}_{j}$ are not irreducible (this happens only if $n$ is integer) they are indecomposable and have a unique irreducible quotient that we denote by $\BX^{b/u}_{j}$. 
We refer to~\cite{MartinWoodcock}  and to the review in~\cite{GJSV}, where it is discussed how to construct the simple blob algebra modules  $\BX^{b/u}_{j}$ out of the standard $\bSt^{b/u}_{j}$ modules.

\subsection{The periodic case}
\label{sec:ATL-def}

When one tries to generalize the TL algebra to periodic models, the most natural thing to do is to simply
add a last generator $e_{N}$ and to define the labels modulo $N$ so that $e_{N+1} = e_1$, $e_{N} e_1 e_{N} = e_1$,
\mbox{{\it etc}}. (The resulting algebra $\PTL{N}(\delta)$ is a quotient of the affine Hecke algebra of $A$-type.) It is infinite dimensional, even for finite $N$, and considerably  too big for our purpose.
In a nutshell, the problem is due to winding through lines and presence of non-contractible loops, and we shall now give precise statements about how to resolve this issue.

In order to define the relevant quotients,  it is useful to go again to a diagram representation, which now involves an annulus instead of a rectangle. We also introduce
 a translation generator $u$ that shifts the label
of the $e_i$ generators, giving rise to the following relations, with integer indices considered modulo $N$ (that is, $i\in\oZ_N$):
\begin{subequations} \label{TLpdef}
\begin{eqnarray}
e_i ^2 &=& \delta e_i\\
e_i e_{i \pm 1} e_i &=& e_i\\
\left[ e_i , e_j \right] &=&0 \quad (\left|i-j \right| \geq 2 )\\
u e_i u^{-1} &=& e_{i+1}\label{TLpdef-d}\\
u^2 e_{N-1} &=& e_{1} \dots e_{N-1}\label{TLpdef-e}\ .
\end{eqnarray}
\end{subequations}
The translation operator has the diagrammatic representation
\begin{equation*}
u = 
\begin{tikzpicture}[scale=1/3, baseline = {(current bounding box.center)},yscale=-1]
	\foreach \r in {1,2,5,6}{
	\draw[black, line width = 1pt] (\r,0) .. controls (\r,1) and (\r-1,2) .. (\r-1,3);
	};
	\draw[black, line width = 1pt] (0,0) .. controls (0,1) and (-1,2) .. (-1,3);
	\draw[black, line width = 1pt] (7,0) .. controls (7,1) and (6,2) .. (6,3);
	\filldraw[white] (-.5,0) rectangle (-1.5,3);
	\filldraw[white] (6.5,0) rectangle (7.5,3);
	\node[anchor = north] at (3,1) {$\hdots$};
%	\foreach \s in {0,3}{
%		\draw[black, line width = 2pt] (-.5,\s) -- (6.5,\s);
%	};
%	\foreach \r in {0,1,2,5,6}{
%			\filldraw[red] (\r,0) circle (2 pt);
%		};
%	\foreach \r in {0,1,4,5,6}{
%			\filldraw[red] (\r,3) circle (2 pt);
%		};
\end{tikzpicture}.
\end{equation*}
Note that the last relation
% $u^2 e_{N-1} = e_{1} \dots e_{N-1}$ 
is easily understood in terms of diagrams, for example for $N=4$,
\begin{equation*}
 e_{1}e_{2}e_{3} = \quad
 \begin{tikzpicture}[scale=1/4, baseline = {(current bounding box.center)},yscale=-1]
	\draw[black, line width = 1pt] (0,0) .. controls (0,1) and (1,1) .. (1,0);
	\draw[black, line width = 1pt] (0,3) .. controls (0,2) and (1,2) .. (1,3);
	\draw[black, line width = 1pt] (2,0) -- (2,3);
	\draw[black, line width = 1pt] (3,0) -- (3,3);
	\draw[black, line width = 1pt] (1,3) .. controls (1,4) and (2,4) .. (2,3);
	\draw[black, line width = 1pt] (1,6) .. controls (1,5) and (2,5) .. (2,6);
	\draw[black, line width = 1pt] (0,3) -- (0,6);
	\draw[black, line width = 1pt] (3,3) -- (3,6);
	\draw[black, line width = 1pt] (2,6) .. controls (2,7) and (3,7) .. (3,6);
	\draw[black, line width = 1pt] (2,9) .. controls (2,8) and (3,8) .. (3,9);
	\draw[black, line width = 1pt] (0,6) -- (0,9);
	\draw[black, line width = 1pt] (1,6) -- (1,9);
\end{tikzpicture}
\quad = \quad
\begin{tikzpicture}[scale=1/4, baseline = {(current bounding box.center)},yscale=-1]
	\draw[black, line width = 1pt] (0,0) .. controls (0,1) and (1,1) .. (1,0);
	\draw[black, line width = 1pt] (2,3) .. controls (2,2) and (3,2) .. (3,3);
	\draw[black, line width = 1pt] (2,0) .. controls (2,1) and (0,2) .. (0,3);
	\draw[black, line width = 1pt] (3,0) .. controls (3,1) and (1,2) .. (1,3);
\end{tikzpicture}
\quad = \quad 
\begin{tikzpicture}[scale=1/4, baseline = {(current bounding box.center)},yscale=-1]
	\foreach \r in {0,...,4}{
		\draw[black, line width = 1pt] (\r,0) .. controls (\r,1) and (\r-1,2) .. (\r-1,3);
		\draw[black, line width = 1pt] (\r,3) .. controls (\r,4) and (\r-1,5) .. (\r-1,6);
	};
	\filldraw[white] (-.5,0) rectangle (-1.5,6);
	\filldraw[white] (3.5,0) rectangle (4.5,6);
	\draw[black, line width = 1pt] (2,6) .. controls (2,7) and (3,7) .. (3,6);
	\draw[black, line width = 1pt] (2,9) .. controls (2,8) and (3,8) .. (3,9);
	\draw[black, line width = 1pt] (0,6) -- (0,9);
	\draw[black, line width = 1pt] (1,6) -- (1,9);
\end{tikzpicture} 
\quad =
u^{2}e_{3}\ .
\end{equation*}

\bigskip

We note that $u^{N}$ is central.
The resulting  algebra is again infinite dimensional, and called the \textit{affine Temperley--Lieb} algebra $\ATL{N}(\delta)$. The infinite dimensionality, can now be understood easily: there are no relations in the algebra that permit us to undo non contractible loops, or through lines that wind around the periodic direction.

 \subsubsection{Standard modules of  $\ATL{N}(\delta)$}
The finite dimensional modules which are relevant to us are obtained by considering the diagrams on the annulus (recall they do not involve crossings), again with $2j$ through lines, and imposing additional relations. By convention, the evolution operator (transfer matrix or Hamiltonian) of the system acts on the outer rim of the annulus. With $j$ fixed, it follows that the arcs that connect the inner rim to itself are immaterial for the action of the evolution operator and we can hence ignore them. It is therefore sufficient consider ``uneven diagrams'' with $N$ points on the outer rim and only the $2j$ end points of the through lines on the inner rim; the latter $2j$ points are then called ``free sites''.  
Note that, for a given
non-zero value of $j$, it is possible using the  action of the algebra,  to cyclically
permute the free sites, that is to say, to rotate the inner rim of the annulus with respect to the outer rim. These permutations give rise to the introduction of a
pseudomomentum $K$, as follows: Whenever $2j$ through lines wind counterclockwise around
the annulus $l$ times, we can decide to unwind them at the price of a factor
${\rm e}^{2ijlK}$; similarly, for clockwise winding, the phase is ${\rm e}^{-i 2jlK}$ \cite{MartinSaleur}.
 More formally,  we impose the relation~\cite{GrahamLehrer}  (for appropriate and unique $k\in\oZ$)
\begin{equation*}
\mu=\mu'\circ u_j^k \equiv z^{k}\mu',\qquad z=\mathrm{e}^{iK}\ ,
\end{equation*} 
where $\mu$ is a given uneven diagram with $2j$ through lines and a certain winding of a number of through lines,  $\mu'$ is the ``unwound'' diagram which has no through
lines winding the annulus and $u_j$ is the
translational operator acting on the $2j$ sites of the inner rim of $\mu'$, or in other words $u_j^k$  stands for the cyclical permutation necessary for obtaining the unwound diagram $\mu'$.
This action gives rise to a generically
irreducible module over $\ATL{N}(\delta)$, which we denote by
$\AStTL{j}{z^2=\mathrm{e}^{2 iK}}$ and call the
\textit{standard} modules. (We are actually going to use the subalgebra in $\ATL{N}$ generated by the $e_i$'s and by $u^2$, so our notation refers to $z^2$ rather than $z$.)
 
 The dimensions of the standard modules $\AStTL{j}{\mathrm{e}^{2 iK}}$   are then given by
 \begin{equation}\label{eq:dj}
 \hat d_{j}=
 \binom{N}{N/2+j},\qquad j>0\ .
 \end{equation}
Note that the dimensions do not depend on $K$, although the representations with
different ${\rm e}^{iK}$ are not isomorphic.

The case $j=0$ is a bit special. There is no pseudomomentum, but representations are still characterized by another parameter, related with  the weight given to  non contractible loops. Parametrizing this weight as $z+z^{-1}$, the corresponding standard module of  $\ATL{N}(\delta)$ is denoted  $\AStTL{0}{z^2}$ and has  dimension $\binom{N}{N/2}$. 
 These modules are irreducible  for generic $z$. 
  As in the case $j>0$, we indicate only the $z^2$ value, though it does not mean that the two standard modules with $\pm z$ are isomorphic. We will indicate the sign of $z$ when it is necessary.

It is well-known~\cite{GrahamLehrer} that when the standard modules $\StJTL{j}{z^2}$ are not irreducible (this happens for particular conditions on the twist $z^2$) they are indecomposable and have a unique irreducible quotient that we denote by 
$\IrrJTL{j}{z^2}$; see also the review in~\cite[Sec. 3.2 and 3.3]{GRSV}. 
In more details, the standard module $\StJTL{k}{y^2}$ has a non-zero homomorphism to another standard module $\StJTL{j}{z^2}$
\begin{equation}\label{cell-emb}
\StJTL{k}{y^2}\hookrightarrow\StJTL{j}{z^2}
\end{equation}
if and only if  $k-j=\ell$ for a non-negative integer $\ell$ and the pairs $(k,y^2)$ and $(j,z^2)$ satisfy
\begin{equation}\label{eq:emb-cond}
y  = z(-\q)^{-\epsilon \ell} \qquad \text{and}\qquad z^2 = (-\q)^{2\epsilon k},\qquad \text{for} \quad  \epsilon=\pm1.
\end{equation}
Note that we then have $z^2 = (-\q)^{2\epsilon \ell} y^2$.
This homomorphism has a trivial kernel  (see the description of injective homomorphisms in~\cite[Thm.~3.4 and  the proof of Thm.~5.1]{GrahamLehrer})
and we thus have embeddings of modules, as it is indicated in the diagram~\eqref{cell-emb}.
For generic values of $\q$, {\em i.e.}, not roots of unity, there is at most one solution of the equations~\eqref{eq:emb-cond} and thus the standard module $\StJTL{j}{z^2}$ is either irreducible (if there is no solution) or contains a unique proper irreducible submodule isomorphic to $\StJTL{k}{y^2}$.
One can then obtain the simple $\ATL{N}(\delta)$-modules  $\IrrJTL{j}{z^2}$ by taking the quotient by this submodule. In particular, the dimension of $\IrrJTL{j}{z^2}$ in this case is the difference $ \hat d_{j} -  \hat d_{k}$ of the two binomials.

 From the above, we have that  when $z=\pm \q^{\pm1}$ (and for generic $\q$)  $\StJTL{0}{z^2}$ is  no longer irreducible and contains a unique simple submodule isomorphic to
$\StJTL{1}{1}$ while the (unique) simple quotient is the module where the diagrams  (with no through lines) are identified if they  connect the same sites, and we denote it by $\bStJTL{0}{\q^2}$. The latter is the standard module for a certain finite-dimensional quotient   of $\ATL{N}(\delta)$ called \textit{Jones-Temperley-Lieb} algebra $\rJTL{N}(\delta)$.
It appears in models such as the ones defined in \cite{ReadSaleur01} (and studied in details in~\cite{GRS1,GRS2,GRS3, GRSV}) and  it is obtained by:
\begin{itemize}
 \item[(i)]  replacing non contractible loops by the same weight $n$ as for the contractible ones,
 \item[(ii)] setting $u^N=1$, which allows one to unwind through lines, and 
 \item [(iii)] identifying non-isotopic (in the annulus) diagrams connecting the same sites.
\end{itemize}
These additional constraints lead to the pseudomomentum $K=\pi \ell/M$ where $M | j$ and with a greatest common divisor $\ell \wedge M =1$, together with $z^2=\q^{\pm 2}$ for $j=0$. Moreover, in the case $j=0$,  the affine TL module $\AStTL{0}{\q^2}$ becomes reducible, and  identifying non-isotopic diagrams corresponds to the quotient $\bAStTL{0}{\q^2}$  of dimension 
\begin{equation}
\overline{d}_0 = \dim \bAStTL{0}{\q^2}= \binom{N}{N/2} - \binom{N}{N/2+1}\ ,
\end{equation}
This quotient is the standard module of $\rJTL{N}(\delta)$ for $j=0$.

\newcommand{\Hilb}{\mathcal{H}}
\subsection{A spin-chain model for affine TL}\label{sec:twistedXXZ}
Although the geometrical setup of loop models is appealing, it is instructive and useful to consider related periodic
models. First, a comment about vocabulary. The name ``periodic'' in this paper arises from  the role played by the periodicized Temperley-Lieb algebra introduced earlier. It does not necessarily imply that the lattice models providing the representations of this algebra will appear  truly periodic in the sense that all their interactions (involved in the transfer matrix or Hamiltonian) will be identical when going around the $N$ sites. The XXZ chain below, for instance, is usually referred to as ``twisted'' rather than periodic. When this distinction is important, we will sometimes called such systems ``closed''.

We thus define here a ``periodic version'' of the XXZ spin chain~\cite{PasquierSaleur} (or equivalently of the 6-vertex model). It is well known
that the 6-vertex model provides a natural representation of the Temperley-Lieb algebra, where the generators read 
\begin{equation}
e_i = \mathbb{I} \otimes \mathbb{I} \otimes \dots \otimes \left( \begin{array}{cccc}
0 & 0 & 0 & 0 \\
0 & \q^{-1} & -1 & 0 \\
0 & -1 & \q & 0 \\
0 & 0 & 0 & 0 \end{array} \right) \otimes \dots \otimes \mathbb{I},
\label{e_eiXXZ}
\end{equation}
acting on the ``Hilbert'' space $\mathcal{H} = (\mathbb{C}^2)^{\otimes N}$.
We define the last generator as
\begin{equation}
e_N = \mathbb{I} \otimes \mathbb{I} \otimes \dots \otimes \left( \begin{array}{cccc}
0 & 0 & 0 & 0 \\
0 & \q^{-1} & -\mathrm{e}^{i\phi} & 0 \\
0 & -\mathrm{e}^{-i\phi} & \q & 0 \\
0 & 0 & 0 & 0 \end{array} \right) \ ,
\label{e_eiXXZ2}
\end{equation}
where the first tensorand $\mathbb{I}$ now refers to the {\em second} space, whereas the non-trivial $4\times 4$ matrix acts on the spins at sites $i$ and $i+1$, here with $i=N$ and the indices
considered modulo $N$. The parameter $\phi$ is called the twist; we shall relate it to the pseudomomentum $K$ below.

The resulting Hamiltonian, $H=-\sum_{i} e_i$, reads, up to an irrelevant additive constant $N\frac{\q+\q^{-1}}{4}$, 
\begin{equation}\label{Ham-twist}
\displaystyle H =  \half  \sum_{i=1}^{N-1} \left( \sigma^x_i \sigma^x_{i+1} + \sigma^y_i \sigma^y_{i+1}  \right) + \mathrm{e}^{i\phi} \sigma_{N}^{+}  \sigma_{1}^{-}
+ \mathrm{e}^{-i\phi} \sigma_{N}^{-}  \sigma_{1}^{+} +  \half\sum_{i=1}^{N} \ffrac{q + q^{-1}}{2} \sigma^z_i \sigma^z_{i+1}\ ,
\end{equation}
where we  assume that $\sigma^z_{N+1}:=\sigma^z_{1}$.
Following the usual terminology, we shall also refer to this model as the \textit{twisted XXZ spin chain}.
Note that it may be useful to redistribute the twist across all nearest neighbor bonds in order to restore
translation invariance.

We also introduce  the translation (by one site to the right) operator
$u$  as 
\begin{equation}\label{XXZ-trans}
u = i^{N} {\rm e}^{-i\frac{\phi}{2}\sigma^z_1}s_1 s_2\dots s_{N-1},
\end{equation}
where   $s_i$ is the permutation operator of the $i$th and $(i+1)$th sites:
\begin{equation}
s_i = \ffrac{1}{2}\bigl(1+  \sigma^x_i \sigma^x_{i+1} + \sigma^y_i \sigma^y_{i+1} + \sigma^z_i \sigma^z_{i+1}\bigr),\qquad 1\leq i\leq N-1.
\end{equation}
Noticing that~\eqref{e_eiXXZ2} can be rewritten as
\begin{equation}\label{eq:eN-XXZ}
e_{N} = - \ffrac{1}{2} {\rm e}^{-i\frac{\phi}{2}\sigma^z_1}\Bigl( \sigma^x_{N} \sigma^x_{1} + \sigma^y_{N} \sigma^y_{1}  + \ffrac{\q + \q^{-1}}{2} \bigl(\sigma^z_{N} \sigma^z_{1}-1\bigr)  + \ffrac{\q - \q^{-1}}{2} \bigl(\sigma^z_{N} - \sigma^z_{1}\bigr)\Bigr) {\rm e}^{i\frac{\phi}{2}\sigma^z_1}
\end{equation}
it is straightforward to check that~\eqref{e_eiXXZ} and~\eqref{eq:eN-XXZ} together with~\eqref{XXZ-trans} satisfy the relations~\eqref{TLpdef} of the affine TL algebra $\ATL{N}(\delta)$ with $\delta = \q + \q^{-1}$. We call this representation \textit{the twisted XXZ} representation 
of $\ATL{N}$.

The total spin operator, $S_z=\half\sum_{j=1}^N \sigma^z_j$, is the symmetry of the affine TL representation, {\em i.e.}, we have  $[e_j,S_z]=0$ for $1\leq j\leq N$, and $[u,S_z]=0$. We thus have a decomposition  $\Hilb = \bigoplus_{j=-N/2}^{N/2} \Hilb_{S_z = j}$ onto a direct sum of 
affine TL representations.

 The choice of the twist 
$\mathrm{e}^{i\phi}=\mathrm{e}^{2iK}$ in~\eqref{e_eiXXZ2} for generic values of $\phi$ allows one to select specific generic
irreducible representations of the affine TL algebra as fixed $S_z$ sectors $\Hilb_{S_z}$. For $j \neq 0$, the  space of states of this model
in the sector with the total spin $S_z=j$ is isomorphic to the standard module
$\aSt_{j,z^2= \mathrm{e}^{\mp 2i(K+N\pi)}}$, where `$-$' is for positive $j$ and  `$+$' is for negative values of $j$. 

We show this relation between $\Hilb_{S_z}$ and the standard modules by first computing the action of the central element $u^N$. We find
\begin{equation}
	u^{k} = i^{kN} {\rm e}^{-i\frac{\phi}{2}\sum_{j=1}^{k} \sigma^{z}_{j}} (s_1s_2\dots s_{N-1})^{k},
\end{equation}
and so we have $u^{N} = i^{N^{2}} {\rm e}^{-i\phi S_z}$. Therefore, in the sector with $S_z=j$, the $z^2$-parameter in the corresponding standard module $\aSt_{|j|,z^2}$ (which is irreducible for generic $\phi$ and $\q$) is thus a $|j|$'th root of $i^{N^{2}}{\rm e}^{-i j\phi}$. To specify precisely which root, we study the spectrum of the operator
\begin{equation}\label{eq:taujdef}
	\tau_{|j|} = e_1e_3\hdots e_{N-2|j|-1}u.
\end{equation}
The only non-zero eigenvalue of this operator on $\aSt_{|j|,z^2}$ is $z$.
For example 
\begin{equation}
	\tau_{1} \quad
	\begin{tikzpicture}[scale=1/3, baseline = {(current bounding box.center)},yscale=-1]
	\draw[black, line width = 1pt] (0,0) .. controls (0,1) and (1,1) .. (1,0);
	\draw[black, line width = 1pt] (2,0) .. controls (2,1) and (3,1) .. (3,0);
	\draw[black, line width = 1pt] (4,0) .. controls (4,1) and (0,2) .. (0,3);
	\draw[black, line width = 1pt] (5,0) .. controls (5,1) and (1,2) .. (1,3);
	%These are the border lines
%	\draw[black, line width = 2pt] (-.5,0) -- (5.5,0);
%	\draw[black, line width = 2pt] (-.5,3) -- (1.5,3);
%	\foreach \r in {0,...,5}{
%		\filldraw[red] (\r,0) circle (2 pt);
%	};
%	\foreach \r in {0,1}{
%		\filldraw[red] (\r,3) circle (2 pt);
%	};
	\end{tikzpicture} 
 \quad = \quad
\begin{tikzpicture}[scale=1/3, baseline = {(current bounding box.center)},yscale=-1]
	\draw[black, line width = 1pt] (0,0) .. controls (0,1) and (1,1) .. (1,0);
	\draw[black, line width = 1pt] (2,0) .. controls (2,1) and (3,1) .. (3,0);
	\draw[black, line width = 1pt] (4,0) .. controls (4,1) and (0,2) .. (0,3);
	\draw[black, line width = 1pt] (5,0) .. controls (5,1) and (1,2) .. (1,3);
	%The tau
	\draw[black, line width = 1pt] (0,-3) .. controls (0,-2) and (1,-2) .. (1,-3);
	\draw[black, line width = 1pt] (2,-3) .. controls (2,-2) and (3,-2) .. (3,-3);
	\draw[black, line width = 1pt] (0,0) .. controls (0,-.5) and (-.25,-1) .. (-.5,-1);
	\draw[black, line width = 1pt] (5,0) .. controls (5,-.5) and (5.15,-1) .. (5.5,-1);
	\draw[black, line width = 1pt] (1,0) .. controls (1,-1) and (2,-1) .. (2,0);
	\draw[black, line width = 1pt] (4,-3) .. controls (4,-2) and (3,-1) .. (3,0);
	\draw[black, line width = 1pt] (5,-3) .. controls (5,-2) and (4,-1) .. (4,0);
	%These are the border lines
%	\draw[black, line width = 2pt] (-.5,0) -- (5.5,0);
%	\draw[black, line width = 2pt] (-.5,-3) -- (5.5,-3);
%	\draw[black, line width = 2pt] (-.5,3) -- (1.5,3);
%	\foreach \r in {0,...,5}{
%		\filldraw[red] (\r,0) circle (2 pt);
%		\filldraw[red] (\r,-3) circle (2 pt);
%	};
%	\foreach \r in {0,1}{
%		\filldraw[red] (\r,3) circle (2 pt);
%	};
	\end{tikzpicture} 
\quad = z \quad
	\begin{tikzpicture}[scale=1/3, baseline = {(current bounding box.center)},yscale=-1]
	\draw[black, line width = 1pt] (0,0) .. controls (0,1) and (1,1) .. (1,0);
	\draw[black, line width = 1pt] (2,0) .. controls (2,1) and (3,1) .. (3,0);
	\draw[black, line width = 1pt] (4,0) .. controls (4,1) and (0,2) .. (0,3);
	\draw[black, line width = 1pt] (5,0) .. controls (5,1) and (1,2) .. (1,3);
	%These are the border lines
%	\draw[black, line width = 2pt] (-.5,0) -- (5.5,0);
%	\draw[black, line width = 2pt] (-.5,3) -- (1.5,3);
%	\foreach \r in {0,...,5}{
%		\filldraw[red] (\r,0) circle (2 pt);
%	};
%	\foreach \r in {0,1}{
%		\filldraw[red] (\r,3) circle (2 pt);
%	};
	\end{tikzpicture}.
	\label{ex-eigval-tau1}
\end{equation}
In $\mathcal{H}_{j}$, we find
\begin{equation}\label{eq:tauj-eigenv}
	\tau_{|j|} w_{j} = \lambda_j w_{j} \qquad \text{with} \qquad \lambda_j = 
			i^{2(N-|j|)} {\rm e}^{-i \frac{\phi}{2}\rm{sign}{(j)}} \ ,
\end{equation}
and we set
\begin{equation}
	w_{|j|} = v \otimes v \otimes \dots \otimes v\otimes |\underbrace{\uparrow \uparrow\dots \uparrow}_{2|j| \text{ spins}} \rangle, \qquad w_{-|j|} = v \otimes v \otimes \dots \otimes v\otimes |\underbrace{\downarrow \downarrow\dots \downarrow}_{2|j| \text{ spins}} \rangle,
\end{equation}
\begin{equation}\label{eq:def.v}
	v = |\uparrow \downarrow \rangle - \q |\downarrow \uparrow \rangle.
\end{equation}
The configuration $v$ was chosen because 
\begin{equation}
	e_{2} (v\otimes|\uparrow\rangle) = e_{2}(|\uparrow\downarrow\uparrow\rangle - \q |\downarrow \uparrow \uparrow \rangle ) =  \q |\uparrow \downarrow \uparrow \rangle - |\uparrow \downarrow \uparrow \rangle = - |\uparrow\rangle \otimes v, 
\end{equation}
so that 
\begin{equation}
	e_{2}e_{4}\hdots e_{N-2|j|}w_{|j|} = (-1)^{\frac{N-2|j|}{2}}|\uparrow\rangle \otimes v\otimes \hdots \otimes v \otimes |\underbrace{\uparrow \hdots \uparrow}_{2|j| - 1}\rangle.
\end{equation}
It follows from \eqref{eq:tauj-eigenv} that  $z = \lambda_j$ in the sector $S_z = j$ and we indeed have this sector isomorphic to $\aSt_{|j|,\mathrm{e}^{\mp 2i(K+N\pi)}}$, where `$-/+$' is for positive/negative $j$. This analysis was done for generic values of $\phi$, {\em i.e.}, when ${\rm e}^{i\phi}$ is not %a certain 
an integer power of~$\q$. For the non-generic values of the twist, we have two cases:
if the integer in question is positive, the sector is still a standard module but might be indecomposable and reducible, while it is the contragredient of a standard module if the integer number is negative. A detailed proof of this result will appear in a future publication.

The case $j=0$ requires more care as one has to be careful about the loops that wrap around the spatial direction. 
Their  weight can be accounted for by studying the spectrum of 
\begin{equation}\label{eq:tau}
\tau_0 = e_1 e_3 \ldots  e_{N-1} u\ .
\end{equation}
The only non-zero eigenvalue of this operator on $\aSt_{0,z^2}$ is the weight $z+z^{-1}$ of non-contractible loops. By a direct calculation we find
\begin{equation}\label{eq:tau0-eigenv}
	\tau_{0} w_{0} =  2 \cos(\phi/2)  w_{0} \ ,
\end{equation}
where $w_0 = v \otimes v \otimes \dots \otimes v$. We have thus the weight of non-contractible loops $z+z^{-1} = 2\cos{\frac{\phi}{2}}$ and 
we get  $\mathcal{H}_{S_z=0} \simeq \StJTL{0}{e^{ i\phi}}$ which is an irreducible representation of $\ATL{N}$ for generic values of the twist $\phi$.
The only degenerate case corresponds to the weight~$z+z^{-1} = \q^n + \q^{-n}$ or for the  twist  such that $\mathrm{e}^{i\phi}=\q^{\pm2n}$. For $\mathrm{e}^{i\phi}=\q^{+2n}$, the sector $S_z=0$ corresponds to the  standard module  $\StJTL{0}{\q^{2n}}$ (which is indecomposable and contains $\StJTL{n}{1}$ as a submodule, recall  Sec.~\ref{sec:ATL-def}), while for $\mathrm{e}^{i\phi}=\q^{-2n}$ the sector  is  not $\StJTL{0}{\q^{2n}}$ but  its contragredient representation, {\em i.e.},
 $\mathcal{H}_{S_z=0} \cong (\StJTL{0}{\q^{2n}})^*$. In the latter case, $\StJTL{n}{1}$ is not a submodule but the (unique) simple quotient.
 We will give a proof of this result in a future publication.

\subsection{Braid translation}
The braid translation is a tool that allows us to generate new periodic models out of the boundary ones, or more formally, new affine TL algebra
representations from representations of the ordinary TL and blob algebras.
Starting from (a representation of) the  blob algebra $\mathcal{B}_N(\delta,y)$
one defines the braid operators
\begin{equation}\label{hecke-gen}
\displaystyle g_i^{\pm 1 } = 1 - \q^{\mp 1 } e_i.
\end{equation}
These operators satisfy the braid relations
\begin{equation}\label{br-rel}
\displaystyle g_i g_{i\pm 1} g_i = g_{i\pm 1} g_i g_{i\pm 1}.
\end{equation}
The key point observed in~\cite{MartinSaleur} is that the generator defined by\footnote{We note that the expression in~\cite{MartinSaleur} appears slightly different, but they are actually equal because of the identity
$(1+\alpha b)(1-\q^{-1}e_1)e_1(1-\q e_1)(1+\beta b) = (1-\q^{-1}e_1)(1+\beta b) e_1 (1+\alpha b)(1-\q e_1)$ in the blob algebra. Our expression is more convenient for expressing $e_{N}$ as $u^{-1}e_1 u$.}
\begin{equation}\label{eq_defbraid}
\displaystyle
e_{N} = \left( \prod_{i=1}^{N-1} g_i \right)^{-1} \left( \alpha b + 1 \right) e_1 \left(1 + \beta b\right)\prod_{i=1}^{N-1} g_i,
\end{equation}
where  $\prod_{i=1}^n g_i = g_1 g_2\ldots g_n$, and where we have set
\begin{equation}\label{alpha-beta}
\displaystyle \alpha \equiv \alpha(\q) =  \frac{\q-\q^{-1}}{\q^{-1}-y}, \qquad \beta =\alpha(\q^{-1}),
\end{equation}
obeys the ``triple'' relations,  $e_{N} e_i e_{N} = e_{N}$ and $e_{i} e_{N} e_{i} = e_i$ for $i=1$ and $N-1$, as well as $e_{N}^2=(\q+\q^{-1}) e_{N}$ and $[e_N,e_i] = 0$ for $2 \le i \le N-2$, of the affine TL algebra $\ATL{N}(\delta)$.

Recall then that the $\ATL{N}$ is generated by the $e_j$'s and the translation element $u$ together with the defining relations~\eqref{TLpdef}. We can go further and define a translation operator as 
\begin{equation}\label{eq_braid_trans}
\displaystyle
u = (-1)^{N/2}\q^{N/2}\sqrt{\frac{y-\q}{y-\q^{-1}}}(1+\beta b)\prod_{i=1}^{N-1}g_i
\end{equation}
which obeys
\begin{equation}
\displaystyle u e_{i} u^{-1} = e_{i+1},\qquad 1\leq i\leq N \mod N, 
\end{equation}
along with the relation
\begin{equation}
u^2 e_{N-1} = e_{1} \dots e_{N-1}. \label{u2eN-1rel}
\end{equation}
The relation $u^{-1}e_{1}u=e_{N}$ readily follows from the expression~\eqref{eq_defbraid}, while the last one (\ref{u2eN-1rel}) can be easily proven by an induction and using the braid relations~\eqref{br-rel}.

We note that there is a useful way to represent the action of the braid operators $g_i$ graphically as 
\begin{equation*}
(-\q)^{1/2}g_{i} = \quad
\begin{tikzpicture}[scale=1/3,baseline={(current bounding box.center)}]
	\foreach \r in {0,3,6,9}{
		\draw[black, line width = 1pt] (\r,0) -- (\r,3);
	};
	\node[anchor = south] at (1.5,1) {$\hdots$};
	\node[anchor = south] at (7.5,1) {$\hdots$};
	\node[anchor = north] at (4,0) {\tiny{i}};
	\node[anchor = north] at (5,0) {\tiny{i+1}};
	\draw[black, line width = 1pt] (4,0) .. controls (4,1) and (5,2) .. (5,3);
	\draw[white, line width = 3pt] (5,0) .. controls (5,1) and (4,2) .. (4,3);
	\draw[black, line width = 1pt] (5,0) .. controls (5,1) and (4,2) .. (4,3);
\end{tikzpicture}\quad, \qquad
(-\q)^{-1/2}g_{i}^{-1} = \quad
\begin{tikzpicture}[scale=1/3,baseline={(current bounding box.center)}]
	\foreach \r in {0,3,6,9}{
		\draw[black, line width = 1pt] (\r,0) -- (\r,3);
	};
	\node[anchor = south] at (1.5,1) {$\hdots$};
	\node[anchor = south] at (7.5,1) {$\hdots$};
	\node[anchor = north] at (4,0) {\tiny{i}};
	\node[anchor = north] at (5,0) {\tiny{i+1}};
	\draw[black, line width = 1pt] (5,0) .. controls (5,1) and (4,2) .. (4,3);
	\draw[white, line width = 3pt] (4,0) .. controls (4,1) and (5,2) .. (5,3);
	\draw[black, line width = 1pt] (4,0) .. controls (4,1) and (5,2) .. (5,3);
\end{tikzpicture}\quad.
\end{equation*}

For instance, we can represent the translation operator $u$ and its inverse,  up to the normalization,  in terms of diagrams as
\begin{equation*}
u \propto (1+ \beta b)\quad
\begin{tikzpicture}[scale=1/3,baseline={(current bounding box.center)},yscale=-1]
	\foreach \r in {1,2,5,6}{
		\draw[black, line width = 1pt] (\r,0) .. controls (\r,1) and (\r - 1, 2) .. (\r - 1,3);
	};
	\node[anchor = north] at (3.5,0) {$\hdots$};
	\node[anchor = south] at (2.5,3) {$\hdots$};
	\draw[white, line width = 3pt] (0,0) .. controls (0,1) and (6,2) .. (6,3);
	\draw[black, line width = 1pt] (0,0) .. controls (0,1) and (6,2) .. (6,3);
\end{tikzpicture}\quad , \qquad
u^{-1} \propto \quad
\begin{tikzpicture}[scale=1/3,baseline={(current bounding box.center)},yscale=-1]
	\foreach \r in {1,2,5,6}{
		\draw[black, line width = 1pt] (\r - 1 ,0) .. controls (\r - 1,1) and (\r, 2) .. (\r,3);
	};
	\node[anchor = north] at (2.5,0) {$\hdots$};
	\node[anchor = south] at (3.5,3) {$\hdots$};
	\draw[white, line width = 3pt] (6,0) .. controls (6,1) and (0,2) .. (0,3);
	\draw[black, line width = 1pt] (6,0) .. controls (6,1) and (0,2) .. (0,3);
\end{tikzpicture}\quad \left(1- \frac{\beta}{1+\beta} b\right).
\end{equation*}
Then the diagrammatic expression for $e_N$ given in~\eqref{eq_defbraid} is coherent with  Figure~\ref{cartoonbraidtrans.}.

This construction can be used to generate a representation of the affine TL algebra $\ATL{N}(\delta)$ starting from a representation of $\mathcal{B}_N(\delta,y)$.
In more mathematical terms, \textit{the braid translation} is an algebra homomorphism\footnote{We will later omit $y$ and write simply $\brt$ when it is clear what value of $y$ is used.}
\begin{equation}\label{eq:brt}
\brt(y)\colon \ATL{N}(\delta) \longrightarrow \mathcal{B}_N(\delta,y)\ ,
\end{equation}
 and thus
having a representation of the blob algebra $\rho :\mathcal{B}_N(\delta,y) \longrightarrow \mathrm{End}(W)$ (with $W$ a vector space, for instance
$W = V^{\otimes N}$ for a spin chain), we get the representation of  the affine TL algebra $\ATL{N}(\delta)$ as the composition $\rho \circ \brt(y): \ATL{N}(\delta) \longrightarrow \mathrm{End}(W)$ on the same vector space $W$. We call this kind of $ \ATL{N}(\delta)$ representation a \textit{braid-translation generated} representation. This is a standard construction in mathematics and is called `pull-back'.

\medskip

It is worth noticing that  $u^{N}$ is central:
 it  acts on irreducible representations proportionally to the identity operator $\one$.
If we parametrize
\begin{equation}\label{y-eta}
\displaystyle y = \frac{\q^{-1}-\q \mathrm{e}^{2 i \eta}}{1- \mathrm{e}^{2 i \eta}},
\end{equation}
one can show that  in the standard representation with $2|j|$ through lines (we use the convention $j>0$ in the blobbed sector, and $j<0$ in the unblobbed one) its action reads \begin{equation}
\displaystyle u^{N} =  ((-\q)^{2j} e^{2i\eta})^j \one.
\end{equation}
In other words, the factor picked by $2j$ through lines  as they wind one time around the annulus is given by
\begin{equation}
\alpha_{2j}=((-\q)^{2j} e^{2i\eta})^j.
\end{equation}

Finally for the sector with zero through lines, $j=0$, the braid translation generates affine TL representations where each non-contractible loop should be replaced by the factor
\begin{equation}\label{weight-loop}
\alpha_0 = e^{i\eta}+e^{-i\eta}.
\end{equation}

The general result from braid translation is thus for the standard modules, for $j\geq 0$,
\begin{gather}
\bSt^b_j \; \xrightarrow{\quad \brt(y)\quad} \;\aSt_{j,(-\q)^{2j}e^{2i\eta}}\,, \label{corr:st}\\
\bSt^u_j \; \xrightarrow{\quad \brt(y)\quad} \;\aSt_{j,(-\q)^{2j}e^{-2i\eta}}\,,
\end{gather}
and for their simple quotients
\begin{equation}
\BX^{b/u}_j \; \xrightarrow{\quad \brt(y)\quad} \;\IrrJTL{j}{(-\q)^{2j}e^{\pm2i\eta}}\label{corr:X}\,, %\\
\end{equation}
where we use a pictorial convention: the arrow $\xrightarrow{\;\, \brt(y)\;\,}$ is not a homomorphism of modules (as both sides are representations over different algebras) but stands for the pull-back along $\brt(y)$ as discussed above. 
We also note that for $j=0$ the $z$ parameter equals $+e^{i\eta}$.
These results   agree with~\cite[Cor. 6.12]{GLDiagramAlgebras}.\footnote{One must be careful when comparing these results because they use a different convention: one must change $\q \to -\q^{-1}$ in our formulas to account for this difference.} 
Proving that the braid translation of $\bSt^{b/u}_j$ is a standard module involves a few subtleties, but if we assume that it is the case, the $z$ parameter can be identified rather easily by using the element $ \tau_{j}$ defined in equation \eqref{eq:taujdef}. In the blob algebra, one has
\begin{align*}
	u \quad
\begin{tikzpicture}[scale=1/3, baseline = {(current bounding box.center)},yscale=-1]
	\draw[black, line width = 1pt] (0,0) .. controls (0,1) and (1,1) .. (1,0);
	\draw[black, line width = 1pt] (2,0) .. controls (2,1) and (3,1) .. (3,0);
	%The blob
	\draw[black, line width = 1pt] (4,0) .. controls (4,1) and (0,2) .. (0,3);
	\draw[decoration = {
		markings,
		mark= at position (1/2) with \coordinate (0);,
		},decorate] (4,0) .. controls (4,1) and (0,2) .. (0,3);
	\filldraw[black] (0) circle (4pt);
	%End of the blob
	\draw[black, line width = 1pt] (5,0) .. controls (5,1) and (1,2) .. (1,3);
	%These are the border lines
%	\draw[black, line width = 2pt] (-.5,0) -- (5.5,0);
%	\draw[black, line width = 2pt] (-.5,3) -- (1.5,3);
%	\foreach \r in {0,...,5}{
%		\filldraw[red] (\r,0) circle (2 pt);
%	};
%	\foreach \r in {0,1}{
%		\filldraw[red] (\r,3) circle (2 pt);
%	};
\end{tikzpicture} 
	 & = (-\q)^{1/2}e^{-i \eta}(1+ \beta b) \quad
\begin{tikzpicture}[scale=1/3, baseline = {(current bounding box.center)},yscale=-1]
	\foreach \r in {1,...,5}{
		%\draw[white, line width = 3 pt] (\r,0) -- (\r-1,3);
		\draw[black, line width = 1 pt] (\r,0) -- (\r-1,3);
	};
	\draw[white, line width = 3pt] (0,0) .. controls (0,1) and (5,2) .. (5,3);
	\draw[black, line width = 1pt] (0,0) .. controls (0,1) and (5,2) .. (5,3);
	\draw[black, line width = 1pt] (0,3) .. controls (-0.2,4) and (0.8,4) .. (1,3);
	\draw[black, line width = 1pt] (2,3) .. controls (1.8,4) and (2.8,4) .. (3,3);
	%This is the blob
	\draw[black, line width = 1pt] (4,3) .. controls (3.8,4) and (0,5) .. (0,6);
	\draw[decoration = {
		markings,
		mark= at position (1/2) with \coordinate (0);,
		},decorate] (4,3) .. controls (4,4) and (0,5) .. (0,6);
	\filldraw[black] (0) circle (4pt);
	%End of the blob
	\draw[black, line width = 1pt] (5,3) .. controls (5,4) and (1,5) .. (1,6);
	%These are the border lines
%	\draw[black, line width = 2pt] (-.5,0) -- (5.5,0);
%	\draw[black, line width = 2pt] (-.5,3) -- (5.5,3);
%	\draw[black, line width = 2pt] (-.5,6) -- (1.5,6);
%	\foreach \r in {0,...,5}{
%		\filldraw[red] (\r,0) circle (2 pt);
%		\filldraw[red] (\r,3) circle (2 pt);
%	};
%	\foreach \r in {0,1}{
%		\filldraw[red] (\r,6) circle (2 pt);
%	};
\end{tikzpicture}\\
& =(-\q)^{1/2} e^{-i \eta}(1+ \beta b)\quad
\begin{tikzpicture}[scale=1/3, baseline = {(current bounding box.center)},yscale=-1]
	\draw[black, line width = 1pt] (1,0) .. controls (1,1) and (2,1) .. (2,0);
	\draw[black, line width = 1pt] (3,0) .. controls (3,1) and (4,1) .. (4,0);
	%The blob
	%\draw[white, line width = 3pt] (5,0) .. controls (5,1) and (0,2) .. (0,3);
	\draw[black, line width = 1pt] (5,0) .. controls (5,1) and (0,2) .. (0,3);
	\draw[decoration = {
		markings,
		mark= at position (.9) with \coordinate (0);,
		},decorate] (5,0) .. controls (5,1) and (0,2) .. (0,3);
	\filldraw[black] (0) circle (4pt);
	%End of the blob
	\draw[white, line width = 3pt] (0,0) .. controls (0,1) and (1,2) .. (1,3);
	\draw[black, line width = 1pt] (0,0) .. controls (0,1) and (1,2) .. (1,3);
	%These are the border lines
%	\draw[black, line width = 2pt] (-.5,0) -- (5.5,0);
%	\draw[black, line width = 2pt] (-.5,3) -- (1.5,3);
%	\foreach \r in {0,...,5}{
%		\filldraw[red] (\r,0) circle (2 pt);
%	};
%	\foreach \r in {0,1}{
%		\filldraw[red] (\r,3) circle (2 pt);
%	};
\end{tikzpicture} \\
&= (-\q)^{1}e^{i \eta} \quad
\begin{tikzpicture}[scale=1/3, baseline = {(current bounding box.center)},yscale=-1]
	\draw[black, line width = 1pt] (1,0) .. controls (1,1) and (2,1) .. (2,0);
	\draw[black, line width = 1pt] (3,0) .. controls (3,1) and (4,1) .. (4,0);
	%The blob
	\draw[black, line width = 1pt] (0,0) -- (0,3);
	\draw[decoration = {
		markings,
		mark= at position (1/2) with \coordinate (0);,
		},decorate] (0,0) -- (0,3);
	\filldraw[black] (0) circle (4pt);
	%End of the blob
	\draw[black, line width = 1pt] (5,0) .. controls (5,1) and (1,2) .. (1,3);
	%These are the border lines
%	\draw[black, line width = 2pt] (-.5,0) -- (5.5,0);
%	\draw[black, line width = 2pt] (-.5,3) -- (1.5,3);
%	\foreach \r in {0,...,5}{
%		\filldraw[red] (\r,0) circle (2 pt);
%	};
%	\foreach \r in {0,1}{
%		\filldraw[red] (\r,3) circle (2 pt);
%	};
\end{tikzpicture},
\end{align*}
where we used the interpretation of standard modules in terms of uneven diagrams (see for instance \cite{GrahamLehrer} for the affine TL version) and the standard braid relations. It follows that
\begin{equation*}
 \tau_{1} \quad \begin{tikzpicture}[scale=1/3, baseline = {(current bounding box.center)},yscale=-1]
	\draw[black, line width = 1pt] (0,0) .. controls (0,1) and (1,1) .. (1,0);
	\draw[black, line width = 1pt] (2,0) .. controls (2,1) and (3,1) .. (3,0);
	%The blob
	\draw[black, line width = 1pt] (4,0) .. controls (4,1) and (0,2) .. (0,3);
	\draw[decoration = {
		markings,
		mark= at position (1/2) with \coordinate (0);,
		},decorate] (4,0) .. controls (4,1) and (0,2) .. (0,3);
	\filldraw[black] (0) circle (4pt);
	%End of the blob
	\draw[black, line width = 1pt] (5,0) .. controls (5,1) and (1,2) .. (1,3);
	%These are the border lines
%	\draw[black, line width = 2pt] (-.5,0) -- (5.5,0);
%	\draw[black, line width = 2pt] (-.5,3) -- (1.5,3);
%	\foreach \r in {0,...,5}{
%		\filldraw[red] (\r,0) circle (2 pt);
%	};
%	\foreach \r in {0,1}{
%		\filldraw[red] (\r,3) circle (2 pt);
%	};
\end{tikzpicture} 
= (-\q)e^{i \eta}\quad
\begin{tikzpicture}[scale=1/3, baseline = {(current bounding box.center)},yscale=-1]
	\draw[black, line width = 1pt] (1,0) .. controls (1,1) and (2,1) .. (2,0);
	\draw[black, line width = 1pt] (3,0) .. controls (3,1) and (4,1) .. (4,0);
	%The blob
	\draw[black, line width = 1pt] (0,0) -- (0,3);
	\draw[decoration = {
		markings,
		mark= at position (1/2) with \coordinate (0);,
		},decorate] (0,0) -- (0,3);
	\filldraw[black] (0) circle (4pt);
	%End of the blob
	\draw[black, line width = 1pt] (5,0) .. controls (5,1) and (1,2) .. (1,3);
	\draw[black, line width = 1pt] (0,-3) .. controls (0,-2) and (1,-2) .. (1,-3);
	\draw[black, line width = 1pt] (0,0) .. controls (0,-1) and (1,-1) .. (1,0);
	\draw[black, line width = 1pt] (2,-3) .. controls (2,-2) and (3,-2) .. (3,-3);
	\draw[black, line width = 1pt] (2,0) .. controls (2,-1) and (3,-1) .. (3,0);
	\draw[black, line width = 1pt] (4,-3) -- (4,0);
	\draw[black, line width = 1pt] (5,-3) -- (5,0);
	%These are the border lines
%	\draw[black, line width = 2pt] (-.5,0) -- (5.5,0);
%	\draw[black, line width = 2pt] (-.5,-3) -- (5.5,-3);
%	\draw[black, line width = 2pt] (-.5,3) -- (1.5,3);
%	\foreach \r in {0,...,5}{
%		\filldraw[red] (\r,0) circle (2 pt);
%		\filldraw[red] (\r,-3) circle (2 pt);
%	};
%	\foreach \r in {0,1}{
%		\filldraw[red] (\r,3) circle (2 pt);
%	};
\end{tikzpicture}
\quad = (-\q)e^{i \eta}\quad
\begin{tikzpicture}[scale=1/3, baseline = {(current bounding box.center)},yscale=-1]
	\draw[black, line width = 1pt] (0,0) .. controls (0,1) and (1,1) .. (1,0);
	\draw[black, line width = 1pt] (2,0) .. controls (2,1) and (3,1) .. (3,0);
	%The blob
	\draw[black, line width = 1pt] (4,0) .. controls (4,1) and (0,2) .. (0,3);
	\draw[decoration = {
		markings,
		mark= at position (1/2) with \coordinate (0);,
		},decorate] (4,0) .. controls (4,1) and (0,2) .. (0,3);
	\filldraw[black] (0) circle (4pt);
	%End of the blob
	\draw[black, line width = 1pt] (5,0) .. controls (5,1) and (1,2) .. (1,3);
	%These are the border lines
%	\draw[black, line width = 2pt] (-.5,0) -- (5.5,0);
%	\draw[black, line width = 2pt] (-.5,3) -- (1.5,3);
%	\foreach \r in {0,...,5}{
%		\filldraw[red] (\r,0) circle (2 pt);
%	};
%	\foreach \r in {0,1}{
%		\filldraw[red] (\r,3) circle (2 pt);
%	};
\end{tikzpicture}\quad. 
\end{equation*}
However in the affine TL algebra, one has instead the diagram \eqref{ex-eigval-tau1}.
Comparing the two then gives $z = (-\q) e^{i \eta} $ when $j=1$. The generic case is obtained by the obvious generalization of the preceding argument.

We can also consider the braid translation in the ordinary TL case, which corresponds to setting $y=\q+\q^{-1} = \delta$ (so $\mathrm{e}^{i \eta} =\q$) as well as setting  the blob operator to the identity\footnote{The additional relation $b=\one$ corresponds actually to a quotient of the blob algebra by the first Jones--Wenzl projector~\cite{MartinWoodcock} $P^u_1$, see also a review in~\cite{GJSV}.}: $b\to\one$.
In this case, \eqref{eq_defbraid} reduces to (note that $(\alpha+1)(\beta+1)=1$) 
\begin{equation}\label{eq_defbraid2}
\displaystyle e_{N} = \left( \prod_{i=1}^{N-1} g_i \right)^{-1} e_1 \left( \prod_{i=1}^{N-1} g_i \right).
\end{equation}
 Similar formulae were used in the construction  of $U_\q(\mathfrak{sl}_2)$ invariant periodic XXZ spin chain in~\cite{Braid1,Braid3}.
 Indeed, note that the  Hamiltonian obtained {\it via} the braid translation $\brt$ of the $U_\q(\mathfrak{sl}_2)$ open spin chain enjoys  the same quantum group symmetry, contrarily to the ordinary (twisted) periodic XXZ chain discussed above in Sec.~\ref{sec:twistedXXZ}. The translation operator $u$ takes then the expression
\begin{equation}\label{eq_braid_trans-TL}
\displaystyle
u = (-1)^{N/2}\q^{N/2+1}\prod_{i=1}^{N-1}g_i,
\end{equation}
  and the factor picked by $2j$ through lines  as they wind one time around the annulus is given by
$\alpha_{2j}=(-\q)^{(2j+2)j}$.

We note that the braid translation~\eqref{eq_defbraid} in the full blob algebra case (without imposing $b=\one$) and $y=\delta$ generates the affine TL representation $\aSt_{j,(-\q)^{2j+2}}$ from the blobbed sector with $2j$ through lines, while the TL standard module $\StTL{j}$ is a quotient of $\StTL{j}^b$.
Therefore, the braid translation~\eqref{eq_defbraid2} and~\eqref{eq_braid_trans-TL} applied to $\StTL{j}$ generates an affine TL representation which is a corresponding quotient of $\aSt_{j,(-\q)^{2j+2}}$. The  representation $\aSt_{j,(-\q)^{2j+2}}$ is degenerate and has a proper submodule $\aSt_{j+1,(-\q)^{2j}}$.
This can also be seen from the blob algebra side because the standard module in the blobbed sector of the blob algebra at $y=\q+\q^{-1}$ with $2j$ through lines is
degenerate~\cite{MartinWoodcock} and has a proper submodule in the unblobbed sector with $2j+2$ through lines.
The quotient modules are isomorphic to the usual standard modules $\StTL{j}$ of the TL algebra.
The  $\ATL{N}(\delta)$ module obtained from the TL standard module by the braid translation thus corresponds to the quotient $\aSt_{j,(-\q)^{2j+2}} / \aSt_{j+1,(-\q)^{2j}}$.
The dimensions match since we have
\begin{equation}\label{eq:hatd}
\displaystyle{ \hat{d}_j = \dim \AStTL{j}{\mathrm{e}^{2iK}} = \binom{N}{N/2- j}},
\end{equation}
while the dimension of the TL standard modules reads
\begin{equation}
\displaystyle{ d_j = \dim \StTL{j} = \binom{N}{N/2 + j} - \binom{N}{N/2+ j +1}}.
\end{equation}

For the  case $j=0$ (note that this can only occur for $N$ even) we have to  calculate  the weight of non-contractible loops. Using~\eqref{eq_defbraid2} and~\eqref{eq_braid_trans-TL}, we get the weight $z+z^{-1} = \q+ \q^{-1}$ for non-contractible loops as the unique non-zero eigenvalue of the operator $\tau_0$ from~\eqref{eq:tau}. One then needs to consider the twist $\mathrm{e}^{2iK}=\q^2$
to account for this, and also,  in the geometrical language of links,  the braid-translated $\StTL{0}$  consists of all the diagrams  on an annulus with no through lines and with the identification of non-isotopic diagrams connecting the same sites. This is the quotient $\bStJTL{0}{\q^2}$ of  $\StJTL{0}{\q^2}$ by its submodule $\StJTL{1}{1}$  (recall the discussion in  Sec.~\ref{sec:ATL-def}).

\section{RSOS and Minimal models}\label{sec:minmodels}

We begin by studying the braid translation of lattice Minimal models -- the open RSOS models.
We shall show that a remarkable phenomenon occurs, as the periodic
Minimal models can be exactly recovered as the braid translation of
various sectors of the corresponding open models. We will
study in details some simple examples like the Ising model to show
that the braid translation generates the usual periodic TL generator
in these cases. A similar discussion can be found in~\cite{Braid3}.
We will then give a more formal description of this phenomenon
using RSOS representations and the blob algebra.
 In this section, we consider even $N$ and set $N=2L$.\footnote{Taking $N=2L$ is not essential but makes matters simpler.} 
 Also, we will denote an expression for $e_N$ obtained using the braid translation by $e_N^\brt$ in order to avoid a confusion with the standard expressions for the periodic generator.

\subsection{The Ising model and the three-state Potts model}\label{subsec:Ising}

In some simple cases, it is possible to work
out explicitly the expression of  $e_N^{\brt}$ as defined in~\eqref{eq_defbraid} and~\eqref{eq_defbraid2}. For the Ising and three-state Potts models, 
it turns out that this last
interaction is local: it  couples the first and the last sites only, in a way that is totally similar to the form  of  the other generators, except for the potential appearance of additional twist terms.

\subsubsection{Ising model}

Let us start with the Ising model, which is defined, in the periodic case,  by the  Hamiltonian
\begin{equation}
H_{\rm Ising}\propto \sum_{i=1}^{L} \sigma_i^z\sigma_{i+1}^z+\sigma_i^x \,, \label{IsingHam0}
\end{equation}
where the $\sigma$ are Pauli matrices\footnote{Since we focus here on general algebraic properties,  the coefficient of proportionality in the definition of the Hamiltonian is not relevant.} and we start by assuming periodic boundary conditions: $\sigma_{L+1}^a \equiv  \sigma_{1}^a$ for $a=x,y,z$. This Hamiltonian can be reformulated in terms of a representation of the periodic Temperley-Lieb algebra 
 as 
 \begin{equation}
 H_{\rm Ising}\propto \sum_{i=1}^{2L}  e_i \,, \label{IsingHam}
 \end{equation}
where the generators $e_i$ are defined by   (note: there are $2L$ generators for $L$ Ising spins)
\begin{subequations}
\begin{eqnarray}
e_{2i-1} & = & \frac{1}{\sqrt{2}} \left( 1 + \sigma_i^x \right), \\
e_{2i} & = & \frac{1}{\sqrt{2}} \left( 1 + \sigma_i^z \sigma_{i+1}^z \right).
\end{eqnarray}
\end{subequations}
They correspond to taking a particular quotient of the Temperley-Lieb algebra with $\q={\rm e}^{i\pi/4}$, or $\delta=\sqrt{2}$:
\begin{equation}\label{eq_quotientIsing}
\displaystyle e_i e_{i+1} + e_{i+1} e_i -\sqrt{2} (e_i + e_{i+1}) +1 =0.
\end{equation}
In (\ref{IsingHam}), the  coupling between the last and first spins on the chain is encoded into 
$$
e_{2L}={1\over\sqrt{2}}(1+\sigma_{L}^z\sigma_1^z).
$$ 

It is also customary to consider the ``anti-periodic'' Ising chain, where the coupling between the last and first spins has the opposite sign, that is,  $\sigma_{L+1}^z \equiv - \sigma_{1}^z$ in (\ref{IsingHam0}). This turns out to be also expressible in the Temperley-Lieb setting: in fact, the Hamiltonian takes exactly the same form~(\ref{IsingHam}) but now $e_{2L}={1\over\sqrt{2}}(1-\sigma_{L}^z\sigma_1^z)$, which also satisfies the periodic Temperley-Lieb relations.

 We now want to investigate whether the periodic (as well as the antiperiodic)  Ising chains can be obtained by  braid translating some open chain with, maybe, non-trivial boundary conditions. We start with the simplest situation, that is the  Ising model with free boundary conditions. Its Hamiltonian is similar to (\ref{IsingHam}) except that there is no coupling between the first and last spins, and thus no $e_{2L}$ generator in the sum. Now, 
straightforward algebra using relations $\sigma^{a}\sigma^{b} = \delta_{ab} + i \epsilon_{abc} \sigma^{c} $ and induction yields, starting from (\ref{eq_defbraid2})
\begin{equation}\label{exp1}
\displaystyle e_{2L}^{\brt}(y=\delta=\sqrt{2})  = \frac{1}{\sqrt{2}}  \left(1 +P\sigma^{z}_{1}  \sigma^{z}_{L} \right).
\end{equation}
Here, the operator $P$ reads 
\begin{equation}
\displaystyle P = \prod_{i=1}^{L} \sigma^{x}_{i} = \left(\begin{array}{cc}  0 & 1  \\ 1 & 0 \end{array} \right) \otimes \left(\begin{array}{cc}  0 & 1  \\ 1 & 0 \end{array} \right) \otimes \dots \otimes  \left(\begin{array}{cc}  0 & 1  \\ 1 & 0 \end{array} \right)
\end{equation}
and  is nothing but the usual $\mathbb{Z}_2$-symmetry operator of the Ising model. We have $P^2=\mathbb{I}$ so that its eigenvalues are $\epsilon = \pm 1$: $\epsilon=1$ corresponds to the $\mathbb{Z}_2$-symmetric sector, and $\epsilon=-1$ to the $\mathbb{Z}_2$-antisymmetric sector.  Note that the operator $P$ acts as well, and with the same meaning, on the closed chain. 
 We thus see that in  the $\mathbb{Z}_2$ symmetric sector, braid translation of the open chain gives rise to the periodic Ising Hamiltonian, while in the $\mathbb{Z}_2$ antisymmetric sector,   braid translation of the open chain gives the twisted (or antiperiodic) Ising Hamiltonian.

It is possible as well to obtain the opposite identification of sector labels. To this end,
we need to perform our braid translation starting from an open chain with a boundary interaction that can be interpreted \cite{JS} as fixed boundary conditions in the Ising model. Specifically, we note that the  operator
 $b=(1+\sigma^z_1)/2$ satisfies the blob algebra relations with the parameter $y=1/\sqrt{2}$, corresponding to giving a weight $y \delta = 1$ to Fortuin-Kasteleyn clusters touching the boundary  \cite[eq.~(3.7)]{JS}. It turns out that  this case gives a  new periodic representation non-isomorphic to the one in~\eqref{exp1}. This is clear because the blob operator now breaks the $\mathbb{Z}_2$ symmetry. For  $b=(1+\sigma^z_1)/2$, we have indeed
  \begin{equation}\label{exp2}
\displaystyle e_{2L}^{\brt}(y=1/\sqrt{2})  = \frac{1}{\sqrt{2}}  \left(1 - P\sigma^{z}_{1}  \sigma^{z}_{L} \right),
\end{equation}
which has just an opposite sign comparing to the expression in~\eqref{exp1}. 

We see therefore that we cannot obtain the periodic Ising chain simply by braid translation of some well chosen open Ising chain.
The proper procedure is more subtle, and hints at a more general picture, where the twisted Ising chain also plays a role. Of course, the correct language for all this is Temperley-Lieb and blob algebras representation theory to which we shall turn soon. 

It is interesting  to perform a similar analysis for a related model, the   Majorana fermions chain. That is, consider a system of $2L$ Majorana fermions  $\Gamma_i$ (note that there are twice as many fermions as there are Ising spins), satisfying $\{ \Gamma_i, \Gamma_j \} = \delta_{ij}$ with open boundary conditions and Hamiltonian $H\propto \sum_{i=1}^{2L-1} e_i$, where
\begin{equation}
\displaystyle e_j = \frac{1}{\sqrt{2}} + i \sqrt{2} \Gamma_j \Gamma_{j+1}\,, \qquad 1\leq j\leq 2L-1 \,.
\end{equation}
It is now straightforward to show by induction that the  generator obtained by braid translation reads
\begin{equation}
\displaystyle e_{2L}^\brt(y=\sqrt{2}) = \left( \prod_{i=2}^{2L-1} g_i \right)^{-1} e_1 \left( \prod_{i=2}^{2L-1} g_i \right) = \frac{1}{\sqrt{2}} + i \sqrt{2} \Gamma_1 \Gamma_{2L} \,.
\end{equation}
In this formulation, there is no $P$ operator like before, and, after braid translation of the open chain,  we  get  a closed  chain that must be considered as having antiperiodic boundary conditions (the periodic chain would have instead the coupling $\Gamma_{2L}\Gamma_1=-\Gamma_1\Gamma_{2L}$). The Majorana spin chain\footnote{Note that the  Majorana and Ising chains in a closed geometry are different, of course.  As we see here, the periodic Ising chain corresponds to periodic or antiperiodic Majorana chains, depending on the  $\mathbb{Z}_2$ sector \cite{SML64}.} $V=\bigl(\mathbb{C}^2\bigr)^{\otimes L}$ is, when considered as an open chain,  decomposed onto a direct sum of two non-isomorphic irreducible representations of dimension $2^{L-1}$, see also~\cite{Nichols} 
(recall the definition of simple modules  $\BX^b_j$ for the blob algebra $\mathcal{B}_N(\delta,y)$ in Sec.~\ref{sec:blob-def} and simple $\aX_{j,z^2}$ for $\ATL{N}(\delta)$ in Sec.~\ref{sec:ATL-def}): 
\begin{equation}
V = \BX^b_0 \oplus \BX^b_1
\end{equation}
(where $y=\sqrt{2}$ and $b=\one$). Meanwhile, considered as a closed  antiperiodic Majorana chain
 we have
\begin{equation}
V= \aX_{0,i} \oplus \aX_{1,-1}
\end{equation}
where now the decomposition is for the affine TL algebra $\ATL{N}$. In other words, the antiperiodic Majorana  chain can be indeed obtained as the result of the braid translation:
\begin{equation}
\BX^b_0\; \xrightarrow{\quad \brt(y=\sqrt{2})\quad}\;\aX_{0,i} \ , \qquad \BX^b_1\; \xrightarrow{\quad \brt(y=\sqrt{2})\quad}\;\aX_{1,-1}
\end{equation}
 from a single open chain, in contrast with the Ising model, where we saw that the $\mathbb{Z}_2$ sectors played a crucial role. Here,
  we used the correspondence~\eqref{corr:X}. We note that in terms of fermions the first summand $\aX_{0,i}$ contains even fermionic states while all the odd ones  are in $\aX_{1,-1}$.

Of course, if we braid translate the chain with the non trivial $b$ operators introduced above, we will get this time Majorana fermions with periodic boundary conditions 
\begin{equation}
\displaystyle e_{2L}^\brt(y=1/\sqrt{2}) = \left( \prod_{i=2}^{2L-1} g_i \right)^{-1} (1-b) e_1(1-b) \left( \prod_{i=2}^{2L-1} g_i \right) = \frac{1}{\sqrt{2}} + i \sqrt{2} \Gamma_{2L} \Gamma_{1} \,.
\end{equation}
The full spin chain gives rise this time to a single irreducible representation $\aX_{0,-1}$ of the affine TL algebra 
%(each non-contractible loop should be replaced by zero in this case)
\begin{equation}
V = \BX^b_0  \cong \aX_{0,-1}:~~ \BX^b_0\; \xrightarrow{\quad \brt(y=1/\sqrt{2})\quad}\; \aX_{0,-1} \,,
\end{equation}
where  the first decomposition is over the blob algebra (now, with $y=1/\sqrt{2}$) and the second one is for the affine TL.
We note that if we  restrict to the subalgebra generated by all $e_j$'s and $u^2$, the  spin chain would now   decompose onto a direct sum of two {\sl isomorphic} representations of dimension~$2^{L-1}$.
 
What we learned here is that, while one has to proceed carefully sector by sector in order to reproduce the periodic (or antiperiodic) Ising model using braid translation, no such subtlety exists for Majorana fermions. Of course, this is compatible with the fact that the relationship between the Ising model and Majorana fermions itself involves a careful consideration of sectors. The point is, that for all models corresponding to local rational CFTs we have studied---of which the Ising model is just one example---the relationship between closed and open geometries involves a complex decomposition into sectors \cite{SML64}.  

\subsubsection{Three-state Potts model}
To prove this point, we now turn to a slightly more complicated example. Let us consider
the three-state Potts model with $\q=\mathrm{e}^{i \pi/6}$ or $\delta=\sqrt{3}$. We use the following expression (see, {\it e.g.}, \cite{Nichols}) of the TL generators 
\begin{subequations}
\begin{eqnarray}
e_{2i-1} & = & \frac{1}{\sqrt{3}} \left( 1 + M_i + M_i^2 \right) \\
e_{2i} & = & \frac{1}{\sqrt{3}} \left( 1 + R_i R_{i+1}^2 + R_i^2 R_{i+1} \right),
\end{eqnarray}
\end{subequations}
with 
\begin{equation}
\displaystyle R = \left(\begin{array}{ccc}  1 & 0 & 0 \\ 0 & \q^4 & 0 \\ 0 & 0 & \q^8 \end{array} \right), \ \ \ \  M = \left(\begin{array}{ccc}  0 & 1 & 0 \\ 0 & 0 & 1 \\ 1 & 0 & 0 \end{array} \right).
\end{equation}
We use the usual convention $R_i = \mathbb{I} \otimes \dots \otimes \mathbb{I}  \otimes R \otimes \mathbb{I}  \otimes \dots \otimes \mathbb{I}$, where $R$ is on $(i,i+1)$ sites.
The key features of these matrices are $R^3=M^3=\mathbb{I}$ along with
\begin{equation}
\displaystyle R M + \q^2 M R = 0.
\end{equation}
Using these properties
 it is straightforward to show by induction that the braid translation~\cite{Braid3} gives the   generator
\begin{equation}
\displaystyle e_{2L}  = \frac{1}{\sqrt{3}}  \left(1 + P^2 R_L R_1^2 + P R_1 R_L^2 \right),
\end{equation}
with the $\mathbb{Z}_3$-symmetry operator  $P=M_1\otimes M_2\otimes\ldots\otimes M_L$ and its eigenvalues $\epsilon^3 = 1$.
Hence what happens is very similar to the Ising case: we get a closed three-state Potts chain under the twisted conditions $R_{L+1}=\epsilon^{n}R_1$, for $n=0,1,2$, depending only on the $\mathbb{Z}_3$ sector (instead of $\mathbb{Z}_2$ for Ising). 
The full spin chain  $V=\bigl(\mathbb{C}^3\bigr)^{\otimes L}$ is decomposed~\cite{Nichols} onto a direct sum of three non-isomorphic irreducible representations with multiplicities $1$ or $2$:
\begin{equation}
V = \BX^b_0 \oplus 2\BX^b_1 \oplus \BX^b_2 \ \cong \  \aX_{0,\q^2} \oplus 2\aX_{1,\q^4}\oplus \aX_{2,-1},
\end{equation}
or 
\begin{equation}
 \BX^b_0\; \xrightarrow{\quad \brt(y=\sqrt{3})\quad}\;\aX_{0,\q^2}\,; \quad \BX^b_1\; \xrightarrow{\quad \brt(y=\sqrt{3})\quad}\;\aX_{1,\q^4}\,; \quad  \BX^b_2\; \xrightarrow{\quad \brt(y=\sqrt{3})\quad}\;\aX_{2,-1} \,,
 \end{equation}
where the first decomposition is over the blob algebra with $y=\sqrt{3}$ and the second one is for $\ATL{N}$ in the braid-translated chain.
Like in the Ising case, other sectors of the periodic models or irreducible representations of the periodic TL algebra can be obtained starting from representations of the blob algebra having different parameter $y$ and thus more complicated blob operator $b$. We shall come back to
this when we turn to the more general setup of RSOS models.

\subsection{The boundary RSOS models}\label{sec:bndRSOS}
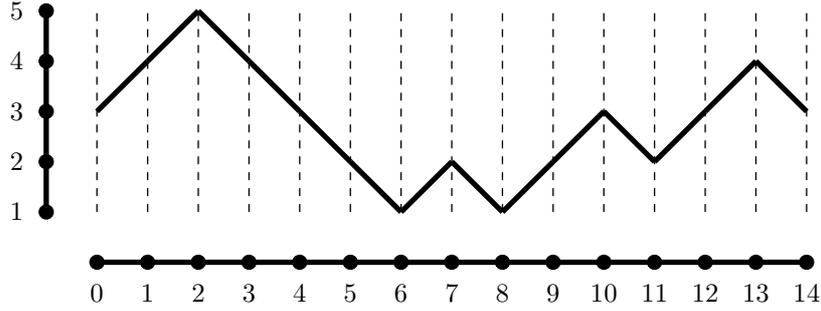
\begin{figure}
\begin{center}
\begin{tikzpicture}[scale = 2/3, baseline = {(current bounding box.center)}]
%Path first
	\foreach \r [count = \i ]in {3,4,5,4,3,2,1,2,1,2,3,2,3,4,3}{
		\coordinate (S\i) at (\i,\r);
	}
	\foreach \i [evaluate=\i as \x using \i+1] in {1,...,14}{
		\draw[black, line width = 2pt] (S\i) -- (S\x);
	};
%Side Graph
\draw[black, line width = 2pt] (0,1) -- (0,5); 
\foreach \r in {1,...,5}{
	\filldraw[black] (0,\r) circle (4 pt);
	\node[anchor = east] at (-.25,\r) {\small{\r}};
	};
%Bottom scale
\draw[black, line width = 2pt] (1,0) -- (15,0); 
\foreach \r in {0,...,14}{
	\draw[black, dashed, line width = .5pt] (\r +1,1) -- (\r +1, 5);
	\filldraw[black] (\r + 1,0) circle (4 pt);
	\node[anchor = north] at (\r +1,-.25) {\small{\r}};
	};
\end{tikzpicture}
\end{center}
  \caption{Example of RSOS path for $p=5$ and $N=14$ with fixed boundary conditions ($h_0=h_N=n=3$). The blob operator $b$  acts on this state
  as the identity as $h_1=n+1=4$.}
  \label{FigRSOS}
\end{figure}

The boundary RSOS model of type $A_p$ is
a height lattice model with heights $h_i \in \lbrace 1,2, \dots, p \rbrace$, where the lattice sites are labeled by $i=0,1,\dots,N$, and the heights are subject to the relation
\be\label{eq:hight-cond}
\left| h_i - h_{i \pm 1} \right| =1\ .
\ee
An example of the height configuration is given in Fig.~\ref{FigRSOS}. Let us fix  $\q=\mathrm{e}^{i \gamma}$ with $\gamma = \frac{\pi}{p+1}$. We define the action of the Temperley-Lieb generators on the height  configurations as (now for $i=1,\ldots,N-1$)
\begin{equation}
\displaystyle e_i \Ket{h_0,h_1, \dots, h_i,  \dots, h_N} = \delta_{h_{i-1},h_{i+1}} \sum_{h'_i = h_{i-1} \pm 1} \frac{\sqrt{\left[ h_i\right]_\q \left[ h'_i\right]_\q}}{\left[ h_{i-1}\right]_\q}\Ket{h_0,h_1, \dots, h'_i, \dots, h_N} \ .
\label{TLRSOSaction}
\end{equation}
To specify the boundary conditions, we 
fix the heights  on the left and right boundaries~\cite{BauerSaleur,Cardybdr} to a value from $1$ to $p$, say, that $h_0 = n$ and $h_N = n + k$,  where $-n+1\leq k \leq p-n$. To implement this algebraically, we consider  a representation of the blob algebra with 
\begin{equation}\label{b-action}
\displaystyle b \Ket{h_0=n,h_1, \dots, h_N=n+k} = \delta_{h_1,n+1} \Ket{h_0=n,h_1, \dots, h_N=n+k}.
\end{equation}
The image space of $b$ corresponds to all the states with $h_{1}=n+1$. 
It is straightforward to check that we obtain a representation of the blob algebra with the  blob parameter given precisely by~\eqref{eqDefr}:
\begin{equation}\label{eq:yn}
\displaystyle y_n := \frac{\sin (n+1) \gamma}{ \sin n \gamma} = \frac{[n+1]_\q}{[n]_\q},\qquad n=1, \dots ,p.
\end{equation}
We will denote such a boundary RSOS representation of $\mathcal{B}_N(\delta,y_n)$ as $\rho_{n,n+k}$.

To make a connection with the standard modules discussed above, we note that for each of these values of $y=y_n$ and $\q$, the standard modules are highly reducible~\cite{MartinWoodcock} and the corresponding
irreducible quotients can  in fact precisely be described in terms of a boundary RSOS model~\cite{JS}.
Our claim is: \textit{the vector space of RSOS paths from $h_0=n$ to $h_N=n+k$ is the irreducible blob algebra module $\BX^b_{k/2}$ for positive $k$ and $\BX^u_{|k|/2}$ for negative $k$:}
\be\label{eq:claim-X}
\rho_{n,n+2j} \cong  \BX^{b/u}_{j} \ .
\ee
 We have not found in literature a proof of this fact (for the given boundary conditions).  However, we will need this statement in the application of the braid translation below. We thus give a simple proof here and discuss several combinatorial aspects of the result. 

\subsubsection{Proof of the claim}\label{sec:number-RSOS-proof}
\newcommand{\bp}{d}
We prove this claim by computing  the number of RSOS paths with fixed boundary heights using the transfer-matrix formalism and then comparing with known dimensions of  the irreducible modules $\BX^{b/u}_j$ of the blob algebra.
We begin with the transfer-matrix formalism.
There is  a closed-form solution to this counting problem that can be extracted from~\cite[eq.~(4.17)]{BauerSaleur}. The basic idea is to consider walks on
the $A_p$ Dynkin diagram, with heights $h_i = 1,2,\ldots,p$, and diagonalise the corresponding
adjacency matrix
${\cal A} = \{{\cal A}_{ij}\}_{i,j=1}^p$ with entries ${\cal A}_{ij} = \delta_{i,j+1} + \delta_{i,j-1}$. Notice that ${\cal A}$ is also the transfer matrix that builds up
RSOS paths of length $N$ from those of length $N-1$. Therefore the number of walks $M_{p,N}(h_0,h_N)$ on $A_p$ going from height $h_0 $ to height $h_N$
in $N$ steps, and satisfying the RSOS constraints, is just the matrix element $\langle h_0 | {\cal A}^N | h_N \rangle$. The eigenvalue problem ${\cal A} |v^{(j)} \rangle = \Lambda_j |v^{(j)} \rangle$
is readily solved: the eigenvalues are $\Lambda_j = 2 \cos \left( \tfrac{\pi j}{p+1} \right)$, and the corresponding orthonormalized eigenvectors $|v^{(j)}\rangle$ % = |v^{(j)}_h\rangle_{h=1}^{p}$
have components $v^{(j)}_h = \sqrt{\tfrac{2}{p+1}} \sin \left( \frac{\pi h j}{p+1} \right)$. Therefore
\begin{equation}\label{eq:Mpnab}
 M_{p,N}(h_0,h_N) = \frac{2}{p+1} \sum_{j=1}^p \sin \left( \frac{\pi h_0 j}{p+1} \right) \sin \left( \frac{\pi h_N j}{p+1} \right) \left[ 2 \cos \left( \frac{\pi j}{p+1} \right) \right]^{N} \,.
\end{equation}

On the other hand, we can compute dimensions of  the irreducible representations $\BX^{b/u}_j$ using known results on the structure of the corresponding standard modules $\BW^{b/u}_j$, see~\cite[App.~C]{GJSV}. For the blob parameter
$y= y_n$ and setting $x=p+1$,  we obtain
 (with $N=2L$ as usual):
\begin{eqnarray}
\dimm \BX^{b/u}_j &=& \dimm \BW^{b/u}_{j} -  \sum_{m\geq0}\bigl(\dimm\BW^{u/b}_{mx + n \pm j} + \dimm\BW^{b/u}_{mx + x-(n\pm j)}\bigr) +   \sum_{m\geq1}\bigl(\dimm\BW^{u/b}_{mx+ j} + \dimm\BW^{b/u}_{mx+ j}\bigr) \nonumber \\
  &=&   \sum_{m= -\infty}^{\infty} \left\{ \binom{2L}{L + mx + j} - \binom{2L}{L + mx + (n \pm j)} \right\} \ . \label{eq:dimXub}
\end{eqnarray}
To compare with  $M_{p,N}(h_0,h_N)$,
we rewrite~\eqref{eq:Mpnab} through the powers of $\q$ as
\begin{eqnarray}
M_{p,N}(h_0,h_N) &=&  - \ffrac{1}{2(p+1)} \sum_{j=1}^p (\q^{h_0 j} - \q^{-h_0 j})(\q^{h_N j} - \q^{-h_N j}) (\q^{j} + \q^{-j})^{2L} \nonumber \\
% [a j]_\q [b j]_\q 
&=&  - \ffrac{1}{2(p+1)} \sum_{k=0}^{2L}  \binom{2L}{k} \sum_{j=-p}^p \left(\q^{2j \left(\frac{h_0+h_N}{2}+L-k\right)} - \q^{2j\left(\frac{h_0-h_N}{2}+L-k\right)}\right) \ .
\end{eqnarray}
For the latter expression we then use the identity $\sum_{j=-p}^p \q^{2jx} = 2(p+1)\delta_{x, m(p+1)} - 1$, where $m\in\mathbb{Z}$, and $\delta$ is the Kronecker symbol.
This gives 
\be\label{eq:MpNab-binom}
M_{p,N}(h_0,h_N) =  \sum_{m=-\infty}^{\infty}  \Bigg\{ \binom{2L}{L + m(p+1) + \frac{|h_0-h_N|}{2}} -  \binom{2L}{L + m(p+1) + \frac{h_0+h_N}{2}} \Bigg\}\ ,
\ee
which indeed equals $\dimm \BX^{b/u}_j$ from~\eqref{eq:dimXub} after noticing that $j = \frac{|h_0-h_N|}{2}$ and $n\pm j =  \frac{h_0+h_N}{2}$ in the blobbed and unblobbed cases, respectively. This finishes our proof of the claim in~\eqref{eq:claim-X}.

We have counted the number of paths in the boundary RSOS model in a  rather formal way.
However, the formula~\eqref{eq:dimXub} for the number of paths as the alternating sum of binomials can be better understood combinatorially in terms of non-restricted paths. We describe it in the next remark, which can be skipped at the first reading, where we also give the generating series for the number of the paths in terms of Chebyshev polynomials.

\subsubsection{Remark on combinatorics and generating series}
Consider now the non-restricted SOS paths, where the heights $h_i \in \mathbb{Z}$ (i.e., without imposing the restriction $h_i \in \{1,2,\ldots,p\} = A_p$, as for the RSOS paths discussed above).
We maintain the nearest neighbor condition~\eqref{eq:hight-cond} and the boundary condition that the paths start from $h_0=n$ and end at $h_N=n+k$.
It is obvious that the number of such SOS paths indeed gives the dimension~\eqref{eqDimStdblob} for $|k|=2j$. 
 Let us consider two horizontal lines called \textit{walls} with constant heights $0$ and $p+1$, respectively.  Note that we have exactly $p$ heights between the walls. These are the walls that a given RSOS path can not touch. Our aim is to count the number of such  restricted RSOS paths in terms of the non-restricted SOS paths.
The idea is that starting with non-restricted paths we have to remove  all paths that cross or touch any of the walls (we call them \textit{forbidden} paths). 
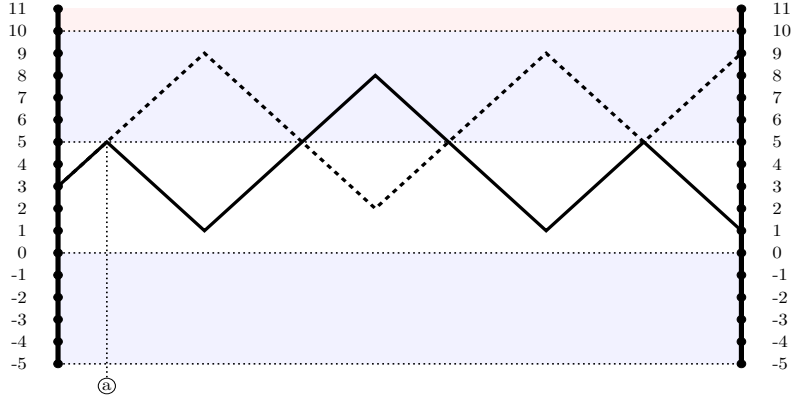
\begin{figure}
\begin{center}
\resizebox{300pt}{150pt}{%
\begin{tikzpicture}[scale = 1/2,every node/.style = {scale = 1}]
%Exlcusion zones
\filldraw[color = blue!5] (-10,4) -- (18,4) -- (18,9) -- (-10,9) -- (-10,4);
\filldraw[color = blue!5] (-10,-1) -- (18,-1) -- (18,-6) -- (-10,-6) -- (-10,-1);
\filldraw[color = red!5] (-10,9) -- (18,9) -- (18,10) -- (-10,10) -- (-10,9);
%Borders
\draw[black, line width = 3pt] (-10,-6) -- (-10,10);
\draw[black, line width = 3pt] (18,-6) -- (18,10);
\foreach \r [evaluate= \r as \i using int(\r +1)] in {-6,...,10}{
	\filldraw[black] (-10, \r ) circle (5pt);
	\filldraw[black] (18, \r ) circle (5pt);
	\node[anchor = east] at (-11,\r) {\i};
	\node[anchor = west] at (19,\r) {\i};
}
%Walls
\foreach \r in {-6,-1,4,9}{
	\draw[black, dotted, line width = 1pt] (-10,\r) -- (18,\r);
}
%The walk
\draw[black, dashed,line width = 2pt] (-10,2) -- (-4,8) -- (3,1) -- (10,8) -- (14,4) -- (18,8);
\draw[black, line width = 2pt] (-10,2) -- (-8,4) -- (-4,0) -- (3,7) -- (10,0) -- (14,4) -- (18,0);
%The marker
\draw[black, dotted, line width = 1pt] (-8,4) -- (-8,-7);
\filldraw[white] (-8,-7) circle (10 pt);
\draw[black, line width = .5pt] (-8,-7) circle (10 pt);
\node at (-8,-7) {a};
\end{tikzpicture}
}
\end{center}
\caption{
Non-restricted SOS paths with $p = 4$. The black path is a forbidden path of type $(i)$; the point $a$ marks the first time which it intersects the upper wall. The dashed path is the one obtained by the reflection in the upper wall.
}\label{fig:refl}
\end{figure}

\begin{figure}\label{fig.walk2}
\begin{center}
\resizebox{300pt}{150pt}{%
\begin{tikzpicture}[scale = 1/2,every node/.style = {scale = 1}]
%Exlcusion zones
\filldraw[color = blue!5] (-10,4) -- (18,4) -- (18,9) -- (-10,9) -- (-10,4);
\filldraw[color = blue!5] (-10,-1) -- (18,-1) -- (18,-6) -- (-10,-6) -- (-10,-1);
\filldraw[color = red!5] (-10,9) -- (18,9) -- (18,10) -- (-10,10) -- (-10,9);
%Borders
\draw[black, line width = 3pt] (-10,-6) -- (-10,10);
\draw[black, line width = 3pt] (18,-6) -- (18,10);
\foreach \r [evaluate= \r as \i using int(\r +1)] in {-6,...,10}{
	\filldraw[black] (-10, \r ) circle (5pt);
	\filldraw[black] (18, \r ) circle (5pt);
	\node[anchor = east] at (-11,\r) {\i};
	\node[anchor = west] at (19,\r) {\i};
}
%Walls
\foreach \r in {-6,-1,4,9}{
	\draw[black, dotted, line width = 1pt] (-10,\r) -- (18,\r);
}
%The walk
\draw[black, dashed, line width = 2pt] (-10,2) -- (-8,0) -- (-6,2) -- (2,-6) -- (8,0) -- (13,-5) -- (17,-1)  -- (18,-2);
\draw[black, line width = 2pt] (-10,2) -- (-8,0) -- (-6,2) -- (-3,-1) -- (2,4) -- (8,-2) -- (13,3) -- (17,-1) -- (18,0);
\draw[black, dash dot, line width = 2pt] (2,4) -- (8,10) -- (13,5) -- (17,9) -- (18,8);
%The markers
\draw[black, dotted, line width = 1pt] (-3,-1) -- (-3,-7);
\filldraw[white] (-3,-7) circle (10 pt);
\draw[black, line width = .5pt] (-3,-7) circle (10 pt);
\node at (-3,-7) {b};
\draw[black, dotted, line width = 1pt] (2,4) -- (2,-7);
\filldraw[white] (2,-7) circle (10 pt);
\draw[black, line width = .5pt] (2,-7) circle (10 pt);
\node at (2,-7) {c};
\end{tikzpicture}
}
\end{center}
\caption{Non-restricted SOS paths with $p = 4$. The black path is a forbidden path of both types $(i)$ and $(ii)$, touching the first lower wall at the point $b$ and the first upper wall at the point $c$; the dashed path is the one obtained by the reflection along the lower wall, and the dash-dotted path is the one obtained by the reflection along the upper wall.}\label{fig:refl-2}
\end{figure}
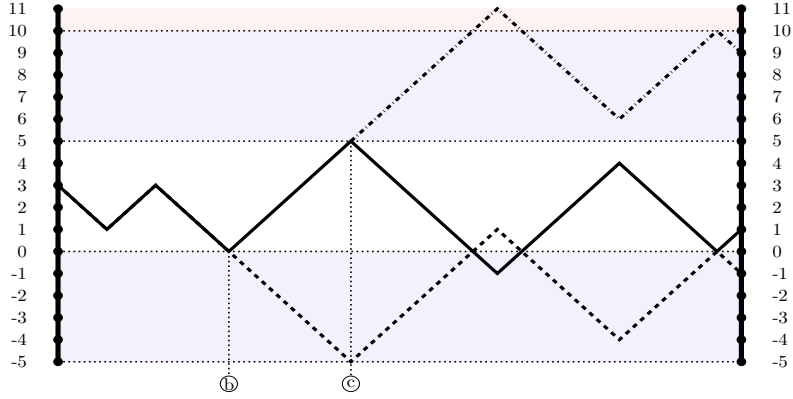

In  counting the forbidden paths,  reflections in the two walls  play a crucial role. 
Indeed, all forbidden paths are of two types:
\begin{itemize}
 \item[(i)] crossing/touching the upper wall (with the height $p+1$) at least once, and
 \item[(ii)] crossing/touching the lower wall (with the height $0$) at least once.
\end{itemize}
Obviously, some paths can be forbidden by both criteria. Those of the type (i) are in one-to-one correspondence with the non-restricted paths going from $h_0= n$ to $h'_N = n + 2(p+1 - n - j)$, where $h'_N$ is the reflection of $h_N=n+2j$ in the upper wall. The correspondence is given by the reflection  in the upper wall of the part of a given forbidden path after the first crossing/touching the upper wall. This is illustrated in Figure~\ref{fig:refl}. The forbidden paths of type (ii) are in one-to-one correspondence with the non-restricted paths going from $h_0= n$ to $h''_N = n - 2(n + j)$, where $h''_N$ is the reflection of $h_N$ in the lower wall.   The correspondence is  given similarly by the reflection  in the lower wall of the part (of a given forbidden path) after the first crossing/touching the lower wall. We have thus to subtract from the total amount $\hat{d}_j = \binom{2L}{L+j}$ of initial non-restricted paths the two binomials $\hat{d}_{p+1 - n - j}$ and $\hat{d}_{n+j}$. This gives the first terms in~\eqref{eq:MpNab-binom} (for $m=0,-1$):
\be\label{eq:Mwithdots}
M_{p,N}(n,n+2j) =    \binom{2L}{L  + j} -  \binom{2L}{L + p+1 - n - j} - \binom{2L}{L  + n + j} +\, \ldots
\ee
where the missing terms in `$\ldots$' are for those forbidden paths that were counted twice and so they should be added.  These are the paths that touch both the upper and lower walls and they are again of two types
\begin{itemize}
 \item[(1)] crossing/touching  the lower wall  after  the upper one, and
 \item[(2)] crossing/touching  the upper wall after the lower one.
\end{itemize}
Again,  some paths can be  of both types. 
 The paths of type (1) reflected in the upper wall are in bijection with  the paths crossing/touching the second upper wall -- the one of height $2p+2$ which is the reflection of the lower wall -- at least once, see the  illustration  in Figure~\ref{fig:refl-2} as well. 
  Therefore, using the reflection argument from the previous step, they are in bijection with all SOS paths  going from $h_0= n$ to $h'''_N =   2(p+1) + n + 2j$, where $h'''_N$ is the reflection of $h'_N =  2(p+1) - n - 2j$ in the second upper wall. As they were already counted using the reflection in the lower wall of the initial forbidden paths, we should thus add their number, the binomial $\hat{d}_{p+1 + n + j}$, in `\dots' in~\eqref{eq:Mwithdots}. We continue similarly with the paths of the type (2) using the lower and second lower walls for the reflections instead of the upper ones. At this point, we have added the two binomials  in `\dots' in~\eqref{eq:Mwithdots}, while again some paths were counted twice -- those that touch/cross the two walls three times in the alternating way:  e.g.\  first the upper wall, then the lower and then the upper one again.
 Repeating recursively the reflection argument from the previous step we obtain for the number of RSOS paths indeed the alternating sum~\eqref{eq:MpNab-binom} of binomials.

Let us remark that the above argument is inspired by Andr\'e's reflection method that he used to provide an elegant solution \cite{Andre1887} of the so-called ballot problem. The latter problem is
in fact equivalent to counting the number of RSOS paths with $h_0 = h_N = 1$, in the limit $p \to \infty$ with no upper wall.
 Finally, we also refer to the book~\cite{feller68} where a similar counting can be found. 

 We give here also a brief remark on generating series for numbers of the boundary RSOS configurations (or equally of dimensions of the irreducible blob representations).
Let us introduce the generating series 
\be\label{eq:path-gen-fun-def}
P^{(n)}_k(z) = \sum_{N\geq0} a_N \cdot z^N \,,
\ee
where the coefficients $a_N$ denote the number of RSOS paths with $h_0=n$ and $h_N = n+ k$ (so we set $k=2j$), and $N$ is the number of sites as usual.
In App.~\ref{app:comb},
we derive a closed form for these generating series in terms of the second-type Chebyshev polynomials $U_k:=U_k(z)$ with 
$U_{-1}=0$, $U_0=1$, $U_1=2z$, $U_2=4z^{2}-1$, {\it etc.} Let us set  $V_k(z):= U_k\big(\tfrac{1}{2z}\big)$ and abbreviate $V_k:=V_k(z)$, then the result is
\begin{equation}\label{eq:path-gen-fun}
P^{(n)}_k(z) = \frac{1}{z} \Biggl(\frac{V_{n-1} V_{p-n-k}}{V_{n-1}V_{p-n+1}-V_{n-2}V_{p-n}}\Biggr),\qquad 0\leq k\leq p-n,
\end{equation}
while the expression for negative values of $k$ is just given by replacing $n\to p-n+1$ and $k\to-k$ in~\eqref{eq:path-gen-fun}.
As discussed above, these are also generating series for 
dimensions of the irreducible representations $\BX^{b/u}_j$ for the blob algebra with $y=y_n$ from~\eqref{eq:yn},
where $j=k/2$ is positive for $\BX^{b}_j$ and negative for~$\BX^{u}_{|j|}$.

\subsection{Periodic RSOS models and braid translation}

The periodic RSOS models are defined physically by introducing  between the first and the last heights the same interaction we had all along the lattice sites in the open case. In other words,   we  simply take the Temperley-Lieb generators defined in (\ref{TLRSOSaction}) (including $e_N$) and let them act on RSOS configurations which are periodic, that is, $h_{N}\equiv h_0$. 
The path configurations will still be denoted as $|h_0,h_1,\ldots,h_N\rangle$.
This produces a finite dimensional representation $\rho_{\rm per}$ of the periodic Temperley--Lieb algebra, which obviously must be some particular quotient. Note that we can choose $h_0$ to be odd or even from the beginning,
and it remains this way under action with the Hamiltonian (or the transfer matrix). It is convenient in what follows to  define the periodic
RSOS models by allowing the height $h_0$ to take all values from $1$ to $p$  ({\em i.e.}, we sum over the two sectors with $h_0$ odd and even).

It is, in fact, known~\cite{PasquierSaleur} 
that  the representation $\rho_{\rm per}$ we just defined decomposes as a direct sum of irreducible representations of the affine TL algebra, where there are no through lines, and non-contractible loops are given the weight $2\cos{n\pi\over p+1}$, with $n=1,\ldots,p$.  These representations are $\aX_{0,\q^{2n}}$ in our standard notations.
The arguments in~\cite{PasquierSaleur} are based only on a calculation of partition (generating) functions and are not  formulated rigorously. Since we were  not able to  find any proof in the mathematics  literature either,  we shall provide one  here.

We begin with noticing that the element (we assume $\q^{2} \neq -1$) 
$$
\omega =(\q+\q^{-1})^{-L} e_1e_3\dots e_{2L-1}
$$ 
is an idempotent of $\ATL{2L}$, i.e.\ $\omega^2 = \omega$. Taking  
the left ideal  $\ATL{2L} \omega$ corresponds to projecting onto affine TL diagrams without through-lines and  without   restriction on non-contractible loops. The module $\ATL{2L} \omega$ is thus infinite-dimensional.
Using our special element introduced in~\eqref{eq:tau}
\begin{equation}
 \tau_0 = e_1e_3\dots e_{2L -1} u =
 \begin{tikzpicture}[scale=1/3, baseline = {(current bounding box.center)},rotate = 180]
	\draw[black, line width = 1pt] (0,0) .. controls (0,1) and (1,1) .. (1,0);
	\draw[black, line width = 1pt] (2,0) .. controls (2,1) and (3,1) .. (3,0);
	\node[anchor = north] at (4.5,0) {$\hdots $};
	\draw[black, line width = 1pt] (6,0) .. controls (6,1) and (7,1) .. (7,0);
	\draw[black, line width = 1pt] (8,0) .. controls (8,1) and (9,1) .. (9,0);
	\draw[black, line width = 1pt] (0,3) .. controls (0,2) and (-1,2) .. (-1, 3);
	\draw[black, line width = 1pt] (9,3) .. controls (9,2) and (10,2) .. (10, 3);
		\filldraw[white] (-.5,0) rectangle (-1.5,3);
		\filldraw[white] (9.5,0) rectangle (10.5,3);
	\draw[black, line width = 1pt] (1,3) .. controls (1,2) and (2,2) .. (2,3);
	\draw[black, line width = 1pt] (3,3) .. controls (3,2) and (4,2) .. (4,3);
	\draw[black, line width = 1pt] (6,3) .. controls (6,2) and (5,2) .. (5,3);
	\filldraw[white] (3.5,2) rectangle (5.5,3); 
	\node[anchor = south] at (4.5,3) {$\hdots $};
	\draw[black, line width = 1pt] (7,3) .. controls (7,2) and (8,2) .. (8,3);
%	\draw[black, line width = 2pt] (-.5,0) -- (9.5,0);
%	\draw[black, line width = 2pt] (-.5,3) -- (9.5,3);
%	\foreach \r in {0,...,3,6,7,8,9}{
%		\filldraw[red] (\r,0) circle (2 pt);
%		\filldraw[red] (\r,3) circle (2 pt);
%	};
	\end{tikzpicture} ,
\end{equation}
chosen such that
\begin{equation}
 \tau_{0}\omega = (\q+\q^{-1})^{-L} 
 \begin{tikzpicture}[scale=1/3, baseline = {(current bounding box.center)}, rotate = 180]
	\draw[black, line width = 1pt] (0,0) .. controls (0,1) and (1,1) .. (1,0);
	\draw[black, line width = 1pt] (2,0) .. controls (2,1) and (3,1) .. (3,0);
	\node[anchor = north] at (4.5,0) {$\hdots $};
	\draw[black, line width = 1pt] (6,0) .. controls (6,1) and (7,1) .. (7,0);
	\draw[black, line width = 1pt] (8,0) .. controls (8,1) and (9,1) .. (9,0);
	\draw[black, line width = 1pt] (0,3) .. controls (0,2) and (-1,2) .. (-1, 3);
	\draw[black, line width = 1pt] (9,3) .. controls (9,2) and (10,2) .. (10, 3);
		\filldraw[white] (-.5,0) rectangle (-1.5,3);
		\filldraw[white] (9.5,0) rectangle (10.5,3);
	\draw[black, line width = 1pt] (1,3) .. controls (1,2) and (2,2) .. (2,3);
	\draw[black, line width = 1pt] (3,3) .. controls (3,2) and (4,2) .. (4,3);
	\draw[black, line width = 1pt] (6,3) .. controls (6,2) and (5,2) .. (5,3);
	\filldraw[white] (3.5,2) rectangle (5.5,3); 
	\node[anchor = south] at (4.5,3) {$\hdots $};
	\draw[black, line width = 1pt] (7,3) .. controls (7,2) and (8,2) .. (8,3);
	\draw[black, line width = 1pt] (0,3) .. controls (0,4) and (1,4) .. (1,3);
	\draw[black, line width = 1pt] (2,3) .. controls (2,4) and (3,4) .. (3,3);
	\draw[black, line width = 1pt] (6,3) .. controls (6,4) and (7,4) .. (7,3);
	\draw[black, line width = 1pt] (8,3) .. controls (8,4) and (9,4) .. (9,3);
	\draw[black, line width = 1pt] (0,6) .. controls (0,5) and (1,5) .. (1,6);
	\draw[black, line width = 1pt] (2,6) .. controls (2,5) and (3,5) .. (3,6);
	\node[anchor = south] at (4.5,6) {$\hdots $};
	\draw[black, line width = 1pt] (6,6) .. controls (6,5) and (7,5) .. (7,6);
	\draw[black, line width = 1pt] (8,6) .. controls (8,5) and (9,5) .. (9,6);
%	\draw[black, line width = 2pt] (-.5,0) -- (9.5,0);
%	\draw[black, line width = 2pt] (-.5,3) -- (9.5,3);
%	\draw[black, line width = 2pt] (-.5,6) -- (9.5,6);
%	\foreach \r in {0,...,3,6,7,8,9}{
%		\filldraw[red] (\r,0) circle (2 pt);
%		\filldraw[red] (\r,3) circle (2 pt);
%		\filldraw[red] (\r,6) circle (2 pt);
%	};
	\end{tikzpicture} \qquad 
= (\q+\q^{-1})^{-L} \quad
\begin{tikzpicture}[scale=1/3, baseline = {(current bounding box.center)}]
	\draw[black, line width = 1pt] (0,0) .. controls (0,1) and (1,1) .. (1,0);
	\draw[black, line width = 1pt] (2,0) .. controls (2,1) and (3,1) .. (3,0);
	\node[anchor = south] at (4.5,0) {$\hdots $};
	\draw[black, line width = 1pt] (6,0) .. controls (6,1) and (7,1) .. (7,0);
	\draw[black, line width = 1pt] (8,0) .. controls (8,1) and (9,1) .. (9,0);
	\draw[black, line width = 1pt] (-.5,2) .. controls (3,3) and (6,1) .. (9.5,2);
	\draw[black, line width = 1pt] (0,4) .. controls (0,3) and (1,3) .. (1,4);
	\draw[black, line width = 1pt] (2,4) .. controls (2,3) and (3,3) .. (3,4);
	\node[anchor = north] at (4.5,4) {$\hdots $};
	\draw[black, line width = 1pt] (6,4) .. controls (6,3) and (7,3) .. (7,4);
	\draw[black, line width = 1pt] (8,4) .. controls (8,3) and (9,3) .. (9,4);
%	\draw[black, line width = 2pt] (-.5,0) -- (9.5,0);
%	\draw[black, line width = 2pt] (-.5,4) -- (9.5,4);
%	\foreach \r in {0,...,3,6,7,8,9}{
%		\filldraw[red] (\r,0) circle (2 pt);
%		\filldraw[red] (\r,4) circle (2 pt);
%	};
	\end{tikzpicture}\,,
\end{equation}
  we can consider the quotients of the module $\ATL{2L}\omega$, for non-zero $z\in \mathbb{C}^{\times}$, 
\begin{equation}\label{eq:Iz}
	(\ATL{2L}\omega)/I_z \simeq  \aSt_{0,z^2}   \ , \quad\text{where}\quad I_z = \ATL{2L}\big(\tau_0\, \omega  - (z+z^{-1}) \omega\big) \ ,
\end{equation}
which are the standard modules introduced in Sec.~\ref{sec:ATL-def}, where again indicating $z^2$ (or the $z$-parameter up to a sign) is just a convention we use in this paper. 
The simple result in~\eqref{eq:Iz} can be rephrased as follows: \textit{a vector $v$ in a given representation  generates through the action of $\ATL{2L}$ (a quotient of)  $\aSt_{0,z^2}$ if \footnote{Note that within $\aSt_{0,z^2}$ there is only one such a state, up to the scalar multiple.}}
$$
\omega v = v \qquad \text{and} \qquad \tau_0 v = (z+z^{-1}) v\ .
$$
(Compare with the use of $\tau_0$ in the analysis of the spin-$0$ sector in XXZ spin-chains above in Sec.~\ref{sec:twistedXXZ} after~\eqref{eq:tau}.)
 Applying this idea to the representation $\rho_{\rm per}$, eigenvectors of $\omega$ are any linear combinations of the $p$ vectors:
	\begin{eqnarray}
		x_{1} &=& |1,2,1,2,\dots ,1\rangle \,, \\
		x_{r} &=& \left(\frac{[r+1]_\q}{[r]_\q} \right)^{-L} \omega\, |r,r+1,r, r+1,\dots ,r\rangle \,, \quad \mbox{for } r= 2,3,\dots, p-1 \,, \\
		x_{p} &=& |p,p-1,p,p-1, \dots ,p\rangle \,.
	\end{eqnarray}
These vectors are linearly independent since $x_s $ only contain linear combinations of paths starting at position $s$; they were chosen such that
	\begin{equation}\label{eq:tau0isatoeplitzmatrix}
		\tau_{0}x_{r} = x_{r-1} + x_{r+1}, \qquad x_{0} \equiv x_{p+1} \equiv 0.
	\end{equation}
We are then looking for those combinations of $x_s$ which are equally the eigenvectors of $\tau_0$, as this will allow to identify the $z$'s. From equation \eqref{eq:tau0isatoeplitzmatrix}, one sees that in the $x_{i}$ basis, $\tau_0$ is the adjacency matrix of the graph $A_{p}$, whose spectrum is well-known. Its eigenvectors are
(see e.g. in~\cite{BauerSaleur})
	\begin{equation}
		v_{n} = \sum_{i=1}^{p} U_{i-1}\Big(\ffrac{\lambda_n}{2}\Big) \, x_i, \qquad \lambda_n = \q^{n} + \q^{-n}, \quad n=1,2,\dots, p, 
	\end{equation}
where $U_{i+1}$ is again the $(i+1)$th Chebyshev polynomial of the second type and $\lambda_{n}$ is the eigenvalue corresponding to $v_{n}$, 
 so it generates a quotient of $\aSt_{0,\q^{2n}}$. Finally, we note that the dimension of $\rho_{\rm per} $ is exactly equal to the sum of the dimensions of the unique simple quotients $\aX_{0,\q^{2n}}$ of $\aSt_{0,\q^{2n}}$ for $n=1,2,\dots, p$.  
To show this we use the previous result on boundary RSOS models. We first note that the periodic RSOS paths are in one-to-one correspondence with the  boundary RSOS paths that start at $h_0=n$ and end at $h_N=n$ for any $1\leq n \leq p$. For fixed $n$, the number of these boundary RSOS paths was computed in Sec.~\ref{sec:number-RSOS-proof} and it agrees with $\dim \aX_{0,\q^{2n}}$, which is the alternating sum of binomials as in~\eqref{eq:dimXub} for $j=0$ (here, we  used the know subquotient structure of $\aSt_{0,\q^{2n}}$, see~\cite{GrahamLehrer} and also~\cite[Sec.~3]{GRSV}).
We thus conclude that the periodic RSOS model has the decomposition
\begin{equation}
	\rho_{\rm per} \simeq  \bigoplus_{n=1}^{p} \aX_{0,\q^{2n}}\ .
\end{equation} 

The point is now that each of these irreducible representations $\aX_{0,\q^{2n}}$ can be obtained by the braid translation~$\brt$, defined in~\eqref{eq_defbraid}, applied to the boundary RSOS representation of  $\mathcal{B}_N(\delta,y_n)$:
\begin{equation}
\rho_{n,n} = \BX^b_0 \;\xrightarrow{\quad \brt(y_n)\quad}\; \aX_{0,\q^{2n}},
\end{equation}
where we used the result in~\eqref{eq:claim-X} and the identification~\eqref{weight-loop}, or~\eqref{corr:X}, and that $e^{i\eta}=\q^n$ for $y_n=\frac{[n+1]_\q}{[n]_\q}$.
After summing over all values of $n$, we finally obtain
\begin{equation}\label{eq:braid-RSOS}
\bigoplus_{n=1}^p\rho_{n,n}\; \xrightarrow{\quad \brt\quad}\; \rho_{\rm per}\ .
\end{equation}

A simple
counting on the left-hand side of~\eqref{eq:braid-RSOS} shows that we obtain a total dimension which is exactly the size of the RSOS
Hilbert space with periodic boundary conditions. The braid translation in this case thus has a natural interpretation as a glueing of boundaries (under the boundary condition $h_0=h_N$) of the strip into a cylinder.  It is unfortunately impossible to show in the original basis from the action of $b$ in~\eqref{b-action} that $e_{2L}$ has the standard periodic expression.
Indeed, using~\eqref{eq_defbraid} directly it is easy to see  that the action of $e_{2L}$ on each component in~\eqref{eq:braid-RSOS} in fact is expressed {\it non-locally}.
 It means we just work in a different basis. Amazingly, if we take a sum over all boundary conditions $n=1,\dots,p$ we obtain then \textit{in a different basis} the locally expressed action of the last generator $e_{2L}$.
 We were able to show this explicitly only in the Ising and three-state Potts models discussed above in Sec.~\ref{subsec:Ising} where we extracted different sectors of the corresponding RSOS models from spin chains having different blob operators.

We note further that  direct summands in~\eqref{eq:braid-RSOS} appear
in isomorphic pairs. Recall on the blob algebra side that we have a symmetry (an automorphism) $b\leftrightarrow 1-b$ and $y_n \leftrightarrow [2]_\q - y_n$ or $n\leftrightarrow p + 1 -n$. Therefore $\rho_{n,n}$ is actually isomorphic to $\rho_{p+1-n,p+1-n}$, as modules over the same blob algebra, of course. Thus, for even values of $p$ we have the decomposition
\begin{equation}
\rho_{\rm per} = 2\bigoplus_{n=1}^{p/2} \aX_{0,\q^{2n}}.
\end{equation}
 For odd values of $p$, as in the Ising case discussed above, the unique irreducible module $\rho_{\frac{p+1}{2},\frac{p+1}{2}}$ or $\aX_{0,-1}$ for the affine TL algebra is split\footnote{It is only the module $\aX_{0,-1}$ which is split onto two, all other modules $\aX_{0,z^2\ne-1}$ are irreducible for this subalgebra, see~\cite{GrahamLehrer} for more details.} onto two isomorphic modules~\cite{GrahamLehrer} for the subalgebra generated by $e_j$'s and the square $u^2$ of the translation generator. Note that on the RSOS configurations we have naturally the action of $u^2$ and the transfer matrix belongs to this subalgebra.   So, we have in particular the same spectrum  and in this case of odd values of $p$ we thus also have the multiplicity $2$.

\subsubsection{Numerical verifications}

We have also checked numerically that the spectrum
of the Hamiltonian or transfer matrices for the periodic model can be obtained using braid translation of the different sectors of the
open case. 

The periodic RSOS model is defined on the space of periodic paths. We obtain the first TL generator $e_1$ from \eqref{TLRSOSaction}, and define a translation operator $u$
simply by cyclically shifting the path configurations. The remaining TL generators, including the periodic one, $e_N$, are then produced by application of \eqref{TLpdef-d}.
We have checked that using \eqref{TLRSOSaction} directly for all $e_i$, including $e_N$, leads to the same result.

In the braid translated cases, we act on the space of open RSOS paths with $h_0 = h_N \equiv n$. The TL generators $e_i$ with $i=1,2,\ldots,N-1$ are then defined by
\eqref{TLRSOSaction}, and the blob operator $b$ by \eqref{b-action}. The periodic generator $e_N$ follows from \eqref{eq_defbraid}.

\begin{table}
\begin{center}
\begin{tabular}{|c|c|c|c|c|c|}
\cline{1-6}
  $f = -\frac{1}{2L} \log \lambda $
    & \multicolumn{1}{|c|}{Periodic} & \multicolumn{4}{|c|}{Braid translation} \\  \cline{3-6}

   & RSOS & $n=1$ & $n=2$ & $n=3$ & $n=4$ \\
  \hline
  -0.77266213 & 2  & 1 &   &   & 1 \\
  -0.74367629 & 2  &   & 1 & 1 &  \\
  -0.69983167 & 2  &   & 1  & 1  &  \\
  -0.46696892 & 2  & 1 &   &   & 1 \\
  -0.46678488 & 4  &   & 2 & 2 &  \\
  -0.45247382 & 4  &   & 2 & 2 &  \\
  -0.40836380 & 2  &   & 1 & 1 &  \\
  -0.32080788 & 4  & 2 &   &   & 2 \\
  -0.29293347 & 4  &   & 2 & 2 &   \\
  -0.20841910 & 2  &   & 1 & 1 &   \\
  -0.18536432 & 2  &   & 1 & 1 &   \\
  -0.07103936 & 4  &   & 2 & 2 &   \\
  -0.04360048 & 2  & 1 &   &   & 1 \\

  \hline
\end{tabular}
\end{center}
\caption{Complete set of levels for the periodic RSOS model, for size $N=2L=6$ and $p=4$ (the tricritical Ising model).
 The left column shows $f=  -\frac{1}{2L} \log \lambda $ (rounded to 8 digits), where $\lambda$ is the eigenvalue of the
 transfer matrix $T$. The other columns show the multiplicities of the eigenvalues
 in the periodic RSOS model and in the braid-translated representations. Blank entries denote absent eigenvalues (zero multiplicity).}
  \label{Tab_BraidRSOS}
\end{table}

As a first example, we study in Table~\ref{Tab_BraidRSOS} the spectrum of the resulting periodic transfer matrix,
\begin{equation}
 T= \prod_{i =1}^L (1+e_{2i}) \prod_{i =1}^L (1+e_{2i-1}) \,,
 \end{equation}
for the case $p=4$ (tricritical Ising model) with size $N=2L=6$.  As expected, all the eigenvalues of the periodic
RSOS model can be recovered from braid translation of the various sectors of the open RSOS model.
Accordingly, the multiplicity for each eigenvalue in the braid translated model, summed over the sector label $n$, equals
the corresponding multiplicity in the periodic RSOS model.

\begin{table}
\begin{center}
\begin{tabular}{|c|c|c|c|c|c|c|c|c|c|}
\cline{1-10}
  $f = -\frac{1}{2L} \log \lambda $
    & \multicolumn{1}{|c|}{Periodic} & \multicolumn{5}{|c|}{Braid translation} & \multicolumn{3}{|c|}{Potts spin} \\  \cline{3-7} \cline{8-10}

   & RSOS & $n=1$ & $n=2$ & $n=3$ & $n=4$ & $n=5$ & Periodic & $\mathbb{Z}_2$ & $\mathbb{Z}_3$ \\
  \hline
  -0.82187352 & 2 & 1 &   &   &   & 1 & 1 &   &   \\
  -0.77284238 & 2 &   & 1 &   & 1 &   &   & 1 &   \\
  -0.69959761 & 2 &   &   & 2 &   &   & 2 &   & 1 \\
  -0.61485446 & 2 &   & 1 &   & 1 &   &   & 1 &   \\
  -0.50252627 &   &   &   &   &   &   &   &   & 2 \\
  -0.45912387 &   &   &   &   &   &   &   & 2 &   \\
  -0.40029661 &   &   &   &   &   &   &   &   & 2 \\
  -0.25126313 & 6 &   & 2 & 2 & 2 &   & 2 & 2 & 1 \\
  -0.18317902 & 2 & 1 &   &   &   & 1 & 1 &   &   \\
  -0.11988196 & 2 &   & 1 &   & 1 &   &   & 1 &   \\
  -0.10222966 &   &   &   &   &   &   &   &   & 2 \\
  -0.05419180 & 2 &   &   & 2 &   &   & 2 &   & 1 \\
  -0.04340240 &   &   &   &   &   &   &   & 2 &   \\
  \ 0.00000000 &   &   &   &   &   &   & 1 &   &   \\
  \hline
\end{tabular}
\end{center}
\caption{Complete set of levels for the periodic RSOS model, for size $N=2L=4$ and $p=5$ (the $3$-state Potts model).
 The left column shows $f=  -\frac{1}{2L} \log \lambda $ (rounded to 8 digits), as in Table~\ref{Tab_BraidRSOS}.
 The middle columns show the multiplicities of the eigenvalues
 in the periodic RSOS model and in the braid-translated representations.
 The rightmost columns give the multiplicities in the Potts spin model with various boundary conditions (see main text).
 Blank entries denote absent eigenvalues.
  All the eigenvalues of the periodic RSOS model can be recovered by braid translation of the open model, and they form
  a proper subset of the union of eigenvalues in the Potts spin model.}
  \label{Tab_BraidRSOS_Q3}
\end{table}

We next consider the case $p=5$ (the three-state Potts model). The transfer matrix spectra are shown in Table~\ref{Tab_BraidRSOS_Q3} for
size $N=2L=4$. We see again that all the eigenvalues of the periodic RSOS model can be recovered by braid translation of the open model,
using all possible sector labels $n=1,2,\ldots,p$.

In addition to this, Table~\ref{Tab_BraidRSOS_Q3} also shows the spectrum of the periodic Potts model in the spin representation.
We have used three different types of boundary conditions. The first one is just the standard periodic boundary condition on the Potts spins.
The other two boundary conditions are obtained by replacing the factor $\delta_{\sigma_L,\sigma_1}$ appearing in the Boltzmann weights
by $\delta_{s(\sigma_L),\sigma_1}$, where $s \in {\cal S}_3$ is some non-trivial permutation of the three spin labels. One choice of $s$,
that we shall call $\mathbb{Z}_2$ boundary conditions, is obtained by choosing $s~: (1,2,3) \to (2,1,3)$, {\em i.e.}, 
by permuting two of the spin labels while keeping the third one unchanged. The other relevant choice of $s$, called
$\mathbb{Z}_3$ boundary conditions, corresponds to a cyclic permutation of the labels, $s~: (1,2,3) \to (2,3,1)$.

We observe from the table that the spectrum with periodic boundary conditions contains the braid translated spectra with $n=1$ and $n=3$,
but there is one more eigenvalue in the spin model spectrum (at this size). Going back to the dimension formula \eqref{eq:Mpnab} we find
that
\begin{equation}
 M_{5,N}(1,1) + M_{5,N}(3,3) + M_{5,N}(1,5) = \frac{3^L+3}{6} + \frac{2}{3} \times 3^L +  \frac{3^L-3}{6} = 3^L \,, \label{dimconject1}
\end{equation}
the correct dimension of the spin transfer matrix. We therefore conjecture that the missing part of the spectrum of the periodic spin-model
transfer matrix can be obtained by braid translating the open RSOS representation with $(h_0,h_N) = (1,5)$.
Similarly, the spectrum with $\mathbb{Z}_2$ boundary conditions contains the braid translated spectrum with $n=2$, but again contains
further eigenvalues. From the observation
\begin{equation}
 M_{5,N}(2,2) + M_{5,N}(2,4) = \frac{3^L+1}{2} + \frac{3^L-1}{2} = 3^L
\end{equation}
we conjecture that the missing part of the spectrum is the braid translation of the $(h_0,h_N) = (2,4)$ representation.
Finally, the spectrum with $\mathbb{Z}_3$ boundary conditions contains {\em half} of spectrum obtained by braid translation of the $n=3$ representation.
This is possible because the latter is a direct sum of two isomorphic representations, as follows from the fact that the corresponding weight of
non-contractible loops is zero \cite{GrahamLehrer,Jacobsen14}. The missing eigenvalues are this time accounted for by observing that
\begin{equation}
 \tfrac{1}{2} M_{5,N}(3,3) + M_{5,N}(3,1) + M_{5,N}(3,5) = 3 \times 3^{L-1} = 3^L \,. \label{dimconject3}
\end{equation}

We have checked explicitly that braid translation of the RSOS representations, in accordance with eqs.~(\ref{dimconject1})--(\ref{dimconject3}), indeed provides all
the eigenvalues of the Potts spin transfer matrix reported in Table~\ref{Tab_BraidRSOS_Q3}.

Summarizing, we have seen that all eigenvalues present in the three-state Potts model with periodic and twisted periodic boundary conditions of 
$\mathbb{Z}_2$ and $\mathbb{Z}_3$ type can be obtained by braid translating open RSOS representations labeled $(h_0,h_N)$, including the
non-diagonal ones with $h_0 \neq h_N$. Moreover, spectra for these three types of boundary conditions on the spin model exhausts those coming
from all possible braid translated representations (up to isomorphisms) in the boundary RSOS model. In this respect, the three boundary conditions imposed on the spin model
can be considered complete (in the case of periodic boundary conditions).%
\footnote{A similar, but less complete set of observations was made in \cite[Section 6.1.2]{Salas17}.}

\bigskip

Note that the RSOS models we have discussed so far can be considered as models  where the heights take values on an $A_p$ Dynkin diagram. Their continuum limit (to be discussed next) corresponds to so called diagonal modular invariants, which involve only scalar primary fields (with conformal weights for the chiral and anti-chiral Virasoro algebras being the same, $h=\bar{h}$). There exists also RSOS models where the heights take values on other Dynkin diagrams: these models also provide representations of the affine Temperley-Lieb algebra, and involve different simple modules than the ones we have discussed here. Braid translation can be generalized to this case as well \cite{PasquierSaleur}. 

\subsection{Continuum limit}
\label{sec:cont-limit-JS-blob}

We shall now discuss what becomes of the correspondence between representations of the blob algebra and representations of the periodic Temperley-Lieb algebra when going to the continuum limit. 
To that end, we need  once again a detour through some technicalities.

We introduce the usual notations for the central charge and the conformal weights
\begin{subequations}
\begin{eqnarray}
\displaystyle c &=& 1 - \frac{6}{x (x+1)}, \\
\displaystyle h_{r,s} &=& \frac{ \left[(x+1)r - xs \right]^2 - 1}{4 x (x+1)},\label{eq_KacChar}
\end{eqnarray}
\end{subequations}
where with this parametrization $\q = \mathrm{e}^{i \pi/ (x+1)}$ and $r,s\in \mathbb{Z}$ are so-called Kac labels.

We now consider the continuum limit of the lattice models studied in the previous subsections. The simplest object to consider is the generating function of  energy levels (eigenvalues) of the Hamiltonian  $H=\sum_{i=1}^{2L} e_i$ that approach the ground state energy like $1/L$ when $L\to \infty$. A well known analysis of this generating function based on the conformal map between the complex plane and the cylinder leads to  \cite{CardyScaling} 
\begin{equation}
 \hbox{Tr}\, e^{-\beta_R(H-N\varepsilon_0)}e^{-i\beta_I P}\;\xrightarrow{\, N\to\infty\,}\; \hbox{Tr}\, q^{L_0-c/24}\bar{q}^{\bar{L}_0-c/24},\label{charform}
 \end{equation}
 where $H$ is the lattice hamiltonian  (normalized such that the velocity of sound is unity), and we have also introduced the lattice momentum  $P$ in order to keep track of the left and right moving Virasoro components, while $\varepsilon_0$  is the ground state energy per site in the thermodynamic limit, and we now  set $q(\bar{q})=\exp\left[-{2\pi\over N}(\beta_R\pm i\beta_I)\right]$ with $\beta_{R,I}$ real and $\beta_R>0$, and $N=2L$ is the length of the chain.

If we now evaluate the generating function in the modules ~$\aSt_{j,\mathrm{e}^{2 i K}}$, e.g. identifying them with the $S_z$ sectors as in Sec.~\ref{sec:twistedXXZ}, we find the well known result~\cite{PasquierSaleur}
\begin{equation}\label{eq_F}
\displaystyle F_{j,\mathrm{e}^{2iK}} \equiv \mathrm{Tr}_{\aSt_{j,\mathrm{e}^{2iK}}}e^{-\beta_R(H-N\varepsilon_0)}e^{-i\beta_I P} \;\xrightarrow{\, N\to\infty\,}\; \frac{q^{-c/24}\bar{q}^{-c/24}}{\Pi(q) \Pi(\bar{q})} \sum_{e \in \mathbb{Z}} q^{h_{e+{K\over\pi},-j}} \bar{q}^{h_{e+{K\over\pi},j}},
\end{equation}
where 
\begin{equation}
\displaystyle \Pi(q) =  \prod_{n=1}^{\infty} (1 - q^n) = q^{-1/24} \eta (q).
\end{equation}

We now discuss the scaling limit for $j=0$ 
and consider  the braid-translated TL representation $\StTL{0}$ 
which is $\bStJTL{0}{\q^2}$, as the affine TL representation.
Following the discussion in Sec.~\ref{sec:twistedXXZ}, the trace over the scaling limit of the representation $\bStJTL{0}{\q^2}$ thus reads
\begin{equation}\label{eq:Fbar}
\displaystyle  \mathrm{Tr}_{\bStJTL{0}{\q^2}} q^{L_0-c/24}\bar{q}^{\bar{L}_0-c/24} = F_{0,\q^2}-F_{1,1}.
\end{equation}

For each of the irreducible representations of the blob algebra with  $j=0$ (no through lines) and $y=y_n$, using $\brt$ we obtain the corresponding irreducible
quotient $\aX_{0,\q^{2n}}$ of the standard affine TL representation with $j=0$ and
$\mathrm{e}^{2iK} = \q^{2n}$. The continuum limit of the generating function of levels for the simple modules of the affine Temperley-Lieb algebra appearing in the RSOS models can be obtained by simple manipulations \cite{PasquierSaleur} using~\eqref{eq_F} and~\eqref{eq:Fbar}:
\begin{equation}\label{eq:X-lim}
\displaystyle  \mathrm{Tr}_{\aX_{0,\q^{2n}}} q^{L_0-c/24}\bar{q}^{\bar{L}_0-c/24}= \displaystyle \sum_{r=1}^{p-1} \chi_{r,n} \cdot \bar{\chi}_{r,n}\ ,
\end{equation}
where $\chi_{r,n}:=\chi_{r,n}(q)$  are characters of the irreducible Virasoro algebra representations $\IrrV{r,n}$ of  the conformal weight $h_{r,n}$, and similarly  $\bar{\chi}_{r,n}:=\chi_{r,n}(\bar{q})$ for the anti-chiral part.

As discussed in our earlier works~\cite{GRS1,GRS3,GRSV}, the correspondence between the  Temperley-Lieb algebra and the left-right Virasoro algebras $\VirN$ goes far beyond the identification of generating functions. For instance,  expressions in terms of the affine Temperley-Lieb generators are conjectured, that reproduce the action of the Virasoro generators when acting on low energy levels as $N\to\infty$  \cite{KooSaleur}, and this was in fact shown in several cases~\cite{KooSaleur,GRS1}. It is also believed (although this has only been worked out in the open or boundary case)  that a  fusion of Temperley-Lieb modules can be defined on the lattice that matches fusion in the conformal continuum limit.  We indicate this deep (and not yet fully explored) correspondence by the symbol $\mapsto$ (see \cite{GRS1,GRS2,GRS3,GRSV}) and thus write the limit on the level of representations and not only generating functions  (see, however, a more rigorous account to this in~\cite{GRS3})
\begin{equation}\label{eq:Xoplusr}
\aX_{0,\q^{2n}}\mapsto \bigoplus_{r=1}^{p-1} \IrrV{r,n}\otimes \IrrVb{r,n}\ ,
\end{equation}
where the summands  are the irreducible representations of the non-chiral Virasoro algebra $\VirN$ of the conformal weights $(h_{r,n},\bar{h}_{r,n})$.

It is natural at this stage to wonder what is the continuum limit of the representations of the blob algebra we have used in the open case. An  analysis similar to what we did in the closed case can be done to relate the blob algebra and the Virasoro algebra, based, in particular, on a relation similar to (\ref{charform})
\begin{equation}
 \hbox{Tr}\, e^{-\beta(H-N\varepsilon_0-\varepsilon_s)}\;\xrightarrow{\, N\to\infty\,}\; \hbox{Tr}\, q^{L_0-c/24},\label{charform1}
 \end{equation}
 where now $q=\exp\left[-{\pi\over N}\beta\right]$ and $\varepsilon_s$ is the surface energy. It is best in this case to take for  Hamiltonian $H=b\sum_{i=1}^{2L-1} e_i$; since only products $be_i$ appear, the underlying algebra is strictly speaking a subalgebra of the blob algebra \cite{JS}, which we shall henceforth refer to as the  ``JS blob algebra''
following~\cite{GJSV}. The RSOS model with $h_0=h_N=n$ then gives an irreducible representation $\rho_{n,n}$ and the usual analysis of the continuum limit\footnote{It is even possible to consider the limit where the heights on the left and on the right are different. In this case one finds
$
\rho_{n_1,n_2}\mapsto \IrrV{n_1,n_2}
$
so all the characters of the corresponding rational CFT can be obtained. }
 gives  \cite{GJSV}
\begin{equation}
\rho_{n,n}\ \mapsto \ \IrrV{n,n}\ .
\end{equation}
The combination of the braid translation $\brt$ and continuum limit can then be sketched in the following diagram:
\begin{equation}\label{commutdiag}
\xymatrix@C=55pt@R=35pt@M=10pt@W=10pt{
\rho_{n,n} \ar@{|->}[r]^{\mbox{}\quad N\to\infty \quad} \ar[d]_{\brt} &  \IrrV{n,n} \ar[d]^{?} \\
\aX_{0,\q^{2n}}  \ar@{|->}[r]^{\; N\to\infty \;}  &  \bigoplus_{r} \IrrV{r,n}\otimes \IrrVb{r,n}
}
\end{equation}%
where the direct sum $\oplus_r$ is as in~\eqref{eq:Xoplusr}.
To conclude this subsection, 
the important message for us is that the algebraic content of the periodic RSOS model is simply related by braid translation to the algebraic content of the RSOS model with conformal invariant boundary conditions. In other words, all the relevant simple modules for the affine Temperley-Lieb algebra are obtained by braid translation from (some of the) relevant simple modules for the blob algebra.  Clearly, the diagram (\ref{commutdiag})
suggests there might be a simple operation, directly in the bulk CFT,  corresponding to the braid translation on the lattice and illustrated by   the right vertical arrow with `?'. But we do not know what it is for now.

\newcommand{\rmi}{\mathrm{i}}
\section{Braid translation of  open SUSY spin chains}\label{sec:susy}

We have seen in the previous sections how the periodic RSOS models can be obtained by braid translation from their open versions, by establishing an explicit correspondence between the  irreducible representations of the affine TL algebra and of the blob algebra.  It is natural to ask whether such a correspondence---which, once again, is not obvious since braid translation produces what seems to be (at least at first sight) a non-local interaction---extends to more complicated models, in particular to models whose continuum limit is a LCFT. One of the simplest set-ups to obtain a LCFT is to take the continuum limit of $\mathfrak{gl}(n|m)$ spin chains with alternating fundamental and conjugate fundamental representations. We now present results for two cases: the $\mathfrak{gl}(1|1)$  chain---whose continuum limit is the well-known symplectic fermions theory---and, somewhat more briefly,  the $\mathfrak{sl}(2|1)$ spin chain, which is a $c=0$ LCFT related with percolation hulls \cite{SaleurPoly,MathieuRidout,SQHE,GRSV}. 

\subsection{The case of the  $\mathfrak{gl}(1|1)$ superspin chain: overview}\label{sec:gl11chain}

The $\mathfrak{gl}(1|1)$ spin chain corresponds to the following TL generators
\begin{equation}\label{gen-TL-ferm}
\displaystyle e_{i} = (f_i+ f_{i+1})(f^\dag_i + f^\dag_{i+1}) \,,
\end{equation}
with $i=1,\dots,N-1$ in the open case, and $\q=\rmi$, whence $\delta=0$. The $f_i,f_i^\dagger$ are fermionic generators, and the alternating sign on the right hand side  of their (anti)commutation relations, $\{ f_i, f_j^{\dagger}\} = (-1)^{i} \delta_{ij}$, originates from the alternating representations in the $\mathfrak{gl}(1|1)$ formulation. The boundary conditions for the fermions are ``free'', and this is known to correspond to conformal boundary conditions in the continuum limit \cite{RS3}. In terms of the blob algebra, we have simply $b=\one$ and the open chain Hamiltonian is as usual $H=-\sum_{i=1}^{N-1} e_i$.

The $\TL{N}$-representations  content of the open $\mathfrak{gl}(1|1)$ spin chain consists of the standard modules~$\StTL{j}$ for $\q=\rmi$, each arising 
with a multiplicity $(2j+1)$. These modules themselves have the subquotient structure 
\begin{equation}\label{eq:Wj-str}
\StTL{j}=~~~
\begin{array}{ccc}
\IrrTL{j}&&\\
&\hskip-.2cm\searrow&\\
&&\hskip-.3cm \IrrTL{j+1}
\end{array} \\,
\end{equation}
where the $\IrrTL{j}$ are the $\TL{N}$ simple  modules. Each of the simples $\IrrTL{j}$ gives rise, after braid translation, to modules 
$ \IrrJTL{j}{(-1)^{j+1}}$ of the affine TL algebra, actually of its finite-dimensional quotient -- the Jones-Temperley-Lieb algebra $\rJTL{N}(\delta)$ introduced in Sec.~\ref{sec:ATL-def}. Hence the content of the closed chain obtained by the braid translation $\brt$ consists of the modules $ \IrrJTL{j}{(-1)^{j+1}}$, each with multiplicity $2j+1+2(j-1)+1=4j$. This  and the result in~\cite{GRS2} imply that the spectrum of the braid translated chain with the Hamiltonian $H^\brt = -\sum_i e_i$, where now $e_N$ is set to $e_N^\brt$, is the same as the one of the genuine periodic $\mathfrak{gl}(1|1)$ chain. 

We can explore the equivalent result in the continuum limit. We have seen above~\eqref{eq:hatd} that the   $\ATL{N}(\delta)$ module generated from $\StTL{j}$ by the braid translation  corresponds to the quotient $\aSt_{j,\q^{2j+2}} / \aSt_{j+1,\q^{2j}}$. Using the result (\ref{eq_F}) for the continuum limit of modules, we find the corresponding spectrum of the braid-translated Hamiltonian $H^\brt$ for $\q=\rmi$ is
\begin{equation}
\displaystyle F_{j,(-1)^{j+1}}-F_{j+1,(-1)^{j}} \ ,
\end{equation}
Meanwhile, since each of the modules $\StTL{j}$ comes with a multiplicity $(2j+1)$, so the generating function of levels in the continuum limit reads\footnote{This can be interpreted as well as a partition function with antiperiodic boundary conditions in the ``imaginary time direction''.}   
\begin{equation}
\displaystyle Z^{\brt}= \sum_{j=0}^{\infty} (2j+1) \left( F_{j,(-1)^{j+1}}-F_{j+1,(-1)^{j}} \right) = F_{0,-1} + 2 \sum_{j=1}^{\infty} F_{j,(-1)^{j+1}}.
\end{equation}
This is in fact the same as the generating function for the spectrum of the ``genuine'' periodic $\mathfrak{gl}(1|1)$ chain --- where the generator $e_N= (f_N+f_1)(f_N^\dagger +f_1^\dagger)$ --- in the continuum limit \cite{GRS3}.

While the content in terms of simples is the same for the braid translated and periodic spin chains, the action of the two images of $\ATL{N}$ is different. 
Indeed, from the results in \cite{RS3} for the open chain, we see that  the standard modules in the braid translated chain get glued according to the usual diamond pattern~\cite{RS3}
\begin{equation}\label{eq:diamond}
 \begin{array}{ccccc}
      &&\hskip-.7cm \IrrJTL{j}{(-1)^{j+1}} &&\\
      &\hskip-.2cm\swarrow&\searrow&\\
     \IrrJTL{j-1}{(-1)^{j}} &&&\hskip-.3cm \IrrJTL{j+1}{(-1)^{j}} \\
      &\hskip-.2cm\searrow&\swarrow&\\
      &&\hskip-.7cm \IrrJTL{j}{(-1)^{j+1}}&&
\end{array}
\end{equation}
These indecomposable modules appear as direct summands in the closed spin-chain obtained by  the braid translation. This structure of the affine TL or rather $\rJTL{N}$ modules is obviously quite different from the structure for the genuine  periodic $\mathfrak{gl}(1|1)$ spin-chain studied  in~\cite{GRS2}. 

This difference carries over to the continuum limit. 
 If we denote by $F^{(0)}_{j,\mathrm{e}^{2iK}}$ the generating functions of conformal weights corresponding to the simple modules  $\IrrJTL{j}{\mathrm{e}^{2iK}}$of the affine TL algebra, we will have
 \begin{equation}
\displaystyle F_{j,(-1)^{j+1}}-F_{j+1,(-1)^{j}} = F^{(0)}_{j,(-1)^{j+1}}+F^{(0)}_{j+1,(-1)^{j}}.
\end{equation}
The operator content of the  simple modules then reads (using the structure of $\StTL{j}$ in~\eqref{eq:Wj-str})
\begin{equation}
\displaystyle  F^{(0)}_{j,(-1)^{j+1}} = \sum_{j'=j}^{\infty} (-1)^{j'-j} \left( F_{j',(-1)^{j'+1}} - F_{j'+1,(-1)^{j'}}\right) = \sum_{j'=j}^{\infty} (-1)^{j'-j} \sum_{r=1}^{\infty} K_{r,1} \bar{K}_{r,1+2j'} \ ,
\end{equation}
where $K_{r,s}$ is  the character  of the so-called Kac module -- it is the quotient of the Verma module $\Verma{r,s}$ by its submodule $\Verma{r,-s}$:
\begin{equation}
K_{r,s}=q^{-c/24}~{q^{h_{r,s}}-q^{h_{r,-s}}\over\Pi(q)} \ .
\end{equation}
One can then readily expand the Kac characters $K_{r,s}$ onto the irreducible Virasoro characters~\cite{Characters} to extract the left and right Virasoro content of the simple modules, as it was done in~\cite{GRS3}. Of course, this left-right Virasoro content matches the content of symplectic fermions with periodic boundary conditions. However, the glueings and arrows between modules will turn out---like in the lattice case---to be different, as we shall now see.

\newcommand{\chib}{\boldsymbol{\chi}}
\newcommand{\chibd}{\boldsymbol{\chi}^{\dagger}}
\newcommand{\etab}{\boldsymbol{\eta}}
\newcommand{\etabd}{\boldsymbol{\eta}^{\dagger}}

\subsection{Braid-translated $\mathfrak{gl}(1|1)$ spin chain: exact spectrum}
We have seen that the continuum partition function of the braid-translated spin chain
is the same as that of the usual periodic chain.
This statement is not only true in the continuum limit, but also in finite size.
Recall the expression of the TL generators~\eqref{gen-TL-ferm}. We choose in what follows the blob operator $b=\one$. For $N=2L$ even or odd,  the braid translation~\eqref{eq_defbraid2} in the ordinary TL case then gives the generator $e^{\brt}_{N}$ which can be simply written as
\begin{equation}\label{eq:e-braid}
\displaystyle e^{\brt}_{N}
= \fbr_N(\fbr_N)^{\dagger} \,,
\end{equation}
where
\begin{equation}
\fbr_N = f_1 + (1-i) \sum_{j=2}^{N-1} f_j + (-1)^N i^{N\,\mathrm{mod}\,2} f_N
\end{equation}
and $(\cdot)^{\dagger}$ is the hermitian conjugation replacing $f_j$ by $f_j^{\dagger}$ and $i$ by $-i$. The derivation of these expressions is given in App.~\ref{app:A1}.
Notice that this last generator $e^{\brt}_{N}$ is highly non-local in terms of the fermions, and couple all
the sites with a mean-field like interaction. This is fundamentally different from the case of
the minimal models, since the braid-translated last generator does not reduce to the usual periodic 
expression $e_{N} = (f^\dag_1 + f^\dag_{N})(f_1+ f_{N})$.

We now define a new Hamiltonian as
\begin{equation}\label{eq:Hbrt-def}
\displaystyle H^{\brt}= -\sum_{i=1}^{N-1} e_i - e^{\brt}_{N} = H - e^{\brt}_{N} + e_{N},
\end{equation}
where $H=-\sum_{i=1}^{N} e_i$ is the 
original  Hamiltonian of the periodic $\mathfrak{gl}(1|1)$ spin-chain. Writing  this way  the Hamiltonian $H^{\brt}$ for the braid-translated open spin-chain allows us
to use results of~\cite[Sec. 4]{GRS1} for diagonalisation of this Hamiltonian. 
We refer for details to App.~\ref{app:A} where we do an analysis similar to~\cite{GRS1}. Introducing particular linear combinations~\eqref{eq-modes} of Fourier modes of the lattice fermions $f_j$ and $f_j^{\dagger}$ --- the fermions  $\chib_p$ and $\etab_p$ ---  we find that the braid-translated Hamiltonian can be written in (almost) diagonal form
\begin{equation}\label{Hbr-chi}
\displaystyle H^{\brt} = 4 \chib_0^\dag \etab_0 + 2 \sum_{\substack{p=\step\\\text{step}=\step}}^{\pi-\step}  \sin p \left( \chib_p^\dag \chib_p - \etab_p^\dag \etab_p \right),
\end{equation}
where $\step= \frac{2 \pi}{N}$.
It has formally the same expression as for the usual periodic spin chain (compare with~\cite[Eq.\ (4.5)]{GRS1}), so that in particular, the spectrum is the same and so is the whole Jordan
cell structure.
This is quite expected since the partition functions were also the same.

\subsection{Lattice Virasoro modes out of the braid translation}
\label{sec:lat-Vir-brt}
Since the spectrum and the Jordan cells are the same as in the usual periodic case, we might be tempted to think that the models are the same. However, the difference appears in  expressions for all other modes $L_n$ and
$\bar{L}_n$ of the stress tensor field in the continuum limit.
To show this, consider the Fourier transform of $e_j$'s following~\cite{GRS1}:
\begin{equation}\label{Hbr-n}
\displaystyle H^{\brt}(n)= -\sum_{j=1}^{N-1}  e^{-iqj}  e_j - e^{-iqN} e^{\brt}_{N},
\quad \mbox{with } q={\ffrac{n\pi}{L}},
\end{equation}
where $n$ is integer and recall that $N=2L$.
Following lines in App.~\ref{app:A3}, the ``higher modes'' $H^{\brt}(n)$  can be  rewritten in terms of the $\chib^{(\dagger)}$s and
$\etab^{(\dagger)}$s fermions as
\begin{multline}\label{lat-Vir-H}
H^{\brt}(n) =
2e^{iq/2}\Biggl((1+i)\sqrt{\sin{q}} \, \chib^{\dagger}_{0}\chib_{q}
+ \sum_{\substack{p=\step\\\text{step}=\step}}^{\pi-q-\step}
\sqrt{\sin{(p)}\sin{(p+q)}} \bigl(\chib^{\dagger}_{p}\,\chib_{p+q} -
\etab^{\dagger}_{p}\,\etab_{p+q}\bigr)\\
+ \sum_{\substack{p=\step\\\text{step}=\step}}^{q-\step}
\sqrt{\sin{(p)}\sin{(q-p)}}\bigl(\chib^{\dagger}_{\pi-p}\,\etab_{q-p} +
\etab^{\dagger}_{\pi-p}\,\chib_{q-p}\bigr) + (1-i) \sqrt{\sin{q}} \, \etab^{\dagger}_{\pi-q}\etab_{0}\Biggr),
\end{multline}
for $0< n<L$ and in~\eqref{lat-Vir-H-m} for negative $n$.
It is interesting to note that the $\chib$-$\etab$ fermionic expression for $H^{\brt}(n)$ and the one for $H(n)$ in the truly periodic spin chain from~\cite[Sec.\ 4.3.3]{GRS1} have differences only in the zero mode contributions. This will modify the Virasoro representation in the continuum limit of our new ``periodic'' spin chain, as it will be discussed in the next section.

\medskip

We now study the momentum operator for our braid-translated spin chain. Recall that the momentum operator for the usual periodic spin chain was written as~\cite[Sec.\ 4.3.2]{GRS1}
\begin{equation}\label{lattmom}
P=\ffrac{i}{2}\sum_{j=1}^{N} [e_j,e_{j+1}].
\end{equation}
This operator is now modified by replacing $e_N\to \ebr_N$:
\begin{equation}
\Pbr = P + \ffrac{i}{2}\Bigl([e_N-\ebr_N,e_{N-1}] - [e_N-\ebr_N,e_{1}]\Bigr),
\end{equation}
and its fermionic expression  is (see App.~\ref{app:A3})
\begin{equation}\label{Pbr}
\Pbr = 4\chibd_0\etab_0 +  \sum_{\substack{p=\step\\\text{step}=\step}}^{\pi-\step} \sin{2p}\bigl(\chibd_p\chib_p + \etabd_p\etab_p\bigr).
\end{equation}
We stress that this expression now differs from the one in the ordinary periodic spin chain~\cite[Eq.\ (4.22)]{GRS1} by the zero modes term $\chibd_0\etab_0$ responsible for non-trivial Jordan blocks. 
This agrees with our observation that the translation operator $u^2$ defined via~\eqref{eq_braid_trans} also admits Jordan blocks, i.e.\ non-diagonalizable, in contrast to the ordinary periodic spin chain where $u^N=\mathrm{id}$.

The Fourier equivalent of the momentum operator $\Pbr$ is introduced as
\begin{equation}\label{Pbr-def}
\Pbr(n) = \ffrac{i}{2}\sum_{j=1}^{N} {\rm e}^{-iqj} [e_j,e_{j+1}]  + \ffrac{i}{2}\Bigl(e^{iq}[e_N-\ebr_N,e_{N-1}] - [e_N-\ebr_N,e_{1}]\Bigr),
\end{equation}
with $q=\frac{n\pi}{L}$. In terms of the modes $\chib_p$ and $\etab_p$, we refer to the expression in~\eqref{Pbr-fer}.

We next study the continuum or scaling limit of the operators $\Hbr(n)$ and $\Pbr(n)$ for any integer~$n$.

\subsection{Scaling limit of the braid-translated $\mathfrak{gl}(1|1)$ spin chain}
\label{sec:gl11-lim}
The scaling limit of the original periodic  $\mathfrak{gl}(1|1)$ spin chain was identified~\cite{GRS1} with the bulk theory of  symplectic fermions. The idea of this section is to compare this result with the scaling limit of the  braid-translated $\mathfrak{gl}(1|1)$ spin chain. In what follows when we talk about the scaling limit, we follow the lines of~\cite{GRS1}, e.g.\ the arrow $\mapsto$ here  denotes the scaling limit of  operators. 

Let us introduce the chiral and anti-chiral  symplectic-fermions modes $\psi^{1,2}_m$ and  $\bar\psi^{1,2}_m$, see e.g.~\cite[Sec.\,5]{GRS1}.
The symplectic fermions are fermionic fields~\cite{Kausch1,Kausch2} of conformal dimensions $(0,0)$:
\begin{equation}\label{eq:Phi}
\Phi^{\alpha}(z,\bz) = \phi^{\alpha}_0 - i\psi^{\alpha}_0\log(z\bz) + i\sum_{n\ne0}\ffrac{\psi^{\alpha}_n}{n} z^{-n}
 + i\sum_{n\ne0}\ffrac{\bar{\psi}^{\alpha}_n}{n} \bz^{-n},\qquad \alpha=1,2,
\end{equation}
and the local action is given by
\begin{equation}
S \sim \int d^2z J_{\alpha\beta}\partial \Phi^{\alpha}\bar{\partial}\Phi^{\beta},
\end{equation}
where $J_{\alpha\beta}$ is the two-dimensional symplectic form. By $L_n$ and $\bar{L}_n$ we denote the Virasoro modes of the chiral and anti-chiral energy-momentum tensors $T(z)$ and $\bar{T}(\bar{z})$, respectively.

The lattice fermions $\chib_p$ and $\etab_p$  used above should be appropriately rescaled before identifying them with the modes $\psi^{1,2}_m$ and  $\bar\psi^{1,2}_m$:
\begin{eqnarray}\label{eq:sferm-lat-def}
\begin{split}
\sqrt{m}\,\chib_p\mapsto\sfermm_m,\qquad\sqrt{m}
\,\bar{\etab}^\dagger_p\mapsto\sfermp_m,\qquad\sqrt{m}\,\bar{\chib}_p\mapsto\bsfermm_m,
\qquad-\sqrt{m}\,\etab^\dagger_p\mapsto\bsfermp_m,\\
-\sqrt{m}\,\bar{\etab}_p\mapsto\sfermm_{-m}, \qquad\sqrt{m}\,\chib^\dagger_p\mapsto\sfermp_{-m},
\qquad\sqrt{m}\,\etab_p\mapsto\bsfermm_{-m}, \qquad\sqrt{m}\,\bar{\chib}^\dagger_p\mapsto\bsfermp_{-m},
\end{split}
\end{eqnarray}
where we consider any finite integer $m>0$ and set $p=\frac{m\pi}{L}$ and  $\bar{\etab}^{(\dagger)}_p=\etab^{(\dagger)}_{\pi-p}$, and
$\bar{\chib}^{(\dagger)}_p=\chib^{(\dagger)}_{\pi-p}$, while  the scaling limit for zero modes is
\begin{eqnarray}\label{eq:sferm-zero-def}
\begin{split}
\sqrt{L/\pi}\,\etab_0&\mapsto \sfermm_0=\bsfermm_0,
&\qquad
 \sqrt{L/\pi}\,\chib_0^{\dagger}&\mapsto\sfermp_0=\bsfermp_0,\\
\sqrt{\pi/L}\,\etab^{\dagger}_{0}&\mapsto i\phip_0,& \qquad
\sqrt{\pi/L}\,\chib_{0}&\mapsto -i\phim_0,
\end{split}
\end{eqnarray}
where the zero modes were introduced in~\eqref{eq-modes}.
Indeed, having this limit we obtain
 that the scaling limit of the Hamiltonian $\Hbr$ from~\eqref{Hbr-chi} is the zero mode of the stress energy tensor $T(z)+\bar{T}(\bz)$ in the symplectic fermions CFT:
\begin{align}\label{Hbr-lim}
\ffrac{L}{2\pi}\Bigl(\Hbr+\ffrac{2N}{\pi}\Bigr)\; \mapsto \; 
& 2\sfermp_{0}\sfermm_0 + \sum_{\substack{m\in\oZ\\m\ne 0}}
\left(\nord{\sfermp_{-m}\sfermm_m}+
\nord{\bsfermp_{-m}\bsfermm_m}\right)
-\ffrac{c}{12} \\
&:= L^{\brt}_0+\bar{L}^{\brt}_0-\ffrac{c}{12}
= L_0+\bar{L}_0-\ffrac{c}{12}
,\nonumber
\end{align}
where $c=-2$
and we used here the  fermionic normal ordering prescription
\begin{equation}
\nord{\sferm^{\alpha}_{m}\sferm^{\beta}_n}
=\begin{cases}
\sferm^{\alpha}_{m}\sferm^{\beta}_n&\textit{for}\;\; m<0,\\
-\sferm^{\beta}_n\sferm^{\alpha}_{m}&\textit{for}\;\; m>0.
\end{cases}
\end{equation}
The result~\eqref{Hbr-lim} agrees with the genuine period spin-chain~\cite[Sec.\,3.4]{GRS1}. The difference is in  the conformal spin operator.
Indeed, the momentum operator $\Pbr$ computed in~\eqref{Pbr} gives in  the scaling limit  the conformal spin operator $L^{\brt}_0-\bar{L}^{\brt}_0$:
\begin{align}\label{Pbr-lim}
\ffrac{L}{2\pi}\Pbr \; \mapsto \;
&2\sfermp_{0}\sfermm_0 + \sum_{\substack{m\in\oZ\\m\ne 0}}
\nord{\sfermp_{-m}\sfermm_m} -
\sum_{\substack{m\in\oZ\\m\ne 0}}
\nord{\bsfermp_{-m}\bsfermm_m}\\
&:= L^{\brt}_0-\bar{L}^{\brt}_0\nonumber
\end{align}
and it involves the zero modes. 
We recall that the quadratic words in the fermionic zero modes are responsible for the Jordan blocks in the Hamiltonian~\cite{GRS1}.
Therefore, the operator $L^{\brt}_0-\bar{L}^{\brt}_0$ is non-diagonalizable,
 due to their presence in the mode expansion, in contrast to the expression for $L_0-\bar{L}_0$.
  It will be shown below to  entail drastic consequences on the (non-)physical properties of correlation functions.

From the limits~\eqref{Hbr-lim} and~\eqref{Pbr-lim} we finally obtain expressions for  zero modes of the left and right  Virasoro algebra:
\begin{eqnarray}
L^{\brt}_0 &=& \sfermp_{0}\sfermm_0 + \sum_{m\in\oZ}\nord{\sfermp_{-m}\sfermm_m}\,, \label{L0-def}\\
\bar{L}^{\brt}_0 &=& \sum_{\substack{m\in\oZ\\m\ne 0}}\nord{\bsfermp_{-m}\bsfermm_m}\,.
\end{eqnarray}
These ones differ from the usual expression in the symplectic fermions. The crucial difference is the absence of zero modes in the anti-chiral Virasoro generator. 
In particular, $\bar{L}^{\brt}_{0}$ is diagonalizable, while ${L}^{\brt}_{0}$ is not.

We can go further and obtain expressions for all left and right Virasoro modes $L^{\brt}_n$ and $\bar{L}^{\brt}_n$, and compare them with the standard symplectic-fermions representation of $L_n$ and $\bar{L}_n$. Recall the Fourier transforms $\Hbr(n)$ of $e_j$'s introduced in~\eqref{Hbr-n} and the Fourier equivalent $\Pbr(n)$ of the momentum operator   introduced in~\eqref{Pbr-def}. They are expressed in $\chib$-$\etab$ fermions in~\eqref{lat-Vir-H} and~\eqref{Pbr-fer}, respectively. Linearizing the dispersion relation, we obtain the contribution corresponding to low-lying
excitations over the ground state, {\em i.e.}, the scaling limit, for non-zero values of $n$:
\begin{equation}\label{scal-lim-Hn}
\ffrac{L}{2\pi}\Hbr(n) \mapsto i \sfermp_{0}\sfermm_{n} -i \sfermp_{n}\sfermm_{0} + \sum_{m\in\oZ}\sfermp_{-m}\sfermm_{m+n} +
\sum_{\substack{m\in\oZ\\m\ne 0,n}}\bsfermp_{-m}\bsfermm_{m-n} :=
L^{\brt}_{n}+\bar{L}^{\brt}_{-n},\qquad n\in\oZ,
\end{equation}
and
\begin{equation}\label{scal-lim-Pn}
\ffrac{L}{2\pi}\Pbr(n) \mapsto i \sfermp_{0}\sfermm_{n} -i \sfermp_{n}\sfermm_{0} + \sum_{m\in\oZ}\sfermp_{-m}\sfermm_{m+n} -
\sum_{\substack{m\in\oZ\\m\ne 0,n}}\bsfermp_{-m}\bsfermm_{m-n} :=
L^{\brt}_{n}-\bar{L}^{\brt}_{-n},\qquad n\in\oZ,
\end{equation}
where the last equalities are by definition. From them we conclude that $\bar{L}^{\brt}_{-n}=\sum_{m\in\oZ}\bsfermp_{-m}\bsfermm_{m-n}$ and contains no zero modes.  Furthermore, the expression for $\bar{L}^{\brt}_{-n}$ coincides with the representation of the anti-chiral Virasoro generators $\bar{L}_{-n}$ in the symplectic fermions CFT but restricted on the socle of the full  space of states---the subspace annihilated by both zero fermionic modes $\sfermp_0$ and $\sfermm_0$. As the kernel of $\sfermp_0$ and $\sfermm_0$ decomposes onto a direct sum of irreducible representations of the left-right Virasoro algebras, we thus see that the anti-chiral Virasoro representation  is fully reducible  in the scaling limit of the braid-translated spin chain (it is not so of course in the scaling limit of the truly periodic chain).  The ``indecomposable  but reducible" part of the theory is thus concentrated in the holomorphic sector only.
This will be also confirmed by  the zero value of the logarithmic coupling $\bar{\beta}$  measured from the lattice below.\footnote{More precisely, the coefficient that would be the logarithmic coupling $\beta$ if we had a Jordan cell is found analytically and numerically to be zero.}

It is interesting to analyze the left Virasoro action
\begin{equation}\label{Ln-def}
L^{\brt}_n=i \sfermp_{0}\sfermm_{n} -i \sfermp_{n}\sfermm_{0} + \sum_{m\in\oZ}\sfermp_{-m}\sfermm_{m+n} \,,
\end{equation}
which looks very similar to the original action $L_n=\sum_{m\in\oZ}\sfermp_{-m}\sfermm_{m+n}$. We first check by a direct calculation that $L^{\brt}_n$ do satisfy the Virasoro relations:
\begin{equation}
[L^{\brt}_n,L^{\brt}_m] = (n-m) L^{\brt}_{n+m} -\ffrac{1}{6}(n^3-n)\delta_{n+m,0}
\end{equation}
with the definition of $L^{\brt}_0$ given in~\eqref{L0-def}. We  thus see that the modified action still gives a representation of the Virasoro at $c=-2$. This representation of the chiral part is actually isomorphic to the original action $L_n$ in the symplectic fermions theory. In particular, we have the same indecomposability parameter $\beta_{1,5}=-1$ for the first excited states in the Jordan cell for $L^{\brt}_0$. The difference we have in the full theory is only in the right or anti-chiral Virasoro action.

\newcommand{\Phim}{\tilde{\Phi}}
\subsubsection{Non-locality}\label{sec:non-loc}
We can describe the scaling limit of our braid-translated spin chain in terms of modified fermionic fields (compare with the original symplectic fermions in~\eqref{eq:Phi})
\begin{equation}
\Phim^{\alpha}(z,\bz) = \phi^{\alpha}_0 + ((-1)^{\alpha} - i)\psi^{\alpha}_0\log(z) + i \sum_{n\ne0}\ffrac{\psi^{\alpha}_n}{n} z^{-n}
 +  i \sum_{n\ne0}\ffrac{\bar{\psi}^{\alpha}_n}{n} \bz^{-n}.
\end{equation}
with the energy-momentum tensor
\begin{equation}
\tilde{T}(z) = -\ffrac{1}{2} J_{\alpha\beta}\nord{\partial\Phim^{\alpha}\partial \Phim^{\beta}}\,,
\qquad \tilde{\overline{T}}(\bz) = -\ffrac{1}{2} J_{\alpha\beta}\nord{\bar{\partial}\Phim^{\alpha}\bar{\partial} \Phim^{\beta}}\,.
\end{equation}
We then indeed obtain the modes $L^{\brt}_n$ from~\eqref{L0-def} and~\eqref{Ln-def} and $\bar{L}^{\brt}_n=\sum_{\substack{m\in\oZ\\m\ne 0,n}}\bsfermp_{-m}\bsfermm_{m+n}$.

The two-point fuction of the modified fermions is
\begin{equation}
\langle\Phim^{\alpha}(z,\bz)\Phim^{\beta}(w,\bar{w})\rangle = - J^{\alpha\beta}(\log{|z-w|^2} +(-1)^{\alpha}i \log{z} - \log{\bz}),
\end{equation}
and its non-locality is expressed by the non-trivial monodromy
\begin{equation}
\langle\Phim^{\alpha}(e^{2\pi i}z,e^{-2\pi i}\bz)\Phim^{\beta}(0,0)\rangle = \langle\Phim^{\alpha}(z,\bz)\Phim^{\beta}(w,\bar{w})\rangle + J^{\alpha\beta}2\pi ((-1)^{\alpha} - i).
\end{equation}

\medskip 

We will show in  Conclusion section, with more details in App.~\ref{app:inv-brt}, how one can solve this problem of non-locality at least on the formal level. 
We can introduce on the lattice a non-locally (in terms of spin matrices) defined blob operator $b$ that gives, after applying the braid translation~$\brt$, the ordinary expression for $e_{N}$ and in the continuum limit the correct bulk Virasoro action in the LCFT of symplectic fermions.

\subsection{Lattice indecomposability parameters}\label{sec:lat-ind-par}
We demonstrate here our calculation of the logarithmic couplings (also called  indecomposability parameters $\beta$) in the diamond-shape modules of the affine TL algebra encountered in the braid-traslated spin chains, recall the structure of these modules in~\eqref{eq:diamond}. In the open case it is known~\cite{RS3,GV} that these modules  in the continuum limit become the so-called staggered Virasoro modules~\cite{KytolaRidout}, and those are characterized by the logarithmic couplings $\beta=\beta_{1,2j+1}$ (couplings between the primary field and its logarithmic partner of conformal dimension $h_{1,2j+1}$). These couplings can be measured directly on the lattice~\cite{DJS,VJS} using a regularized expression for the Virasoro generators.
We use the results from the previous sections, in particular fermionic expressions for $H^\brt(n)$ that we will also denote as $(L^{\brt}_n + \bar{L}^{\brt}_n)^{(N)}$, for a given number of sites $N$, in finding the analytical expressions for these couplings.

\begin{table}
\begin{center}
\begin{tabular}{|c|c|c|}
  \hline
  $N = 2L$ & $\beta_{1,5}^{(N)}$ & $\bar{\beta}_{1,5}^{(N)}$ \\
  \hline
  10 &  -0.935489 & 0 \\
  12 &  -0.954930 & 0  \\
  14 &  -0.966766 & 0 \\
  16 &  -0.974495 & 0 \\
  18 &  -0.974495 & 0 \\
  20 &  -0.983632 & 0 \\
  22 &  -0.986461 & 0 \\
  \hline
  $\infty$ &  -1.00000 $\pm$ 0.00005 &  0  \\
  \hline
  Conjecture & $-1$ & $0$ \\
  \hline
\end{tabular}
\end{center}
\caption{Measure of $\beta_{1,5}$ and $\bar{\beta}_{1,5}$ for the braid translated  $\mathfrak{gl}(1|1)$ spin chain. Results are different in the holomorphic and antiholomorphic sectors.}
\label{Fig_BraidSymplecticFermions}
\end{table}

We review the approach of measuring the indecomposability parameters on the lattice in App.~\ref{app:gl11} where we also present our numerical and analytical calculation of the $\beta$ parameters in the ordinary periodic $\mathfrak{gl}(1|1)$ spin-chain. Here, we present our result of a similar calculation of these parameters in the braid translated   $\mathfrak{gl}(1|1)$ spin-chain  from Sec.~\ref{sec:gl11chain}, in the Jordan cells for $H^\brt$ corresponding to the first excited states.
Following the lines in  App.~\ref{app:gl11}, we  define the lattice indecomposability parameters for such a Jordan cell in both holomorphic and anti-holomorphic sectors as\footnote{We note that this expression for $ \beta^{(N)}_{1,5}$ slightly differs from the one~\eqref{eq:beta-App} in App.~\ref{app:gl11} -- here, it is modified by the action with $(\bar{L}^{\brt}_{+1})^{(N)}$ which has zero effect in the limit $N\to\infty$ because $\bar{L}^{\brt}_1|\xi\rangle = 0$, as well as $L^{\brt}_1|\xi\rangle = 0$. But this modification allows to use directly $H^\brt(-1)$ from~\eqref{scal-lim-Hn} in the calculation of the matrix elements.}
\begin{subequations}
\begin{eqnarray}
 \beta^{(N)}_{1,5} &=& \frac{\left|\Braket{\psi^{(N)}| (L^{\brt}_{-1}+\bar{L}^{\brt}_{+1})^{(N)}|\xi^{(N)}}\right|^2}{\Braket{\psi^{(N)}| \phi^{(N)}}}, \\
 \bar{\beta}^{(N)}_{1,5} &=& \frac{\left|\Braket{\bar{\psi}^{(N)}| (\bar{L}^{\brt}_{-1}+L^{\brt}_{+1})^{(N)}|{\xi}^{(N)}}\right|^2}{\Braket{\bar{\psi}^{(N)}| \bar{\phi}^{(N)}}},
\end{eqnarray}
\end{subequations}
where  the state $|\xi^{(N)}\rangle$
is the groundstate in the sector $S_z=1$, normalized such that $\Braket{\xi^{(N)} | \xi^{(N)}}=1$, and
 the two vectors $|\phi^{(N)}\rangle$ and $|\psi^{(N)}\rangle$ are the states corresponding to conformal weights $(h,\bar{h})=(1,0)$. The latter two are mixed by the Hamiltonian into a Jordan cell. They belong to the bottom and top nodes, respectively, in the diagram~\eqref{eq:diamond} for $j=1$, while the state $|\xi^{(N)}\rangle$
 is in the left node. And similarly for $|\bar{\phi}^{(N)}\rangle$ and $|\bar{\psi}^{(N)}\rangle$ -- they correspond  to conformal weights $(h,\bar{h})=(0,1)$.
These states are also eigenvectors of the translation operator given by~\eqref{eq_braid_trans}, with the appropriate
eigenvalue consistent with the conformal spin $s=h-\bar{h}=\pm 1$. (We refer to App.~\ref{app:momentum-spin} where it is explained how to identify  conformal spins on the lattice.)

We first give the
numerical results for the $\beta$-parameters, these are shown in Tab.~\ref{Fig_BraidSymplecticFermions}. There are two striking facts: we find $\bar{\beta}^{(N)}_{1,5}=0$ for any size,
and the limit of $\beta_{1,5}^{(N)}$ is $-1$.
This result is clearly surprising, as one usually expects the indecomposability parameters to be the same in the holomorphic and antiholomorphic
sectors.

Further, it is possible to understand this property analytically using free fermions calculation. This is rather technical and
we shall not give the details here.
The path one has to follow is quite clear though, and consists in expressing the Koo-Saleur generators~\eqref{eqKooSaleur} (for the lattice regularisations of $L_n$ and $\bar{L}_n$)
in terms of the fermionic modes~\eqref{eq-modes}. This was done above in Sec.~\ref{sec:lat-Vir-brt}. The indecomposabilty parameters can then be computed analytically, we find
\begin{subequations}
\begin{eqnarray}
 \beta^{(N)}_{1,5} &=& - \ffrac{N}{2\pi} \sin \ffrac{2\pi}{N}, \\
 \bar{\beta}^{(N)}_{1,5} &=& 0,
\end{eqnarray}
\end{subequations}
in  agreement with our numerical results. This confirms the asymmetry between the holomorphic and antiholomorphic sectors.
Another non-physical feature is that $L^{\brt}_0-\bar{L}^{\brt}_0$ is not diagonalizable -- it is due  to presence of the zero modes in~\eqref{Pbr-lim}.
Therefore, the bulk theory obtained by braid translating the open $\mathfrak{gl}(1|1)$ spin chain has several unphysical features.

\subsection{One-parameter family of chiral theories} \label{specialadvert}
As an interesting observation coming out of our previous analysis is  a fermionic construction of Virasoro staggered modules of any possible indecomposability parameter $\beta$.
We first note that  in general if we take
\begin{equation}\label{Vir-gen}
L'_n = a_n \sfermp_{0}\sfermm_{n} + b_n \sfermp_{n}\sfermm_{0} + L_n,
\end{equation}
 with non zero $n$, then the Virasoro algebra relations on $L'_n$ give the only conditions $a_n=a$ and $b_n=b$ for any complex numbers $a$ and $b$, and
 \begin{equation}\label{L0-gen}
 L'_0 = ab\sfermp_{0}\sfermm_{0} + L_0.
 \end{equation}
It is easy to check that the modified action $L'_n$ is isomorphic to the original action $L_n$ whenever $a=-b$ (our case for $L_n^{\brt}$ corresponds to $a=-b=i$). We can see it after rescaling appropriate states or by rescaling $\psi^1_0$ by $(1+b)$ and $\psi^2_0$ by $(1+a)$.  If one of the numbers $a$ or $b$ is $-1$ we cannot do such a rescaling of course and  we obtain degenerate representations discussed below.

 For the cases $a\ne -b$, we actually obtain non-isomorphic representations of the Virasoro algebra at $c=-2$. In particular, the staggered module $\mathcal{P}_{1,5}$, which has the $\beta$-parameter $\beta_{1,5}=-1$ for $a=b=0$, is modified to a non-isomorphic staggered module with
 \begin{equation}\label{beta-mod}
\beta'_{1,5}=-1 - \ffrac{a+b}{ab+1}.
\end{equation}
\textit{ So we can formally construct a one-parameter family of different chiral theories parametrized for example by the $\beta'_{1,5}$.} This family contains chiral symplectic fermions as a special case when $a=-b$ and $\beta'_{1,5}=\beta_{1,5}$. The other members of this family might be interesting for several applications. For example, it is believed that the abelian sand-pile model is described by $\beta$-parameters which differ from the standard ones in the symplectic fermions theory, see~\cite[Sec.\,5]{Ruelle13}. 

Then, we have few degenerate cases if one of the numbers $ab$, $a$, or $b$ is $-1$.
\begin{itemize}
\item If $ab=-1$  we have non-semisimple action for $L'_n$'s and no Jordan cells for $L'_0$:
\begin{enumerate}
\item $a=1$ and $b=-1$ give the representation in the kernel of $\psi^1_0$. It is isomorphic to a direct sum of Feigin--Fuchs modules. So, the action of $L'_0$ is diagonal and we have diagrams for the subquotient structure like $\bullet\rightarrow\bullet\leftarrow\bullet\rightarrow\ldots$
\item  $a=-1$ and $b=1$ give a similar representation but in the kernel of $\psi^2_0$.
\item $a=c$ and $b=-1/c$ for some $c\in\mathbb{C}$ ({\em e.g.}, $a=b=i$) give representations of diamond shape but without Jordan cells for $L'_0$.
\end{enumerate}
\item If $ab\ne-1$  we have several cases with non-trivial Jordan cells for $L'_0$:
\begin{enumerate}
\item The case $a\ne-1$ and $b\ne-1$ was already discussed above with the formula for the $\beta$-parameters in~\eqref{beta-mod}.
\item If $a=b=-1$ we obtain no glueings by $L'_n$ but Jordan cells for $L'_0$. So the diagrams here can be depicted as  $\bullet\xrightarrow{L'_0}\bullet$.
\item If $a=-1$ and $b\ne -1$ (and similarly for $a\ne-1$ and $b=-1$) we obtain two types of glueings: (i) those given by only positive modes $L'_{n>0}$ and  (ii) those given by only negative modes $L'_{n<0}$.
\end{enumerate}
\end{itemize}
There are no other cases. It is interesting if these abstractly constructed Virasoro representations can be described in field theoretic terms.
Indeed, we can formally introduce the operator valued generating functions
\begin{equation}
\Phi^{\alpha}(z,\bz) = \phi^{\alpha}_0 - i(x^{\alpha} + 1)\psi^{\alpha}_0\log(z\bz) + i \sum_{n\ne0}\ffrac{\psi^{\alpha}_n}{n} z^{-n}
 +  i \sum_{n\ne0}\ffrac{\bar{\psi}^{\alpha}_n}{n} \bz^{-n},
\end{equation}
with $x^1=b$ and $x^2=a$.
Then, the Fourier modes of the field
\begin{equation}
T(z) = -\ffrac{1}{2} J_{\alpha\beta}\nord{\partial\Phi^{\alpha}\partial \Phi^{\beta}}
\end{equation}
give us the general fermionic Virasoro representation~\eqref{Vir-gen} and~\eqref{L0-gen}. So we can formally  describe any staggered module, with any $\beta$ parameter (see~\eqref{beta-mod}), by the new fermionic fields. The problem is the two-point functions for such theories
\begin{equation}
\langle\Phi^{\alpha}(z,\bz)\Phi^{\beta}(w,\bar{w})\rangle = - J^{\alpha\beta}(\log{|z-w|^2} +x^{\alpha} \log{|z|^2}).
\end{equation}
which show unusual dependency on $z,w$. We probably can conclude at this step that
 \textit{the only physical $\beta$ parameters are those corresponding to symplectic fermions.}

\subsection{The braid-translated  $\mathfrak{sl}(2|1)$ spin chain}
The simple lesson we have learned with the $\mathfrak{gl}(1|1)$ spin chains appears to extend to all the known more complicated spin chains whose continuum limit is a LCFT. This is, ultimately, because the indecomposable modules of the lattice algebras that appear in the periodic lattice models are considerably more complicated than those appearing in lattice models with open, local boundary conditions. In contrast, for RSOS models, the  restriction to irreducible modules leaves very little choice, and braid translation is enough to relate the open and periodic cases. 

We will not discuss these points in great detail here, but illustrate them a little further. To start, we comment briefly on the case of the  $\mathfrak{sl}(2|1)$ spin chain, which is related with the percolation model~\cite{ReadSaleur01, GRSV}. This chain has been studied in detail in the open case, and was more recently considered in the periodic case as well~\cite{GRSV}.

Using our knowledge of the representation theory of the TL algebra at $\q=\mathrm{e}^{i\pi/3}$~\cite{Martin1,Martin2,Martin3,RS3} (see also recent works~\cite{GV,Ridout:2012gg}), we can deduce the subquotient
structure of the affine TL modules braid-translated  from the TL standard modules $\StTL{j}$:
\begin{equation}
\brt(\StTL{j})= \aSt_{j,\q^{2j+2}} / \aSt_{j+1,\q^{2j}} =~~~~~\left\{\begin{array}{cl}
\begin{array}{ccc}
\IrrJTL{j}{\q^2}&&\\
&\hskip-.2cm\searrow&\\
&&\hskip-.3cm\IrrJTL{j+2}{1}
\end{array} &\hbox{$j\equiv0 \ ($mod$ 3)$}\nonumber\\
\IrrJTL{j}{1} &\hbox{$j\equiv1 \ ($mod$ 3)$}\nonumber\\
\begin{array}{ccc}
\IrrJTL{j}{1} &&\\
&\hskip-.2cm\searrow&\\
&&\hskip-.3cm\IrrJTL{j+1}{\q^2}
\end{array} &\hbox{$j\equiv2 \ ($mod$ 3)$}\nonumber\\
\end{array}\right.
\end{equation}
We note that these modules are also modules over the JTL algebra $\rJTL{N}(1)$ introduced in Sec.~\ref{sec:ATL-def}.
We see that the braid-translated $\StTL{j}$ modules have the subquotient structure (or the arrows pattern) identical to the standard modules of the TL algebra.
In terms of the generating functions~\eqref{eq_F} for energy levels, this implies for example that
\begin{subequations}
\begin{eqnarray}
F_{3n,\q^2}-F_{3n+1,1} &=& F^{(0)}_{3n,\q^2}+F^{(0)}_{3n+2,1} , \\
F_{3n+2,1}-F_{3n+3,\q^{4}} &=& F^{(0)}_{3n+2,1}+F^{(0)}_{3n+3,\q^2}.
\end{eqnarray}
\end{subequations}
The operator content of the simple modules is then readily obtained, for example
\begin{align}
 F^{(0)}_{3n,\q^2} &= \sum_{p=n}^{\infty} \left( F_{3p,\q^2}-F_{3p+1,1} \right) - \sum_{p=n}^{\infty} \left( F_{3p+2,1}-F_{3p+3,\q^4} \right)\notag \\
 &=  \sum_{p=n}^{\infty} \sum_{r=0}^{\infty} K_{r,1} \left( \bar{K}_{r,6p+1} - \bar{K}_{r,6p+5}\right) .
\end{align}
Once again, the Virasoro content follows from a decomposition of the Kac characters $K_{r,s}$ onto Virasoro simple characters.

The indecomposable  JTL modules that appear in the decomposition of the braid-translated spin chain
have also the (subquotient) structure parallel to the TL modules in the open chain -- so-called tilting  modules, see e.g.~\cite{GRS3,GV} for the decomposition over the TL algebra. For instance, the Jordan cell of the stress energy tensor corresponds
to the diamond module
\begin{equation}
 \begin{array}{ccccc}
      &&\hskip-.7cm \IrrJTL{2}{1} &&\\
      &\hskip-.2cm\swarrow&\searrow&\\
     \IrrJTL{0}{\q^2} &&&\hskip-.3cm \IrrJTL{3}{\q^2} \\
      &\hskip-.2cm\searrow&\swarrow&\\
      &&\hskip-.7cm \IrrJTL{2}{1}&&
\end{array},
\end{equation}
which is a cousin of the Jordan cell that arises in the open chains~\cite{RS3}.
Note that this JTL module is in fact much simpler than the one that contains the stress energy tensor
in the ``physical'' closed $\mathfrak{sl}(2|1)$ spin chain~\cite{GRSV}, which  is much more intricate.

Recall  also that for minimal models, we had  to combine braid translation of different boundary conditions for the open chains to obtain the periodic chain.  Even if the structure of modules in the genuine periodic and braid translated chains are different in the case of non minimal models such as super spin chains, we can still ask whether at least the spectrum of the Hamiltonian can be reproduced.  The $\mathfrak{gl}(1|1)$  case was an extreme example in this regard, since a single boundary condition was enough to obtain the full spectrum of  the periodic chain.  The $\mathfrak{sl}(2|1)$ case seems extreme in the opposite way: the spectrum in the periodic case involves complicated multiplicities obtained from arithmetic formulas \cite{GRSV} which are totally absent from the natural boundary conditions one can think of imposing on the open chain. For now, we simply do not know the set of  conformal boundary conditions (if any) that would give the correct periodic spectrum.

\begin{table}
\begin{center}
\begin{tabular}{|c|c|}

  \hline
  $N = 2L$ & $\beta^{(N)}$ \\
  \hline
  10 & -0.598912 \\
  12 & -0.607285 \\
  14 & -0.612164 \\
  16 & -0.615264 \\
  18 & -0.617357 \\
  20 & -0.618839 \\
  22 & -0.619927 \\
  \hline
  $\infty$ &  -0.62500 $\pm$ 0.00001 \\
  \hline
  Conjecture & $-5/8$ \\
  \hline
\end{tabular}
\end{center}
\caption{Measure of the indecomposability parameter $\beta$ 
 in the braid
translated  $\mathfrak{sl}(2|1)$ spin chain.}
\label{FigPercoBraid}
\end{table}

A simple way to see  that the modules obtained from braid translation 
 in the periodic case are not the ones that occur in the periodic $\mathfrak{sl}(2|1)$ spin chain is to measure  the indecomposability parameter $\beta$
associated with the stress energy tensor of the theory, along the lines discussed in Sec.~\ref{sec:lat-ind-par}. As usual, we define the lattice version of $\beta$ as
\begin{equation}
 \beta^{(N)} =\frac{\left|\Braket{t^{(N)}| (L^\brt_{-2}+\bar{L}^\brt_{+2})^{(N)}|0^{(N)}}\right|^2}{\Braket{t^{(N)}| T^{(N)}}}.
\end{equation}
where $|T^{(N)}\rangle$ and $|t^{(N)}\rangle$ are states in the finite spin chain approximating the states corresponding to the 
stress-energy tensor $T$ and its logarithmic partner $t$ in the continuum limit \cite{VJS}. In the holomorphic sector,
we find (Tab.~\ref{FigPercoBraid}) now  $\beta^{(N)}=-\frac{5}{8}$. This  is the value known to occur in  the open case \cite{DJS} -  a reasonable fact given
the construction of the theory.
 The point of course, is that this value is not the one relevant for the periodic 
$\mathfrak{sl}(2|1)$ theory, which is known to be $\beta=-5$~\cite{VJSprl}. More detailed study shows that the braid translated theory is not really physical, with, in particular, a strong dissymmetry between the left and the right moving sectors, generalizing what was observed earlier for the 
$\mathfrak{gl}(1|1)$ case.

\section{Conclusion}

In conclusion, we saw that for RSOS models---whose continuum limit is given by Rational CFTs---the representations of the  lattice algebras in the open and closed cases are related by  braid translation.  This remarkable result is a sort of a lattice equivalent of the 
close relationship between bulk and boundary rational CFTs~\cite{Fuchsetali}, and raises many  questions. In particular, it would be interesting to understand better how left and right chiral sectors appear in the process, and whether the diagrams (\ref{commutdiag}) can be given a more abstract mathematical meaning. Extensions of our result to other models---such as the $Z_n$ parafermions, or coset models for higher level or higher rank algebras, see e.g.~\cite{ISZ89}---should be possible without much difficulty. Another easy extension would be to cover the non-diagonal cases, corresponding, in the simplest case, to the $D$ and $E$ series of modular invariants, and RSOS models defined on $D$ and $E$-type Dynkin diagrams. 

In contrast, we saw that for the simplest lattice models with Logarithmic CFTs as continuum limits---the  alternating supersymmetric spin chains with $\mathfrak{gl}(1|1)$ and $\mathfrak{sl}(2|1)$ symmetry---the relevant representations of the lattice algebras are quite different. Braid translation of the open spin-chain with local, conformally invariant boundary conditions, gives rise to modules which are not relevant to the analysis of the periodic spin chain. The braid-translated spin chain---even though it may have the right content in terms of dimensions of fields---appears in fact non-physical, with a profound dissymmetry between the left and the right moving sectors. This suggests  that the relationship between bulk and boundary CFTs, familiar in the rational case, must, at best  be considered with caution in  the logarithmic case\footnote{It is important to comment here that examples of such a relationship  were worked out in the cases  $c=-2$ and $c=0$  \cite{GaberdielRunkel,WoodGaberdielRunkel}. These works however relied on the presence of an additional $W$ symmetry.  Moreover, we believe that the bulk $c=0$ theory built in \cite{WoodGaberdielRunkel}  suffers from physical flaws, and does not match  the various $c=0$ theories that are known to occur in physical models \cite{GRSV}. However, for $c=-2$ case the construction~\cite{GaberdielRunkel} using the  $W$ symmetry gives the correct structure of the bulk symplectic fermions.}.

This conclusion must be elaborated a bit. On the one hand,  there are representations of the blob algebra which correspond to conformal boundary conditions for the RSOS models, but whose braid translation does not appear in the periodic models. In fact,  we  will show elsewhere that they are relevant for the periodic model in the presence of topological defects \cite{toppaper}. Meanwhile, still for RSOS models, there are also representations of the blob algebra (and corresponding conformal boundary conditions) whose braid translation gives the periodic model and, once we demand that the spectrum of the periodic model is recovered, these blob representations are uniquely determined, and by construction will give rise to  the full algebraic structure in the periodic case. In contrast, our observation for the supersymmetric chains is that we know representations of the blob algebra (and corresponding conformal boundary conditions) that produce the correct spectrum after braid translation, but the algebraic structure (of indecomposable modules) is not the correct one. 

At this stage, we are handicapped by the fact that conformal boundary conditions for LCFTs are not under total control. It could be, for instance, that there exists yet other conformal boundary conditions associated with a different representation of the blob algebra, and whose braid translation would produce the correct periodic model. This is certainly hard to believe in the $\mathfrak{gl}(1|1)$ case, whose continuum limit is, after all, a free symplectic fermion theory, and for which all conformal boundary conditions seem to be known. But to our knowledge, however, there is no proof the classification is complete without invoking higher $W$ symmetry \cite{Bredthauer}, the so-called triplet $W_2$ algebra~\cite{Kausch1}.  On top of this, we cannot exclude that the object obtained by braid translation of the open boundary conditions for the $\mathfrak{gl}(1|1)$  is not some sort of version of symplectic fermions with topological defects: this requires further study \cite{toppaper}. 

To muddy the waters some more, a crucial point in our discussion is that we have insisted on braid translating models with open, {\sl local} boundary conditions. In the RSOS case, this was all that was needed to understand the periodic models. It turns out, however, that the relationship between representations of the periodic Temperley--Lieb algebra and the blob algebra is much more general~\cite{GLDiagramAlgebras}. One can, for instance, imagine starting from the representation of the affine TL algebra  in the periodic $\mathfrak{gl}(1|1)$ spin chain, and try to ``invert'' the braid translation (we provide details for this in  App.~\ref{app:inv-brt}) in order to obtain a representation of the blob algebra on the open spin chain, with much more complicated modules than the ones encountered so far.  This representation would then be one that, like in the RSOS case, becomes relevant to the understanding of the periodic model. But  in doing so one would get a very  non-local expression for the blob generator, see the derivation in App.~\ref{sec:XX}:
\begin{equation}\label{concl:b-nonlocal}
b = \ffrac{i^{N+1}}{2}e^{i\frac{\phi}{2}\sigma^z_1}s_1 s_2\dots s_{N-1}g^{-1}_{N-1}\dots g^{-1}_1 +\ffrac{1}{2},
\end{equation}
where, recall, the braid generators $g^{\pm1}_j = 1 \pm i e_j$ and $s_j$ are the ordinary permutation operators, and the twist $\phi$ here is chosen such that $e^{i\phi}=(-1)^{S_z+1}$. The representation~\eqref{concl:b-nonlocal} of $b$ indeed gives after applying the braid translation~$\brt$ the standard local expression for $e_{N}$ in the periodic $\mathfrak{gl}(1|1)$ spin chain:
\begin{equation}
\displaystyle e_{N}^{\brt} = e_N=  (f_{N}+ f_{1})(f^\dag_{N} + f^\dag_{1}) .
\end{equation}
Whether the non-local expression for the boundary operator $b$ is nonetheless ``physical'---and corresponds to a conformal boundary condition---is not clear at this stage. Of course, based on representation theory, and since we start with a blob algebra module we know that the Hamiltonian will have a spectrum given by characters of the Virasoro algebra. But this does not guarantee that the boundary state is an acceptable conformal boundary state.  
This problem is discussed in more detail in App.~\ref{app:inv-brt}.

Despite all these caveats, the main point remains that ``natural'' boundary conditions for the supersymmetric spin chains, which correspond to well defined representations of the lattice algebras,   do not (seem to) have anything to do with the genuine periodic model after braid translation, in sharp contrast with what happens for RSOS models. 

We finally note in conclusion that one could imagine other ways to build bulk theories out of boundary theories using lattice models.  For instance, one could imagine  constructing bimodules over two copies of TL algebra---{\em e.g.}, by taking two copies of XXZ spin-chains with $U_\q sl(2)$ symmetry, where TL acts on the left in one and on the right in the second, and then passing to the tensor product of them over $U_\q sl(2)$. This would eliminate the quantum symmetry, but give a ``spin chain'' with two TL actions that goes in the continuum limit to chiral-antichiral Virasoro. We have not explored  this  construction, although it seems  natural from the algebraic point of view, due to its lack of actual physical interpretation.

{\bf Acknowledgments.} \hspace{5pt} We acknowledge interesting discussions with J. Dubail, M. Gaberdiel, P. P. Martin, N. Read, D. Ridout, I. Runkel and S. Wood.
We also thank D. Ridout for careful reading of the manuscript and important comments.
This work was supported by the Agence Nationale de la Recherche (grant ANR-10-BLAN-0414), the Institut
Universitaire de France, the European Research Council (advanced grant NuQFT),
and the Department of Energy (LDRD program of LBNL). The work of AMG was supprted by CNRS. AMG is also grateful to Mathematics Department of Hamburg University for kind hospitality.

\appendix
\section{Calculations for the braid translated $\gl(1|1)$ spin chains}\label{app:A}
In this appendix, we gather all lengthy calculations used in Sec.~\ref{sec:gl11chain}.

\subsection{Braid translation}\label{app:A1}
 For $N=2L$ even or odd,  the braid translation~\eqref{eq_defbraid2} in the ordinary TL case (i.e. having for the blob generator the relation $b=\one$) gives the generator $e^{\brt}_{N}$ which satisfy the recursion relation
\begin{equation}
e^{\brt}_{N+1} = (1-i e_{N})e^{\brt}_{N} (1+i e_{N}), \qquad e^{\brt}_{2} = e_1,
\end{equation}
with $e_N$ given by~\eqref{gen-TL-ferm}, with indices taken modulo $N$. Assuming then a bilinear in fermions expression for  $e^{\brt}_{N}=\sum_{j,k}a^{(N)}_{jk}f_j f_k^{\dagger}$ and using the identity
\begin{equation*}
\Bigl[e_i,\sum_{j,k}a_{jk}f_j f^{\dagger}_k\Bigr] = \sum_k(-1)^i(a_{ik}-a_{i+1,k})(f_i+f_{i+1})f^{\dagger}_k
- \sum_j(-1)^i(a_{ji}-a_{j,i+1})f_j(f_i^{\dagger}+f_{i+1}^{\dagger})
\end{equation*}
we obtain the recursion for the coefficients $a^{(N)}_{jk}$, for $1\leq j,k\leq N-1$:
\begin{equation}
\small
\begin{pmatrix}
a^{(N+1)}_{jk} = a^{(N)}_{jk} & a^{(N+1)}_{jN}=(1+(-1)^Ni)a^{(N)}_{jN} & a^{(N+1)}_{j,N+1} = (-1)^N i a^{(N)}_{jN}\\
a^{(N+1)}_{Nk} = (1-(-1)^Ni)a^{(N)}_{Nk} & a^{(N+1)}_{NN} = 2 a^{(N)}_{NN} & a^{(N+1)}_{N,N+1} = (1+(-1)^Ni)a^{(N)}_{NN}\\
a^{(N+1)}_{N+1,k} = -(-1)^N i a^{(N)}_{Nk} & a^{(N+1)}_{N+1,N} = (1-(-1)^Ni) a^{(N)}_{NN} & a^{(N+1)}_{N+1,N+1} = a^{(N)}_{NN}
\end{pmatrix}
\end{equation}
with initial conditions $a^{(2)}_{ij}=1$ for $i,j=1,2$.
This recursion has a unique solution which gives a simple final expression~\eqref{eq:e-braid}.

\subsection{Spectrum of the braid translated $\gl(1|1)$ spin chain}
Introducing canonical fermions and Fourier transforms $\theta_p$ and $\theta_p^\dag$ (in notations from~\cite[Sec.\,3.2]{GRS1}), we find that the Hamiltonian reads
\begin{multline}\label{H-theta}
\displaystyle H^{\brt}= 2 \sum_p (1+ \sin p) \theta_p^\dag \theta_{p+\pi} + 4 L \theta_{\frac{\pi}{2}}^\dag \theta_{\frac{3 \pi}{2}} \\-(1-i) \sum_p (1 + i \mathrm{e}^{ip}) \theta_{\frac{\pi}{2}}^\dag \theta_p - (1+i) \sum_p (1 + i \mathrm{e}^{-ip})\theta_p^\dag \theta_{\frac{3\pi}{2}},
\end{multline}
where the sums are taken over 
 the set of allowed momenta
\begin{equation}\label{momenta-set}
p_m=
\begin{cases}
\frac{2\pi m}{N},\qquad &L-\text{even},\\
\frac{(2m-1)\pi}{N}, &L-\text{odd},
\end{cases}
\qquad\quad 1\leq m\leq N,
\end{equation}
and the fermions satisfy the usual anti-commutation relations
\begin{equation*}
\{\ferm_{p_1},\fermd_{p_2}\} = \delta_{p_1,p_2}, \qquad \{\ferm_{p_1},\ferm_{p_2}\} = \{\fermd_{p_1},\fermd_{p_2}\} = 0.
\end{equation*}
It can be then rewritten as
\begin{multline}
 H^{\brt}= 2 \sum_{\substack{p=\step\\\text{step}=\step}}^{\pi-\step} \left[ (1- \cos p) \theta_{p-\frac{\pi}{2}}^\dag \theta_{p+\frac{\pi}{2}} + (1+ \cos p) \theta_{p+\frac{\pi}{2}}^\dag \theta_{p-\frac{\pi}{2}} \right] + 4 L \theta_{\frac{\pi}{2}}^\dag \theta_{\frac{3 \pi}{2}} \\
- (1-i) \sum_{\substack{p=\step\\\text{step}=\step}}^{\pi-\step} \left[ (1 -  \mathrm{e}^{ip}) \theta_{\frac{\pi}{2}}^\dag \theta_{p+\frac{\pi}{2}} +(1 + \mathrm{e}^{ip}) \theta_{\frac{\pi}{2}}^\dag \theta_{p-\frac{\pi}{2}} \right] \\
- (1+i) \sum_{\substack{p=\step\\\text{step}=\step}}^{\pi-\step} \left[ (1 -  \mathrm{e}^{-ip}) \theta_{p-\frac{\pi}{2}}^\dag \theta_{\frac{3\pi}{2}}  +(1 + \mathrm{e}^{-ip}) \theta_{p+\frac{\pi}{2}}^\dag \theta_{\frac{3\pi}{2}} \right]
\end{multline}
where $\step= \frac{2 \pi}{N} = \frac{\pi}{L} $.
We then introduce the fermionic modes\footnote{We note that similar  fermionic modes for the periodic spin chain were introduced in~\cite{GRS1} and are denoted there by $\chi_p$ and $\eta_p$. The difference is only in the zero-mode contributions in $\chib^{(\dagger)}_{p}$ and $\etab^{(\dagger)}_{p}$.}
\begin{subequations}\label{eq-modes}
\begin{eqnarray}
\etab_{p} & = & \frac{1}{\sqrt{2}} \left( \sqrt{\cot \frac{p}{2}} \theta_{p - \frac{\pi}{2}} -\sqrt{\tan \frac{p}{2}} \theta_{p + \frac{\pi}{2}} \right) - \frac{e^{-i p/2}}{\sqrt{\sin p}} \theta_{\frac{3 \pi}{2}} ,\\
\etab_{p}^\dag & = & \frac{1}{\sqrt{2}} \left( \sqrt{\tan \frac{p}{2}} \theta_{p - \frac{\pi}{2}}^\dag -\sqrt{\cot \frac{p}{2}} \theta_{p + \frac{\pi}{2}}^\dag \right) + \frac{e^{i p/2}}{\sqrt{\sin p}}  \theta_{\frac{\pi}{2}}^\dag ,\\
\chib_{p} & = & \frac{1}{\sqrt{2}} \left( \sqrt{\cot \frac{p}{2}} \theta_{p - \frac{\pi}{2}} +\sqrt{\tan \frac{p}{2}} \theta_{p + \frac{\pi}{2}} \right) - \frac{i e^{-i p/2}}{\sqrt{\sin p}} \theta_{\frac{3 \pi}{2}}, \\
\chib_{p}^\dag& = & \frac{1}{\sqrt{2}} \left( \sqrt{\tan \frac{p}{2}} \theta_{p - \frac{\pi}{2}}^\dag +\sqrt{\cot \frac{p}{2}} \theta_{p + \frac{\pi}{2}}^\dag \right) + \frac{i e^{i p/2}}{\sqrt{\sin p}} \theta_{\frac{\pi}{2}}^\dag, \\
\chib_{0}^\dag& = & \theta^\dag_{\frac{\pi}{2}}, \ \ \ \ \ \ \etab_{0} = \theta_{\frac{3 \pi}{2}},\\
\etab_{0}^\dag& = & \theta^\dag_{\frac{3\pi}{2}}, \ \ \ \ \ \ \chib_{0} = \theta_{\frac{ \pi}{2}}.\label{eq-modes-last}
\end{eqnarray}
\end{subequations}
which give one-particle states of the energy $\pm2\sin p$ for the Hamiltonian $H^{\brt}$. We also give the backward transformations to the original fermionic modes $\theta_p$ and $\theta^\dagger_p$:
 \begin{subequations}\label{eq-modes-back}
\begin{eqnarray}
\theta_{p-\frac{\pi}{2}} & = & \sqrt{\frac{\tan \frac{p}{2}}{2}}\bigl(\chib_p + \etab_p\bigr) + \frac{(1+i)e^{-i p/2}}{2\cos \frac{p}{2}} \etab_0 ,\\
\theta_{p+\frac{\pi}{2}} & = & \sqrt{\frac{\cot \frac{p}{2}}{2}}\bigl(\chib_p - \etab_p\bigr) + \frac{(i-1)e^{-i p/2}}{2\sin \frac{p}{2}} \etab_0 ,\\
\theta_{p-\frac{\pi}{2}}^\dag & = & \sqrt{\frac{\cot \frac{p}{2}}{2}}\bigl(\chib_p^\dag + \etab_p^\dag\bigr) - \frac{(1+i)e^{i p/2}}{2\sin \frac{p}{2}} \chib_0^\dag ,\\
\theta_{p+\frac{\pi}{2}}^\dag & = & \sqrt{\frac{\tan \frac{p}{2}}{2}}\bigl(\chib_p^\dag - \etab_p^\dag\bigr) + \frac{(1-i)e^{i p/2}}{2\cos \frac{p}{2}} \chib_0^\dag.
\end{eqnarray}
\end{subequations}
In terms of the new modes $\chib_p$ and $\etab_p$, we have
\begin{multline}
 H =  4 L \chib_0^\dag \etab_0 + 2 \sum_{\substack{p=\step\\\text{step}=\step}}^{\pi-\step} \sin p \left( \chib_p^\dag \chib_p - \etab_p^\dag \etab_p \right) \\
 - 2 \sum_{\substack{p=\step\\\text{step}=\step}}^{\pi-\step} \mathrm{e}^{-\frac{ip}{2}} \sqrt{\sin p} \left( \etab_p^\dag \etab_0 -  i \chib_p^\dag \etab_0 \right)
 - 2 \sum_{\substack{p=\step\\\text{step}=\step}}^{\pi-\step} \mathrm{e}^{\frac{ip}{2}} \sqrt{\sin p} \left( i \chib_0^\dag \chib_p - \chib_0^\dag \etab_p \right)
\end{multline}
and
\begin{multline}\label{eq:e-e}
e^{\brt}_{2L} - e_{2L}=   4(L-1) \chib_0^\dag \etab_0 - 2 \sum_{\substack{p=\step\\\text{step}=\step}}^{\pi-\step} \mathrm{e}^{-\frac{ip}{2}} \sqrt{\sin p} \left( \etab_p^\dag \etab_0 -  i \chib_p^\dag \etab_0 \right) \\
 - 2 \sum_{\substack{p=\step\\\text{step}=\step}}^{\pi-\step} \mathrm{e}^{\frac{ip}{2}} \sqrt{\sin p} \left( i \chib_0^\dag \chib_p - \chib_0^\dag \etab_p \right),
\end{multline}
so that the full braid translated Hamiltonian $H^\brt$ in~\eqref{eq:Hbrt-def}  reads as in~\eqref{Hbr-chi}.

\subsection{Higher modes}\label{app:A3}

We rewrite the higher modes $H^{\brt}(n)$  (Fourier transform of $e_j$'s) in~\eqref{Hbr-n} as
\begin{equation}\label{app:Hbr-n}
\displaystyle H^{\brt}(n) = H(n) - e^{\brt}_{2L} + e_{2L},
\quad q={\ffrac{n\pi}{L}},
\end{equation}
where $n$ is integer and
\begin{multline}
H(n) = -\sum_{j=1}^{2L} e^{-iqj} e_j=\sum_{p}
\left[1+e^{iq}+ie^{-ip}-ie^{i(p+q)}\right] \fermd_{p}\,
\ferm_{p+q+\pi}\\
=4e^{iq/2}\Bigl(\sum_{\substack{p=\step\\\text{step}=\step}}^{\pi-\step}
\bigl(\sin{\ffrac{p+q}{2}}
\sin{\ffrac{p}{2}}\,\fermd_{p-\frac{\pi}{2}}\ferm_{p+q+\frac{\pi}{2}} +
[p\to p+\pi]\bigr) +
\cos{\ffrac{q}{2}}\,\fermd_{\frac{\pi}{2}}\ferm_{q-\frac{\pi}{2}}
\Bigr), \quad q={\ffrac{n\pi}{L}},\label{inter}
\end{multline}
is known expression from the usual periodic model~\cite{GRS1}.
 We first consider the case $0< n< L$. We  repeat the calculations similar to those in~\cite{GRS1} and
using the formulas~\eqref{eq-modes-back} 
 and~\eqref{eq:e-e} the $H^{\brt}(n)$ can be finally rewritten as in~\eqref{lat-Vir-H}.
We also have expression for negative values of the modes:
\begin{multline}\label{lat-Vir-H-m}
\Hbr(-n) =
2e^{-iq/2}\Biggl(-(1+i)\sqrt{\sin{q}} \, \chib^{\dagger}_{0}\etab_{\pi-q}
+ \sum_{\substack{p=\step\\\text{step}=\step}}^{\pi-q-\step}
\sqrt{\sin{(p)}\sin{(p+q)}} \bigl(\chib^{\dagger}_{p+q}\,\chib_{p} -
\etab^{\dagger}_{p+q}\,\etab_{p}\bigr)\\
- \sum_{\substack{p=\step\\\text{step}=\step}}^{q-\step}
\sqrt{\sin{(p)}\sin{(q-p)}}\bigl(\chib^{\dagger}_{q-p}\,\etab_{\pi-p} +
\etab^{\dagger}_{q-p}\,\chib_{\pi-p}\bigr) + (1-i)\sqrt{\sin{q}} \,\chib^{\dagger}_{q}\etab_{0}\Biggr), \quad 1\leq n\leq L.
\end{multline}

\medskip

For the momentum operator for our braid translated spin-chain 
\begin{equation}
\Pbr = P + \ffrac{i}{2}\Bigl([e_N-\ebr_N,e_{N-1}] - [e_N-\ebr_N,e_{1}]\Bigr),
\end{equation}
we have in terms of fermion Fourier variables for the commutators
\begin{equation}
[e_N-\ebr_N,e_{1}] = (1+i)\sum_p(ie^{-ip} - e^{-2ip})\fermd_p\ferm_{\frac{3\pi}{2}} \\
- (1+i)\sum_p(e^{ip} + ie^{2ip})\fermd_{\frac{\pi}{2}}\ferm_p
\end{equation}
and
\begin{equation}
[e_N-\ebr_N,e_{N-1}] = (1+i)\sum_p(ie^{ip} - 1)\fermd_p\ferm_{\frac{3\pi}{2}} \\
- (1+i)\sum_p(i+e^{-ip})\fermd_{\frac{\pi}{2}}\ferm_p.
\end{equation}
Finally, the fermionic expression for $\Pbr$ is
\begin{multline}
\Pbr = -\sum_p \sin{2p}\, \fermd_p\ferm_p + \ffrac{i-1}{2}\sum_p\Bigl[(i e^{ip}- ie^{-ip} + e^{-2ip}-1)\fermd_p\ferm_{\frac{3\pi}{2}} \\
- (i+e^{-ip} - e^{ip} - ie^{2ip})\fermd_{\frac{\pi}{2}}\ferm_p\Bigr]
\end{multline}
and in terms of the modes $\chib_p$ and $\etab_p$ we obtain~\eqref{Pbr}.

The higher modes
$\Pbr(n)$ introduced in~\eqref{Pbr-def}
 are expressed in the Fourier variables as
\begin{multline*}
\Pbr(n) = 4e^{iq}\sum_p \sin{(p+q/2)}\sin{\ffrac{p+q-\pi/2}{2}}\cos{\ffrac{p-\pi/2}{2}}\, \fermd_p\ferm_{p+q}\\ + \ffrac{i-1}{2}\sum_p\Bigl[(i e^{i(p+q)}- ie^{-ip} + e^{-2ip}-e^{iq})\fermd_p\ferm_{\frac{3\pi}{2}} \\
- (ie^{iq}+e^{-i(p-q)} - e^{ip} - ie^{2ip})\fermd_{\frac{\pi}{2}}\ferm_p\Bigr]
\end{multline*}
and in terms of the modes $\chib_p$ and $\etab_p$ we finally obtain
\begin{multline}\label{Pbr-fer}
\Pbr(n) = 2e^{iq}\Biggl((1+i)\cos{\ffrac{q}{2}}\sqrt{\sin{q}}\, \chib^{\dagger}_{0}\chib_{q} + \sum_{\substack{p=\step\\\text{step}=\step}}^{\pi-q-\step}
\cos{\bigl(p+\ffrac{q}{2}\bigr)}\sqrt{\sin{(p)}\sin{(p+q)}} \bigl(\chib^{\dagger}_{p}\,\chib_{p+q} +
\etab^{\dagger}_{p}\,\etab_{p+q}\bigr)\\
- \sum_{\substack{p=\step\\\text{step}=\step}}^{q-\step}
\cos{\bigl(p-\ffrac{q}{2}\bigr)}\sqrt{\sin{(p)}\sin{(q-p)}}\bigl(\chib^{\dagger}_{\pi-p}\,\etab_{q-p} -
\etab^{\dagger}_{\pi-p}\,\chib_{q-p}\bigr)
+ (1-i)\cos{\ffrac{q}{2}}\sqrt{\sin{q}}\, \etab^{\dagger}_{\pi-q}\etab_{0}\Biggr)
\end{multline}
and similarly for $\Pbr(-n)$.

\section{Momentum and conformal spin}\label{app:momentum-spin}

In this appendix, we briefly recall how one can measure conformal spins $s=h-{\bar h}$ directly
from the lattice in the periodic case using Fourier decomposition.
First of all, recall that the central charge and the critical exponents $\Delta=h+{\bar h}$ of a CFT defined on a torus can be extracted using finite-size $N=2L$ corrections to the Hamiltonian eigenvalues~\cite{CardyScaling}
\begin{equation}
\displaystyle E_0(N) = E_0(\infty)N -\frac{\pi v_F c}{6 N} + \dots
\end{equation}
\begin{equation}
\displaystyle E_{\phi}(N) - E_{0}(N) = \frac{2 \pi v_F }{N} \Delta_{\phi} + \dots
\end{equation}
 with $E_0(\infty)$ is the bulk groundstate energy density and $v_F$ is the Fermi velocity.
The Hamiltonians that we consider are invariant under translation, that is, we have $[H,Q]=0$
where $Q=u^2$ is the (two-sites) right translation operator. The eigenvalues of $Q$ are $\mathrm{e}^{-ik}$
where $k=2 \pi q /L$ is the momentum. The corresponding eigenvectors span the subspace $V(k)$, and the
projector onto this subspace reads
\begin{equation}
\displaystyle p(k) = \frac{1}{L} \sum_{n=0}^{L-1} \mathrm{e}^{ik m} Q^m.
\end{equation}
$H$ and $Q$ can be simultaneously diagonalized so that we can label the eigenvalues of $H$ with the momentum $k$. The energies corresponding to $s=h-{\bar h}$ are to be found in the sector with $k=2 \pi s/L$, this means\footnote{We assume here that the groundstate is in the sector $k_0=0$. However, if $k_0 \neq 0$ the states with conformal spin~$s$ are in the sector of momentum $k=k_0+2 \pi s/L$.} that if we want to consider operators with a conformal spin $s$ only, we can just diagonalize $H$ in the subspace $V(k)$ of momentum $k=2 \pi s/L$. Therefore,
states with conformal weights $(h,\bar{h})$ can be readily identified using exact diagonalizations and finite size scaling.

\section{Lattice indecomposability parameters for $\mathfrak{gl}(1|1)$ spin chain}\label{app:gl11}

In this appendix, we use notations and conventions from the previous work on the periodic $\mathfrak{gl}(1|1)$ spin-chain~\cite{GRS1}. We review an approach~\cite{VJS} of determining the Virasoro  indecomposability parameters or logarithmic couplings $\beta$ directly from the lattice models. We  apply it for the periodic $\mathfrak{gl}(1|1)$ spin-chains~\cite{GRS1} and determine both $\beta$ and $\bar\beta$ numerically and analytically using the fermions. 

\subsection{Ordinary periodic $\mathfrak{gl}(1|1)$ spin chain: numerics}
\label{section_gl11}

Even though the values of the indecomposability parameters in the bulk symplectic fermions CFT are the same as for the open chiral case,
it remains instructive to show how one can extract these values from a lattice model. The method we shall
use is very similar to the one used in the case of lattice models with open (free) boundary conditions~\cite{VJS}
and relies on a regularization of the Virasoro algebra on the lattice~\cite{KooSaleur}.

We first recall the expression of the Hamiltonian in the periodic $\mathfrak{gl}(1|1)$ spin chain (for even~$N$)
\begin{equation}
\displaystyle H = - \sum_{j=1}^{N} e_j,
\end{equation}
where the (J)TL generator is nothing but the map onto the singlet in the tensor product of the fundamental and the dual of $\mathfrak{gl}(1|1)$ for $i$ odd, and the other way around for $i$ even
\begin{equation}
\displaystyle e_j = (f_j^\dag + f_{j+1}^\dag) (f_j + f_{j+1}),
\end{equation}
with the fermionic operators $\lbrace f_i,f_j \rbrace=0$ and $\lbrace f_i,f^\dagger_j \rbrace=(-1)^{i+1} \delta_{i,j}$.
Note that $H$ commutes with the number of fermions, so we can label the different sectors with the quantum number $S_z$
which is $N/2$ minus the number of fermions. (See also a relation of the $e_j$'s to  the spin representation in~\cite[Sec.\,2.2]{GRS1}.)
The vacuum state of $H$ is in $S_z=0$ and is denoted by $|\Omega\rangle$. The vacuum energy is four times degenerate:  $|\Omega\rangle$ has its ``logarithmic'' partner $|\omega\rangle$ (so, the two states span a Jordan cell for $H$), and there are two fermionic states $|\xi^{\pm}\rangle$ in the sectors $S_z=\pm1$, see the definitions in~\cite[eqs.\,(4.10) and (4.11)]{GRS1}.\footnote{In~\cite{GRS1}, one have to replace $\phi^2$ by $|\xi^{+}\rangle$ and $\phi^1$ by $|\xi^{-}\rangle$.} Their scalar products read
\begin{equation}\label{eq:xipm}
\displaystyle \Braket{\Omega|\Omega}=\Braket{\omega|\omega} = 0 \ \ \ \Braket{\Omega|\omega}=\Braket{\xi^+|\xi^+}=-\Braket{\xi^{-}|\xi^{-}}=1.
\end{equation}

 We now turn to the measure of the logarithmic couplings directly from the lattice, using the periodic $\mathfrak{gl}(1|1)$ spin chain.
We would like to measure the coefficient $\beta_{1,5}$ that corresponds to the Jordan cell for first excited states.
The first step is to identify the states with conformal weights $(h,\bar{h})=(1,0)$ mixed by the Hamiltonian into a Jordan cell.
Such states are readily found in the sector $S_z=1$. We shall denote them $\Ket{\phi^{(N)}}$ and $\Ket{\psi^{(N)}}$, and we denote by $E_{\phi}(N)$ their energy.
We normalize $\Ket{\psi^{(N)}}$ so that in the basis ($\Ket{\phi^{(N)}}$,$\Ket{\psi^{(N)}}$), the Hamiltonian reads
\begin{equation}\label{eq_Jordancell_SP}
\displaystyle H - E_{0}(N) \one =  \frac{2 \pi v_F}{N} \left( \begin{array}{cc} \Delta^{(N)} & 2  \\ 0 & \Delta^{(N)}  \end{array} \right),
\end{equation}
where $\Delta^{(N)}= \frac{N}{2\pi v_F} (E_{\phi}(N)-E_{0}(N))$ is the lattice scaling dimension of the states, $v_F=2$ is the Fermi velocity and $ E_{0}(N)$ is the energy of the groundstate. The matrix expression in~\eqref{eq_Jordancell_SP} corresponds to $L_0+\bar{L}_0$ in the scaling limit, with $\Delta^{(\infty)}=1$. In the scaling limit, recall that $\beta$ is defined using the Virasoro bilinear form: $\beta_{1,5}=\Braket{\psi|\phi}$ (for the states appropriately normalised).
Therefore, we will also need a `scalar product' that goes to the Virasoro form in the scaling limit. This is provided by the usual Fock space scalar product,
that is, taking the $\dag$ symbol on the fermions $f^\dagger_j$ corresponds to taking the adjoint operator.\footnote{Note that there are negative norm states,
for example, the state $\Ket{\alpha} = f_{2}^\dag \Ket{0}$ has norm $-1$ because $\lbrace f_2, f_{2}^\dag\rbrace = -1$.}

\begin{table}
\begin{center}
\begin{tabular}{|c|c|c|c|c|c|}
  \hline
  $N = 2 L $ & $c/2$ & $\beta_{1,5}/2$ \\
  \hline
  6 & -0.341959 &  \\
  8 & -0.573159 &  \\
  10 & -0.708003 & -0.434245 \\
  12 & -0.789720 & -0.453477 \\
  14 & -0.842079 & -0.465432 \\
  16 & -0.877354 & -0.473339 \\
  18 & -0.902141 & -0.478828 \\
  20 & -0.920177 & -0.482789 \\
  22 & -0.933474 & -0.485738 \\
  \hline
  $\infty$ & -1.00 $\pm$ 0.02 & -0.500 $\pm$ 0.001 \\
  \hline

\end{tabular}
\end{center}
\caption{Measure of the logarithmic coupling $\beta_{1,5}$ and of the central charge $c/2=\Braket{0|[L_{+2}^{(N)},L_{-2}^{(N)}]|0}$  in the periodic $\mathfrak{gl}(1|1)$ spin chain.}
  \label{tab_Bulkgl11}
\end{table}

The last step is probably the most important, since it has to do with the proper normalization of the states.
Notice first that the Jordan cell in the Hamiltonian is invariant under a simultaneous rescaling of the states $\Ket{\phi^{(N)}}$ and
$\Ket{\psi^{(N)}}$. Hence, the scalar product $\Braket{\psi^{(N)}| \phi^{(N)}}$ is meaningless if we do not normalize properly
the states. In the scaling limit, we want $\Ket{\phi} = L_{-1} \Ket{\xi}$ where $\xi$ is the only primary field with conformal weights $(0,0)$
in the sector $S_z=1$, normalized such that $\Braket{\xi|\xi}=1$. Unfortunately, there is no trivial way to normalize $\Ket{\phi}$
since it is a singular Virasoro state, and thus $\Braket{\phi|\phi}=0$. Similarly on the lattice: we have $\Braket{\phi^{(N)}|\phi^{(N)}}=0$. Therefore, we need another way to carefully normalize $\Ket{\phi^{(N)}}$ so
that it goes precisely to $\Ket{\phi} = L_{-1} \Ket{\xi}$ in the scaling limit.
Like in the chiral case~\cite{VJS}, this is achieved using a regularised expression of the Virasoro
generators on the lattice~\cite{KooSaleur}. Let us introduce
$H^{(N)}_n = L^{(N)}_n + \bar{L}^{(N)}_{-n}$ and $P^{(N)}_n =  L^{(N)}_n - \bar{L}^{(N)}_{-n}$.
These lattice versions of Virasoro generators, so-called Koo-Saleur generators, can be expressed in terms of the periodic TL generators as
\begin{subequations}\label{eqKooSaleur}
\begin{eqnarray}
\displaystyle H^{(N)}_n &=& -\frac{L}{\pi v_F} \sum_{j=1}^{2L} \mathrm{e}^{inj \pi /L} \left( e_i - e_{\infty} \right) + \frac{c}{12} \delta_{n,0}, \\
\displaystyle P^{(N)}_n &=& -\frac{i L}{\pi v_F^2} \sum_{j=1}^{2L} \mathrm{e}^{inj \pi /L}  \left[ e_i, e_{i+1} \right].
\end{eqnarray}
\end{subequations}
Here, $e_{\infty}$ is the average of the generator $e_i$ in the ground state in the thermodynamic limit. It is proportional to the bulk energy density $\varepsilon_0$ introduced in~\eqref{charform}, the difference arising from the normalization of the Hamiltonian.  In general, $e_\infty$ can be computed using Bethe Ansatz~\cite{PottsBethe};
in the case of the $\mathfrak{gl}(1|1)$, it follows from free fermions computations that $e_{\infty} = \frac{2}{\pi}$.
It was shown in~\cite{GRS1} that  expressions~\eqref{eqKooSaleur} provide a natural analogue of the Virasoro generators on the lattice.
As an example, we show in Tab.~\ref{tab_Bulkgl11} the calculations of the central charge from a non-trivial Virasoro commutator.
The result is of course consistent with $c=-2$ as expected. It is in fact  shown that in this case the commutators of the scaling limit
of the lattice Virasoro generators $L_n^{(N)}$ are equal to the scaling limit of the commutators of the $L_n^{(N)}$'s~\cite{GRS1}. 

We are now ready to formulate a correct version of the indecomposability parameter $\beta^{(N)}_{1,5}$, invariant under a rescaling of the states $\Ket{\phi^{(N)}}$ and
$\Ket{\psi^{(N)}}$, which goes precisely to $\beta_{1,5}$ in the scaling limit. In analogy with the chiral case~\cite{VJS}, we define
\begin{equation}\label{eq:beta-App}
\displaystyle \beta^{(N)}_{1,5}= \frac{ \left|\Braket{\psi^{(N)}| L^{(N)}_{-1}|\xi^{(N)}} \right|^2}{\Braket{\psi^{(N)}| \phi^{(N)}}},
\end{equation}
where $|\xi^{(N)}\rangle=|\xi^+\rangle$ is the groundstate in the sector $S_z=1$, normalized such that $\Braket{\xi^{(N)} | \xi^{(N)}}=1$.
It is straightforward to show that this expression indeed goes to $\Braket{\psi|L_{-1}|\xi}$ in the scaling limit.
Results are shown in Tab.~\ref{tab_Bulkgl11}, and are in very good agreement with $\beta_{1,5}=-1$.
Obviously, all $\beta=\beta_{1,j}$ for more excited Jordan cells at $c=-2$ could be measured numerically along the same lines from the $\mathfrak{gl}(1|1)$ spin chain.
However in this case, such coefficients can even be computed analytically using free fermions. We now turn to such calculation.

\subsection{Ordinary periodic $\mathfrak{gl}(1|1)$ spin chain: exact results}
We will show in this section that the lattice indecomposability parameters can be computed analytically,
using the fact that the spin chain can be (almost) diagonalized in terms of fermionic modes.
We begin with the ground-state energy 
\begin{equation}
\displaystyle E_0^{(N)} = - 2 \sum_{n=1}^{L-1} \sin \left(\ffrac{n \pi}{L}\right)= - 2\, \mathrm{cotan} \left(\ffrac{\pi}{2L}\right) = - \frac{4L}{\pi} + \frac{\pi}{3L} + \mathcal{O}(L^{-2}) .
\end{equation}
This is of course consistent with the values $e_{\infty} = 2/\pi$, $v_F=2$ and $c=-2$ discussed above. 

The Koo-Saleur generators~\eqref{eqKooSaleur} can be reexpressed in terms of  fermionic modes $\eta_p$ and $\chi_p$.
The final expressions are quite cumbersome so we refer the interested reader to~\cite[Sec.\,4.3.3]{GRS1} for more details.
This step is purely technical so we choose not to go into too much detail here.
Using all these elements, we are now ready to compute  $\beta_{1,5}^{(N)}$  exactly.
We define
 \be
 \Ket{\phi^{(N)}} = \eta_{p=\epsilon} \Ket{\Omega}\qquad \text{and}
  \qquad \Ket{\psi^{(N)}} = \epsilon\, \eta_{p=\epsilon} \Ket{\omega},
  \ee
   where $\epsilon = \pi/L$ (recall $N=2L$). One can then express exactly the Hamiltonian and the conformal momentum on the lattice in this basis
\begin{subequations}
\begin{eqnarray}
(L_0 + \bar{L}_{0})^{(N)} &=& \left(\begin{array}{cc}  \tilde{\Delta}_N & 2  \\ 0 & \tilde{\Delta}_N \end{array} \right) \\
(L_0 - \bar{L}_{0})^{(N)} &=& \left(\begin{array}{cc}  \frac{L}{2 \pi} \sin \frac{2 \pi}{L}  & 0  \\ 0 & \frac{L}{2 \pi} \sin \frac{2 \pi}{L} \end{array} \right)
\end{eqnarray}
\end{subequations}
where
\begin{align}
\displaystyle \tilde{\Delta}_{(N)} &= \frac{2 L}{2 \pi v_F} \left( E^{(N)}_1 + 2L e_{\infty} \right) + \frac{c}{12} \notag \\
                                  &= \frac{L}{\pi} \sin \frac{\pi}{L} + \left[ \frac{2 L^2}{\pi^2} - \frac{1}{6} -\frac{L}{ \pi}
                                 \mathrm{cotan} \frac{\pi}{ 2 L}\right]=1+\mathcal{O} \left(\frac{1}{L}\right).
\end{align}
It is important to notice that the action $(L_0 - \bar{L}_{0})^{(N)}$ is closed on the states $(\Ket{\psi^{(N)}},\Ket{\phi^{(N)}})$.\footnote{Unfortunately, this is not the case in more complicated situation like $\mathfrak{sl}(2|1)$ spin chains, and the operator $(L_0 - \bar{L}_{0})^{(N)}$ couple these two states to many others. It is believed that these contributions vanish in the scaling limit.}
Note that these states have energy
\begin{equation}
\displaystyle E_1^{(N)} - E_0^{(N)}= 2 \sin \frac{\pi}{L} = \frac{2 \pi v_F}{2L} 1+ \mathcal{O}(L^{-3}),
\end{equation}
so they belong to the holomorphic sector $(h, \bar{h}) = (1,0)$.

Let us now construct $\beta_{1,5}^{(N)}$ explicitly. First, we note that
\begin{equation}
\displaystyle \Braket{\psi^{(N)}|\phi^{(N)}} = - \epsilon.
\end{equation}
We then recall the fermionic ground state $\Ket{\xi^{+}}$ introduced above~\eqref{eq:xipm}.
Some algebra using the fermionic expressions for~\eqref{eqKooSaleur} yields 
\begin{subequations}
\begin{eqnarray}
(L_{-1} - \bar{L}_{1})^{(N)} \Ket{\xi^{+}} &=& - \cos \frac{\epsilon}{2} \frac{\sqrt{\sin \epsilon}}{\epsilon} \mathrm{e}^{i \epsilon} \eta_{p=\epsilon} \Ket{\Omega} \\
(L_{-1} + \bar{L}_{1})^{(N)} \Ket{\xi^{+}} &=& - \frac{\sqrt{\sin \epsilon}}{\epsilon} \mathrm{e}^{i \epsilon/2} \eta_{p=\epsilon} \Ket{\Omega}
\end{eqnarray}
\end{subequations}
from which we deduce
\begin{equation}
\displaystyle \Braket{\psi^{(N)}|L_{-1}^{(N)}|\xi^{+}} = \frac{\mathrm{e}^{i \epsilon/2}}{4} \sqrt{\sin \epsilon} \left[ 1 + \cos \frac{\epsilon}{2} \mathrm{e}^{i \epsilon/2} \right].
\end{equation}
The lattice indecomposability parameter $\beta_{(N)}^{(1,5)}$ introduced in~\eqref{eq:beta-App} can then be expressed as
\begin{equation}
\displaystyle \beta^{(N)}_{1,5}= \frac{|\Braket{\psi^{(N)}| L_{-1}^{(N)}|\xi^{+}}|^2}{\Braket{\psi^{(N)}| \phi^{(N)}}} = - \frac{\sin \epsilon}{4 \epsilon} \left[ (1+\cos^2 \frac{\epsilon}{2})^2 + \cos^2 \frac{\epsilon}{2} \sin^2 \frac{\epsilon}{2}\right].
\end{equation}
Here, the state $\Ket{\xi^{+}}$ is $|\xi^{(N)}\rangle$ from the previous subsection.
Some trigonometric algebra finally yields the result
\begin{align}
\displaystyle \beta^{(N)}_{1,5} &= - \frac{L}{8 \pi} \sin \frac{\pi}{L} \left( 5+3 \cos \frac{\pi}{L} \right) \notag \\
                                  &= -1 + \frac{17}{48} \left( \frac{\pi}{L} \right)^2 + \mathcal{O}(L^{-4}).
\end{align}
This formula is of course in perfect agreement with the numerical values analyzed above.
A similar calculation holds for anti-holomorphic excitations and  we obtain the same indecomposability parameters in both sectors.

 We remark that one can obtain explicit lattice formulae for the indecomposability parameters for other excited states,
  however, the algebra becomes much more complicated. For example, for the states that  in the continuum limit  become primary fields and their logarithmic partners of the chiral conformal dimension $h_{1,7}$, we find
\begin{align}
\displaystyle \beta^{(N)}_{1,7} &= - \frac{L}{256 \pi^3} \sin \frac{2\pi}{L} \left[ 16 \sin^2 \frac{\pi}{L} \Bigl(4\pi +L\bigl(1+\cos \ffrac{\pi}{L}\bigr)\sin \frac{\pi}{L}\Bigr)^2 \right. \notag \\
 & \left. +\Bigl(32 \pi \cos \frac{\pi}{L} + 5 L \sin \frac{2\pi}{L} + 2L \sin \frac{3 \pi}{L}\Bigr)^2 \right].
\end{align}
This coefficient has the following asymptotic behavior
\begin{equation}
\displaystyle \beta^{(N)}_{1,7} = -18 +\frac{125}{4} \left(\frac{\pi}{L}\right)^2 -\frac{29209}{1502} \left(\frac{\pi}{L}\right)^4 + \mathcal{O}(L^{-6}),
\end{equation}
which agrees with the CFT result.
In principle, we can compute lattice expressions of all the other logarithmic couplings in a similar fashion but we omit details of the computation.

\section{Generating function of RSOS paths}
\label{app:comb}
In this appendix, we obtain the generating function $P^{(n)}_k(z)$ from~\eqref{eq:path-gen-fun} for the number of paths in the open RSOS model for different boundary conditions.
The  compact expression we get seems to be new.

 It will more convenient in this section to shift the heights  in the RSOS model by $-1$, i.e.\ the new heights $s_i:=h_i-1$, for $i=0,\ldots, N$, take  values  from $0$ to $l=p-1$.
In other words, each hight configuration correspond to a walk between two horizontal lines, the floor at height $0$ and the
ceiling at height $l$. For an elementary piece of the walk or the step we associate a weight. Each up step has a weight $z t$, and each down step has a weight $z/t$.
In other words, $z$ is conjugate to the length and $t$ is conjugate to the height of the walk.

We attack the problem of finding the generating function of the walks in three stages. Consider first walks starting and ending at height $s_0=s_N=0$.
Call their generating function $f_l(z)$. A walk is either empty (one walk for $N=0$) or possesses a first step, which is
necessarily up. This step is ``balanced'' by the first down step that takes the walk back to height $0$.
In between this pair of balancing steps one has another walk with floor at height $1$, and the
balancing down step is followed by a walk with the original specifications. Here's an example of such a decomposition:
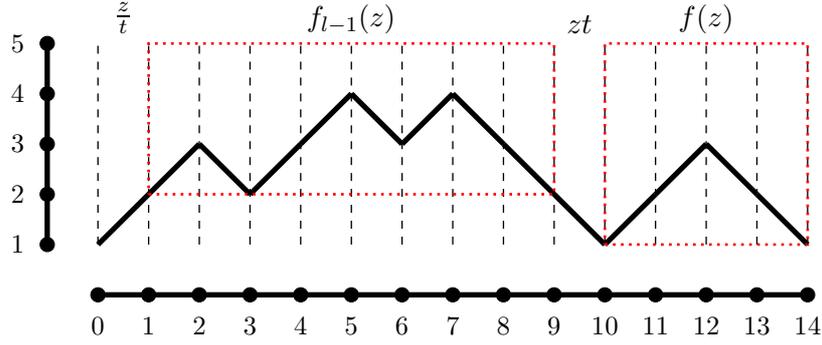
\begin{equation*}
\begin{tikzpicture}[scale = 2/3]
%Path
	\foreach \r [count = \i ]in {0, 1, 2, 1, 2, 3, 2, 3, 2, 1, 0, 1, 2, 1, 0}{
		\coordinate (S\i) at (\i,\r + 1 );
	}
	\foreach \i [evaluate=\i as \x using \i+1] in {1,...,14}{
		\draw[black, line width = 2pt] (S\i) -- (S\x);
	};
%Side Graph
\draw[black, line width = 2pt] (0,1) -- (0,5); 
\foreach \r in {1,...,5}{
	\filldraw[black] (0,\r) circle (4 pt);
	\node[anchor = east] at (-.25,\r) {\small{\r}};
	};
%Bottom scale
\draw[black, line width = 2pt] (1,0) -- (15,0); 
\foreach \r in {0,...,14}{
	\draw[black, dashed, line width = .5pt] (\r +1,1) -- (\r +1, 5);
	\filldraw[black] (\r + 1,0) circle (4 pt);
	\node[anchor = north] at (\r +1,-.25) {\small{\r}};
	};
%Boxes
	\draw[red,dotted, line width = 1pt] (S2) -- (S10) -- (10,5) -- (2,5) -- (S2); 
	\draw[red,dotted, line width = 1pt] (S11) -- (S15) -- (15,5) -- (11,5) -- (S11);
%Math
	\node[anchor = south] at (1.5,5) {$\frac{z}{t}$};
	\node[anchor = south] at (6,5) {$f_{l-1}(z)$};
	\node[anchor = south] at (10.5,5) {$z t$};
	\node[anchor = south] at (13,5) {$f(z)$};
\end{tikzpicture}
\end{equation*}

Note that the first up step combined with its balancing down step give the weight $zt\cdot z/t = z^2$. Altogether, we get a recursion in $l$:
% for the generating functions:
%
\begin{equation}
 f_l(z) = 1 + z^2 f_{l-1}(z) f_l (z)
\end{equation}
with solution
\begin{equation}
 f_l(z) = \frac{1}{1 - z^2 f_{l-1}(z)} \quad \mbox{ for } l \ge 1 \quad 
 \mbox{,\quad and }  f_0(z) = 1 \,.
 \label{sol_f}
\end{equation}

In the second stage we consider walks starting at height $0$ and ending at any height $0\leq s_N\leq l$.
Let $g_l(z,t)$ be the corresponding generating function. If the walk is non-empty, there is
a first up step. This step can either be balanced (if the walk returns to height $0$ at some point)
or unbalanced (if the remainder of the walk stays above height $1$). Therefore, we have the recursion
\begin{equation}
 g_l(z,t) = 1 + z^2 f_{l-1}(z) g_l(z,t) + z t g_{l-1}(z,t)
\end{equation}
with solution
\begin{equation}
 g_l(z,t) = \frac{1 + z t g_{l-1}(z,t)}{1 - z^2 f_{l-1}(z)} \quad  \mbox{ for } l \ge 1
 \quad \mbox{, \quad and } g_0(z,t) = 1 \,.
 \label{sol_g}
\end{equation}

In the third and last stage we consider walks starting at height $s_0=m$ and ending at any height $0\leq s_N\leq l$.
Let $h_{l,m}(z,t)$ be their generating function. We have the boundary conditions
\begin{equation}
 h_{l,0}(z,t) = g_l(z,t) \quad  \mbox{ and } \quad 
 h_{l,l}(z,t) = g_l(z,t^{-1}) \,,
 \label{init_h}
\end{equation}
where in the second identity we have made a reflection so that up and down steps are interchanged
(i.e., $t \to t^{-1}$). If the walk is non-empty, and if $m \neq 0,l$, there are now four possibilities for the first step: it can be
up and balanced, up and unbalanced, down and balanced, or down and unbalanced. For an unbalanced
up step, the floor for the remainder of the walk is at height $m+1$, hence the reminder is described by $g_{l-m-1}(z,t)$.
For an unbalanced down step, the ceiling for the remainder of the walk is at height $m-1$, and using the
reflection argument  $t \to t^{-1}$,  the remainder  is described by $g_{m-1}(z,t^{-1})$. Therefore
\begin{eqnarray}
 h_{l,m}(z,t) &=& 1 + z^2 f_{l-m-1}(z) h_{l,m}(z,t) + z t g_{l-m-1}(z,t) \nonumber \\
 & & \ \; \, + z^2 f_{m-1}(z) h_{l,m}(z,t) + z t^{-1} g_{m-1}(z,t^{-1})
\end{eqnarray}
with solution
\begin{equation}
 h_{l,m}(z,t) = \frac{1 + z \left[ t g_{l-m-1}(z,t) + t^{-1} g_{m-1}(z,t^{-1}) \right]}
  {1 - z^2 \left[ f_{l-m-1}(z) + f_{m-1}(z) \right]} \quad \mbox{ for }  m \neq 0,l \,.
 \label{sol_h}
\end{equation}

\newcommand{\gk}[1]{g_{l}^{(#1)}}
\newcommand{\gkN}[2]{g_{l,#2}^{(#1)}}
In order to find a closed expression for~\eqref{sol_h} we express the generating functions $g_l(z,t)$ and $f_l(z)$ in terms of Chebyshev polynomials.
Let us introduce   the generating functions
\begin{equation}
g_l(z,t) =  \sum_{k=0}^{l}\gk{k}(z) t^k = \sum_{k=0}^{l} \sum_{N\geq0}\gkN{k}{N} z^N t^k\ ,
\end{equation}
where the numbers $\gkN{k}{N}$ in the expansion are by definition the numbers of RSOS paths with $s_0=0$ and $s_N=k$. 
 They satisfy the recursion relation
\begin{equation}\label{recurs-dim}
\gkN{k}{N} =\gkN{k+1}{N-1} + \gkN{k-1}{N-1}  
\end{equation}
with the  boundary conditions
\begin{gather}
\gkN{l+1}{N}=0,\qquad \gkN{-1}{N}=0,\qquad N\geq0,\label{in-cond-1}\\
 \gkN{0}{0}=1, \qquad \gkN{k}{0}=0,\qquad k>0.\label{in-cond-2}
\end{gather}
 The recursion relation~\eqref{recurs-dim}
 is equivalent to
\begin{equation}\label{recurs-pol}
z^{-1} \gk{k}(z) = \gk{k-1}(z) + \gk{k+1}(z)\ .
\end{equation}

Any function satisfying~\eqref{recurs-pol} with~\eqref{in-cond-1}-\eqref{in-cond-2} should be a solution of our problem. We set  
\be
V_k(z):= U_k\big(\tfrac{1}{2z}\big)\ ,
\ee
 where  $U_k(z)$ are the second-type Chebyshev polynomials  with 
$U_{-1}=0$, $U_0=1$, $U_1=2z$, $U_2=4z^{2}-1$, {\it etc.}, and we set $V_k:=V_k(z)$, i.e.\   as above~\eqref{eq:path-gen-fun}. The recursion~\eqref{recurs-pol} determines the Chebyshev polynomials $V_k(z)$.
Then,
  using the conditions~\eqref{in-cond-1} and~\eqref{in-cond-2} we can fix $\gk{k}(z)$ as the ratio of the Chebyshev polynomials
\begin{equation}\label{TL-gen-func}
\gk{k}(z) = \frac{V_{l-k}(z)}{zV_{l+1}(z)},\qquad k\geq0.
\end{equation}
Note that $f_l(z) = \gk{k}(0)$. Then the expression in~\eqref{sol_h} takes the final closed form 
 in terms of the Chebyshev polynomials:
\begin{equation}
h_{l,m}(z,t) = \frac{1}{z} \frac{\sum_{k=-m}^{-1}V_{l-m}V_{m+k}t^k + \sum_{k=0}^{l-m}V_m V_{l-m-k}t^k}{V_{l-m+1}V_{m}-V_{l-m}V_{m-1}}.
\end{equation}
Then we note that the generating functions $P^{(n)}_k(z)$ introduced in~\eqref{eq:path-gen-fun-def} in terms of the original heights $1\leq h_i\leq p$ are then expressed as
\begin{equation}
h_{l,m}(z,t) = \sum_{k=-m}^{l-m}P^{(m+1)}_k(z) t^k
\end{equation}
and so we finally obtain for any $0\leq m\leq l=p-1$ (or $1\leq n\leq p$)
\begin{equation}\label{dim-blob-fun}
P^{(m+1)}_k(z) = \frac{1}{z} \Biggl(\frac{V_m V_{l-m-k}}{V_{m}V_{l-m+1}-V_{m-1}V_{l-m}}\Biggr),\qquad 0\leq k\leq l-m,
\end{equation}
while the expression for negative values of $k$ is just given by replacing $m\to l-m$ and $k\to-k$ in~\eqref{dim-blob-fun}.
 This is the expression claimed in~\eqref{eq:path-gen-fun}.

We note that the term
\begin{equation}
 [ z^N t^{h_N-h_0} ] \, h_{l,m}(z,t)
\end{equation}
of the combinatorial problem considered above in Sec.~\ref{sec:bndRSOS} is equal to $M_{p,N}(h_0,h_N)$ from~\eqref{eq:Mpnab}, with $p=l+1$ and $h_0=m+1$ (because of the convention in this
 section that the ceiling is situated at height 0, not 1).
The equivalence between the two approaches provides some rather interesting identities on generating functions.

We can finally give the generating function for the total number of periodic paths in the RSOS model. It is given by the sum of open RSOS paths with $s_0=s_N=m$:
\begin{equation}\label{per-fun}
\text{\# of periodic RSOS paths} = \frac{1}{z}\sum_{m=0}^{l} \frac{V_m V_{l-m}}{V_{m}V_{l-m+1}-V_{m-1}V_{l-m}},
\end{equation}
where each term is not changed under the transformation $m\to l-m$, which is obvious from the blob algebra point of view. The results in~\eqref{dim-blob-fun} and~\eqref{per-fun} seem to be new.

\section{Inverse braid translation}\label{app:inv-brt}
Here, we remark on a relation between the braid translation of open $\gl(1|1)$ 
and XXZ spin-chains. It becomes clear in this section that the locality issues observed in Sec.~\ref{sec:gl11-lim} can be solved paying certain price for locality of boundary operators. Though our blob operators~\eqref{XXZ-blob} for each sector in the open XXZ chain, as we will see below, are expressed non-locally in terms of spin matrices they give {\sl local} expressions for the periodic generator $e_{N}$, with  correct chiral and anti-chiral actions after the braid translation. \textit{What we also learn from the example of XX models treated below that this non-locality of $b$ is crucial only for indecomposability and for logarithmic structure of the models.}

\subsection{Twisted XXZ chains}\label{sec:XXZ}
Recall our definition of the  XXZ twisted spin-chain in~\eqref{e_eiXXZ} and~\eqref{e_eiXXZ2}, with the Hamiltonian~\eqref{Ham-twist}.
We  now introduce an open spin-chain, as a representation of the blob algebra, that would give us the closed twisted chain after the
braid translation
$\brt: \ATL{N}(\delta) \longrightarrow \mathcal{B}_N(n,y)$ introduced in~\eqref{eq_defbraid}. Using the equation~\eqref{eq_braid_trans}, we define a representation of the blob algebra on the spin-chain, strictly speaking within a given $S_z$ sector, as
\begin{equation}\label{XXZ-blob}
b = \ffrac{1}{\beta}\Biggl((-1)^{N}\q^{-N}\sqrt{\frac{y-\q^{-1}}{y-\q}}e^{i\frac{\phi}{2}\sigma^z_1}s_1 s_2\dots s_{N-1}g^{-1}_{N-1}\dots g^{-1}_1 -1 \Biggr),
\end{equation}
where we used our translation operator~\eqref{XXZ-trans} and $g_i$'s were introduced in~\eqref{hecke-gen} and $\beta$ in~\eqref{alpha-beta}. The blob algebra relations~\eqref{Blobdef} with $b$ given by~\eqref{XXZ-blob} are satisfied in each $S_z=j$ sector for appropriate $y=y_j$. This parameter for the blobbed loops can be easily computed using the correspondence between the standard modules for even $N$, and so $j$ is integer (compare with~\eqref{corr:st})\footnote{The odd $N$ case can be treated similarly.}:
\begin{align}
\bSt^b_j \;& \xrightarrow{\quad \brt\quad} \;\aSt_{j,\q^{2j}e^{2i\eta}}\\
\bSt^u_j \;& \xrightarrow{\quad \brt\quad} \;\aSt_{j,\q^{2j}e^{-2i\eta}}\ .
\end{align}
This gives us for any $S_z=j$
\begin{equation}
y_j = \frac{\q^{-1} - \q^{1-2j}e^{i\phi}}{1-\q^{-2j}e^{i\phi}}.
\end{equation}
So the full twisted, or in particular periodic, XXZ model can be obtained by braid translating $S_z=j$ sectors in the open chains corresponding to different ``boundary conditions'' labeled by $y_j$, or the corresponding sectors of the the blob algebra $\mathcal{B}_N(n,y_{j})$ representations with the blob generator $b$ from~\eqref{XXZ-blob}.
Conversely, we can define the twist in the $S_z$-dependent way
\begin{equation}\label{twist-Sz}
e^{i\phi} = \q^{2S_z}e^{2i\eta}.
\end{equation}
Then, just one blob algebra  with the fixed $y$ given by~\eqref{y-eta} and the blob $b$ given by~\eqref{XXZ-blob}  gives an open spin-chain, where its blobbed standard modules form  $S_z>0$ sectors and unblobbed ones are at $S_z<0$. By the construction, such open chain is mapped by $\brt$ to the twisted XXZ representation 
of $\ATL{N}$ defined in Sec.~\ref{sec:twistedXXZ}. In particular $e_N^{\brt}$ is given by $e_N$ from~\eqref{eq:eN-XXZ}. We therefore see that one can at least formally define an open spin chain with non-local (in terms of the Pauli matrices) expression for the blob generator $b$ that is braid translated to  the twisted closed chain~\eqref{Ham-twist} with the twist $\phi$ from~\eqref{twist-Sz}. 

We also note that at roots of unity $\q=e^{\frac{i\pi}{x+1}}$ we will have only $(x+1)$ different $y_j$'s. 
In terms of structure of representations, at roots of unity each $S_z$-sector  is an indecomposable module with a complicated structure that differs very much from the one of the  standard modules. We will discuss this point below for the free fermionic case. 

\subsection{Twisted XX chains}\label{sec:XX}
We note that the periodic  $\gl(1|1)$ spin chain is equivalent to the twisted XX spin chain (or twisted XXZ at $\q=\rmi$)  with the $S_z$-dependent twist~\cite{GRS1}:
 the twisted  conditions are  $e^{i\phi}=\q^{2(S_z+1)}=(-1)^{S_z+1}$. The equation~\eqref{twist-Sz} then gives $e^{2i\eta}=\q^2$ or $y=0$. It tells us that this twisted closed XX chain can be obtained as the braid translation of an open XX chain considered as a blob algebra representation with the parameter $y=\q+\q^{-1}=0$. The action of the blob operator~\eqref{XXZ-blob} is now given by the expression
\begin{equation}\label{b-XX}
b = \ffrac{\rmi^{N+1}}{2}e^{\rmi\frac{\phi}{2}\sigma^z_1}s_1 s_2\dots s_{N-1}g^{-1}_{N-1}\dots g^{-1}_1 +\ffrac{1}{2},
\end{equation}
where we recall $g^{\pm1}_j = 1 \pm \rmi e_j$. (Note that the first term here ``measures'' the difference of braid permutations from being the usual permutations.) It is then straightforward to see that $\brt$ indeed gives the standard expression for $e_{N}$, {\it i.e.}
\begin{equation}\label{gen-per-TL-ferm}
\displaystyle e_{N}^\brt = e_{N} = (f_{N}+ f_{1})(f^\dag_{N} + f^\dag_{1})
\end{equation}
in terms of free fermions used previously in Sec.~\ref{sec:gl11chain}. The price we paid is non-locality of the expression of the blob operator in terms of the spin-matrices or lattice fermions. This non-locality is  in contrast to the case with RSOS models discussed in Sec.~\ref{sec:minmodels}.

In Sec.~\ref{sec:gl11chain},
 we took the $\gl(1|1)$ spin-chain with the trivial action of $b=\one$ and the same parameter $y=0$. It was a different representation (from the one in~\eqref{b-XX}) of the same blob algebra that factorizes through the ordinary TL algebra.
 We see now that in our present case we also have $y=0$ but $b$ is not the identity. The spectrum in both cases is the same and even the contents of simple blob algebra modules are identical. The crucial point is that the non-triviality or non-locality of the new expression for $b$ gives those additional arrows in the structure of indecomposable modules that were missing in Sec.~\ref{sec:gl11chain}. In the limit this structure, after applying $\brt$ of course, corresponds to the  correct non-semisimple action of the (anti)chiral Virasoro $\VirN$. We can thus conclude here that the non-local action of the blob operator we introduced is  restoring the locality of the bulk theory obtained by the braid translation, compare with Sec.~\ref{sec:non-loc}. It is an interesting question if such non-locally defined boundary/blob operator has a physical meaning.

\bibliographystyle{siam}

\end{document}